\documentclass[sigconf]{acmart}

\newcommand{\beginFigureNarrow}{
  \ifx\singleCol\undefined
  \else
    \begin{minipage}{3.30in}
  \fi
}
\newcommand{\finishFigureNarrow}{ %
  \ifx\singleCol\undefined
  \else
    \end{minipage}
  \fi
}

\graphicspath{{figures/}{pictures/}{images/}{../}} %

\usepackage{soul}
\usepackage{enumitem}
\setlist{leftmargin=0.25in, rightmargin=0.2in}

\usepackage{xcolor}
\definecolor{p5pink}{HTML}{DA3B5F}
\definecolor{y1Color}{HTML}{66C2A5}
\definecolor{y2Color}{HTML}{FC8D62}

\usepackage{array}

\AtBeginDocument{%
  }

\copyrightyear{2023}
\acmYear{2023}
\setcopyright{acmlicensed}\acmConference[CHI '23]{Proceedings of the 2023 CHI Conference on Human Factors in Computing Systems}{April 23--28, 2023}{Hamburg, Germany}
\acmBooktitle{Proceedings of the 2023 CHI Conference on Human Factors in Computing Systems (CHI '23), April 23--28, 2023, Hamburg, Germany}
\acmPrice{15.00}
\acmDOI{10.1145/3544548.3580683}
\acmISBN{978-1-4503-9421-5/23/04}

\begin{document}

\title
{A Study of Editor Features in a Creative Coding Classroom}

\author{Andrew McNutt}
\orcid{0000-0001-8255-4258}
\affiliation{%
  \institution{University of Chicago}
  \city{Chicago}
  \state{IL}
  \country{USA}
  \postcode{60615}
}
\author{Anton Outkine}
\affiliation{%
  \institution{University of Chicago}
  \city{Chicago}
  \state{IL}
  \country{USA}
  \postcode{60615}
}
\author{Ravi Chugh}
\affiliation{%
  \institution{University of Chicago}
  \city{Chicago}
  \state{IL}
  \country{USA}
  \postcode{60615}
}

\renewcommand{\shortauthors}{McNutt et al.}

\newcommand{\toolName}[1]{#1}
\newcommand{\sns}{\toolName{Sketch-n-Sketch}}

\newcommand{\osf}{\url{https://osf.io/cture/}}

\newcommand{\etal}
{et al.}
\newcommand{\etals}
{et al.'s}

\newcommand{\ie}{{i.e.}}
\newcommand{\eg}{{e.g.}}
\newcommand{\etc}{{etc.}}
\newcommand{\cf}{{cf.}}

\ifx\hideComments\undefined
  \newcommand\rc[1]{{\color{blue}[RC: #1]}}
  \newcommand\am[1]{{\color{red}[AM: #1]}}
  \newcommand{\todo}[1]{\textcolor{magenta}{\textsf{{#1}}}}
\else
  \newcommand\rc[1]{{}}
  \newcommand\am[1]{{}}
  \newcommand{\todo}{{}}
\fi

\newcommand{\authorVersion}{HIDE EM} %
\ifx\authorVersion\undefined
  \newcommand\spaceit[0]{\vspace{-0.2in}}
\else
  \newcommand\spaceit[0]{{}}
\fi

\newcommand{\theAppendix}{the appendix}

\newcommand{\secref}[1]{\hyperref[#1]{Sec.~\ref*{#1}}}
\newcommand{\appendixref}[1]{\hyperref[#1]{Appendix.~\ref*{#1}}}
\newcommand{\figref}[1]{\hyperref[#1]{Fig.~\ref*{#1}}}
\newcommand{\eqnref}[1]{\hyperref[#1]{Eqn.~\ref*{#1}}}
\newcommand{\tabref}[1]{\hyperref[#1]{Table ~\ref*{#1}}}

\newenvironment{mycenter}[1][\topsep]
{\setlength{\topsep}{#1}\par\kern\topsep\centering}%
{\par\kern\topsep}

\newcommand{\footnoteWithIndent}[1]
{\footnote{\mbox{}\hspace{0.01in}{#1}}}

\newcommand{\imm}{\emph{su21}}
\newcommand{\csp}{\rc{REMOVE \emph{su21-lite}}}
\newcommand{\spq}{\emph{sp21}}
\newcommand{\wiq}{\emph{wi22}}
\newcommand{\sumtwo}{\emph{su22}}

\newcommand{\nEquals}[1]
{($n$=#1)}
\newcommand{\numYearOneLog}{39}
\newcommand{\numYearOneSurvey}{25}
\newcommand{\numYearTwoLog}{39+}
\newcommand{\numYearTwoSurvey}{23}

\newcommand{\pFiveOne}
{\textbf{\color{y1Color}{p5/y1}}}
\newcommand{\pFiveTwo}
{\textbf{\color{y2Color}{p5/y2}}}
\newcommand{\pA}[1]
{{\color{y1Color}{\textbf{A#1}}}}
\newcommand{\pB}[1]
{{\color{y1Color}{\textbf{B#1}}}}
\newcommand{\pC}[1]
{{\color{y2Color}{\textbf{C#1}}}}
\newcommand{\pD}[1]
{{\color{y2Color}{\textbf{D#1}}}}

\newcommand{\pQuote}[1]
{\textit{``#1''}}

\newcommand{\parahead}[1]
{%
  \vspace{0.07in}%
  \noindent%
  \textbf{\textit{#1.}}%
}

\def\subsubsec#1
{\subsubsection{#1}}

\newcommand{\inlineFig}[1]{%
  \begingroup\normalfont
  \includegraphics[height=1.2\fontcharht\font`\B]{#1}%
  \endgroup
}

\newcommand{\takeaway}[1]
{  \begingroup\normalfont
  \includegraphics[height=1.3\fontcharht\font`\B]{#1}%
  \endgroup}

\newcommand{\takeawayStatic}
{\takeaway{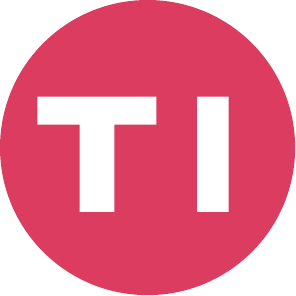}}

\newcommand{\takeawayLive}
{\takeaway{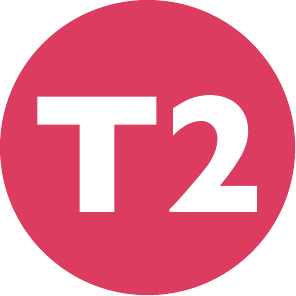}}

\newcommand{\takeawayClutter}
{\takeaway{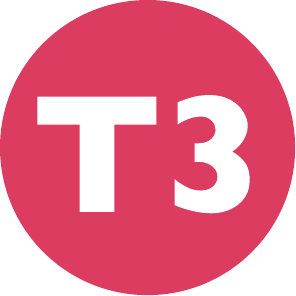}}

\newcommand{\takeawaySkeptic}
{\takeaway{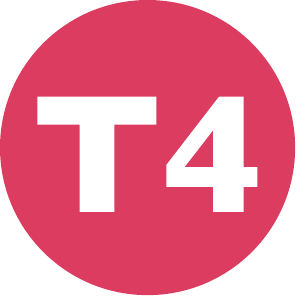}}

\newcommand{\takeawayGeneric}
{\takeaway{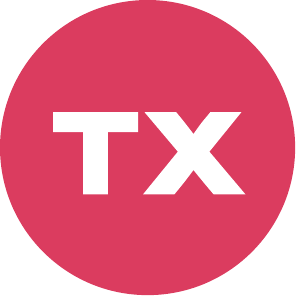}}

\setuldepth{Berlin}
\definecolor{linkColor}{HTML}{257E98}
\newcommand\asLink[2]{\textcolor{linkColor}{\href{#1}{\ul{#2}}}}

\begin{abstract}%
      Creative coding is a rapidly expanding domain for both artistic expression and computational education. Numerous libraries and IDEs support creative coding, however there has been little consideration of how the environments themselves might be designed to serve these twin goals. To investigate this gap, we implemented and used an experimental editor to teach a sequence of college and high-school creative coding courses. In the first year, we conducted a log analysis of student work  \nEquals{\numYearOneLog} and surveys regarding prospective features  \nEquals{\numYearOneSurvey}. These guided our implementation of common enhancements (e.g. color pickers) as well as uncommon ones (e.g. bidirectional shape editing). In the second year, we studied the effects of these features through logging \nEquals{\numYearTwoLog} and survey \nEquals{\numYearTwoSurvey} studies. Reflecting on the results, we identify opportunities to improve creativity- and novice-focused IDEs and highlight tensions in their design---as in tools that augment artistry or efficiency but may be perceived as hindering learning.
\end{abstract}



\begin{teaserfigure}
    \includegraphics[width=\linewidth]{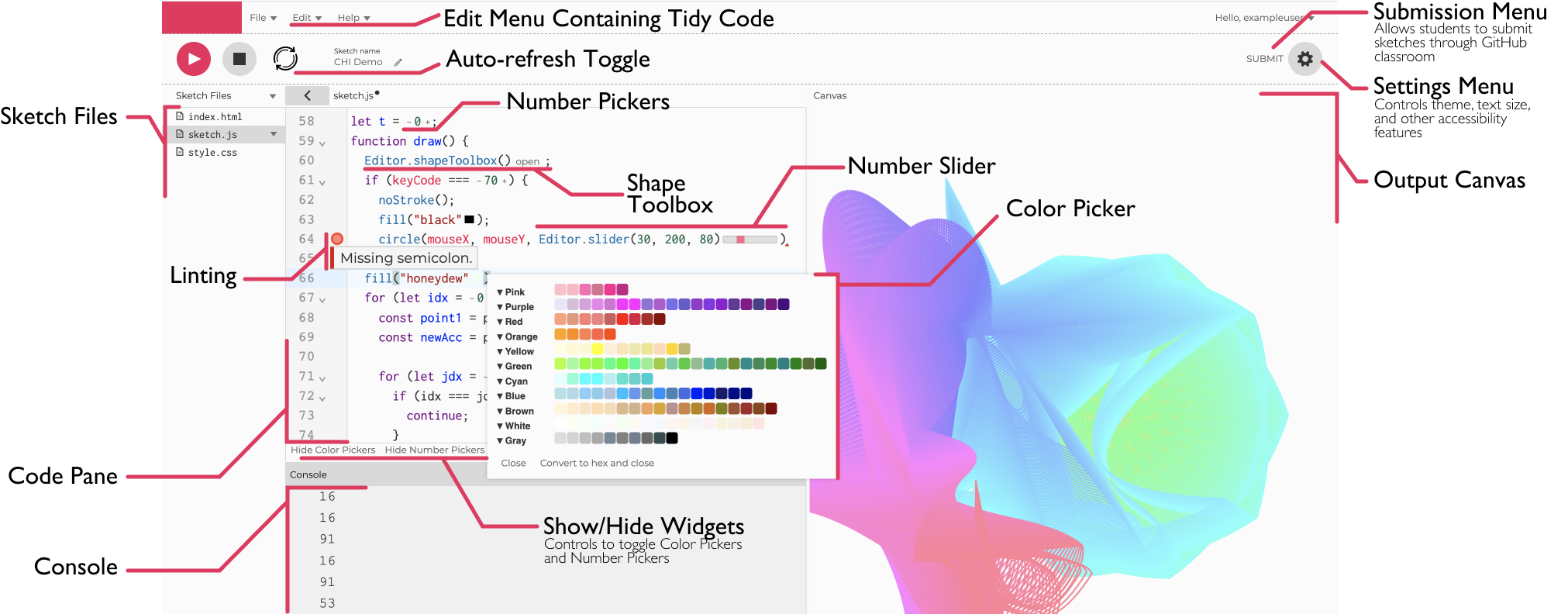}
    \caption{
        This modified p5 editor (dubbed \pFiveTwo{}) was used in a creative coding course to study how students use and perceive various editor features including standard ones, such as linting and auto-formatting (``Tidy Code''), as well as more experimental features, such as live reloading (``Auto-refresh'') and a toolbox for bidirectionally manipulating shapes.
    }
    \Description{An annotated screenshot of the augmented p5 editor. Annotations highlight editor features, including Code Pane, Console, Output Canvas, Color Picker, and others. The editor is running code displayed in the Code Pane and shows a multi-colored shape.
    }
    \label{fig:annotated-ui}
\end{teaserfigure}

\maketitle

\section{Introduction}
\label{sec:intro}

Creative coding is a rapidly expanding computational domain.
It generally refers to programming work that ``blur(s) the distinction between art and design and science and engineering''~\cite{levin21Code},
encompassing pursuits such as generative art, embedded computing, audio editing, performative live programming, and countless others.
Many libraries and languages have arisen to support this programming genre.
Some are tuned to domain-specific purposes---such as Orca~\cite{Orcas} or Tracery~\cite{compton2015tracery} which support creating procedural music and Twitter bots, respectively.
Others simplify the process of many common artistic tasks (such as drawing and interactivity) without specializing in a specific area---as in openFrameworks~\cite{openFrameworksForward} or Processing~\cite{reas2007processing}.
Among these general-purpose tools, those in the Processing family---such as Processing itself and p5.js~\cite{p5}---are particularly well known, having attracted large and active communities, exemplified by the prevalence of artist- and novice-focused educational media, like the Coding Train~\cite{CodingTrain}.
Beyond the potential for creative or artistic expression, this genre of work has long been embraced as means by which to teach introductory programming~\cite{levin21Code, peppler2005creative, wood2016computational, Guzdial2005}---%
an approach often referred to as \emph{media computation}~\cite{MediaCompTeach} within CS departments---%
as it may be easier for students to engage with material that interests them~\cite{ambrose2010learning}, and creative or artistic tasks may be more engaging to students~\cite{Malita20From} not invested in the more common CS Ed topics.
\citet{greenberg2012creative} argue that creative coding-based introductions to computer science are more appealing to women, and create a more inclusive environment than traditional introductory CS curricula.
Despite the potential utility for both artistic expression and learning to code, there has been relatively little consideration of how to enhance creative coding environments to facilitate these goals.
Following a trend exemplified by the development environment bundled with the Processing library, a number of creative coding toolchains come with their own environments, which are often tailored specifically for artists in their domain, as in Orca~\cite{Orcas} or Tweakable~\cite{Tweakable}.
For example, the p5 editor---a browser-based editor maintained by the p5 community that acts as a gateway to coding for often non-technical users---is intentionally simple, and has limited ``features and frills'' to make it easier to jump right into coding~\citep{p5editor}.
By definition, however, this guiding principle forgoes potential benefits of many standard IDE features (such as autocomplete), standard GUI features (such as color pickers), and more experimental features explored in research communities (such as bidirectional editing).
To shed light on the gap between creative coding tools and their goals for users, this work considers the following questions:
\emph{How might we refine and enhance standard tools to extend the creative reach of novices?
      What sorts of features do novices perceive to be beneficial in a creative coding environment?}
We consider these questions in the setting of the p5 editor because of its ubiquity~\cite{levin21Code} in creative coding contexts,
as well as for its simple and mostly standard form, which may inform the design of enhancements to more general-purpose programming tools.

\begin{figure}[t]
\beginFigureNarrow
\includegraphics[width=\linewidth]{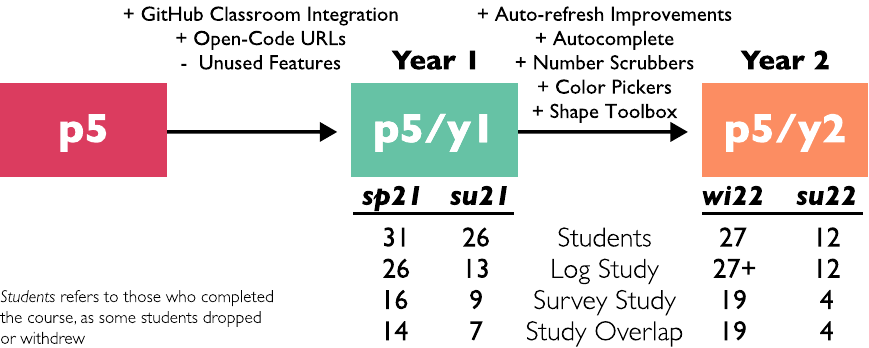}
\caption{
    We modified the p5 editor before each year of a creative coding course.
    We conducted studies to observe student usage and perception of existing and modified features.
}
\Description{ A flow chart showing the progression of the three editor versions. Each line is annotated with medications made from the previous version. Beneath the Year 1 and Year 2 editors is a table describing the number of students in the course and their participation in each of the studies.}
\label{fig:timeline}
\finishFigureNarrow
\end{figure}

\paragraph{Paper Summary}

We conducted a series of studies that considered how students use a modified version of the p5 code editor in an introductory programming and creative coding course at
a private research university in the US.
\figref{fig:timeline} displays an overview of our work, involving four course offerings spanning two academic years
taught to college students (\spq{}, \wiq{}) and to high-school students (\imm{}, \sumtwo{}).
While our studies are situated in a classroom, our work is not about pedagogy per se---rather, we focus on understanding the needs and perceptions of creative coding novices as exhibited across the length of a full programming course.
We used a modified version of the p5 editor (referred to as \pFiveOne{}) in the first-year courses (\spq{} and \imm{}) and ran two studies.
The first was a log analysis based on capturing code executions during the course \nEquals{\numYearOneLog}.
The second was a long-form survey that sought to understand student opinions and expectations about existing and hypothetical editor features \nEquals{\numYearOneSurvey}.

The results of the first-year studies revealed opportunities to improve existing editor features as well as interest in several hypothetical features---including direct manipulation widgets for modifying colors and a bidirectional shape drawing system.
Thus, we further augmented the p5 editor (\pFiveTwo{}) in the second-year offerings (\wiq{} and \sumtwo{}).
Analogous to the studies in the first year, we monitored student behavior \nEquals{\numYearTwoLog} through an anonymized tracking scheme,\footnoteWithIndent{We did not collect user identifiers in the Year 2 log study. Thus, ``\numYearTwoLog{}'' indicates that the log data includes students who did not complete the course.} and solicited their opinions through an abbreviated version of our previous survey \nEquals{\numYearTwoSurvey}.

We identify several key themes based on the results of the studies.

\newcommand{\itemTakeaway}[2]
{\vspace{0.05in} %
      \subparagraph*{\makebox[0.18in][l]{#1}\textbf{#2}}
}

\newcommand{\iconSep}{\hspace{0.01in}}

\begin{description}[leftmargin=0cm, rightmargin=0cm, itemsep=0.4em]
      \item[\takeawayStatic{}\iconSep{} Simple static analysis seen as supportive.]

            Tools supporting basic automated formatting and analysis---such as code ``tidying'' or  linting---are well received by our novices. However, impolite~\cite{whitworth2005polite} designs (which are those that do not respect user agency or act in an otherwise irritating manner) can lead to frustration.

      \item[\takeawayLive{}\iconSep{} Overeager evaluation can overwhelm.]

            Live programming can give immediate feedback on code changes---potentially beneficial for tightening art-making cycles---but it often does so too quickly or in an irritating manner.

      \item[\takeawayClutter{}\iconSep{} Students appreciate avoiding clutter.]

            Like all programmers, novice creative coders are sensitive to inherent tradeoffs between minimal and feature-rich coding environments.

      \item[\takeawaySkeptic{}\iconSep{} Useful features may be ``too useful.'']

            Students were receptive to integrating art-specific and other sophisticated tools into their programming environment. Yet
            such features can inspire skepticism---even by novices---about their effect on learning.

\end{description}

Next, we situate our study within related work (\secref{sec:related}),
and then we describe our creative coding course (\secref{sec:context}).
After describing our methodology (\secref{sec:methods}), we analyze the results and consider our primary themes (\secref{sec:results}).
Through these studies we identify design implications for subsequent creativity- and novice-focused IDEs.
\section{Related Work}
\label{sec:related}

This paper investigates how to
integrate advanced editor techniques into tools focused on novices and creative purposes based on
observation and analysis of novice programmers. %
Given the broad range of related works, we frame the discussion around our primary design decisions:
to start with the p5 editor and its existing feature set (\secref{sec:creative-coding-environments}),
to add a suite of more advanced features (\secref{sec:advanced-features}), and
to evaluate these adaptations in a classroom setting (\secref{sec:classroom-studies}).

\subsection{Creative Coding Environments}
\label{sec:creative-coding-environments}

Creative coding is a multifarious category of work encompassing diverse approaches and topics.
One common element is the use of editing environments that have been customized to address the particular domain of consideration.

One prominent example is p5.js~\cite{p5} which (like its predecessor Processing~\cite{reas2007processing}) can be used as a standalone library, but is made substantially more approachable by novices though the availability of a simple development environment specific to doing work with that library.
The p5 editor~\cite{p5editor} simulates a simple web server in the browser by combining each of the files in a ``sketch'' (synonymous with program in this context) and executing them as a standalone web page in an isolated component.
The existing feature set in this editor is a particularly intriguing object for study for several reasons.
First, it is lightweight, web-based, and supports cloud-based saves and shares---a good fit for an introductory programming class, as it does not have potentially intimidating baggage of a heavyweight IDE.
Second, the text editor (based on CodeMirror~\cite{codeMirrorSix}) supports a number of contemporary IDE features---such as linting~\cite{jsHint} and auto-formatting~\cite{jsBeautify}---that make our findings potentially generalizable.
Furthermore, it contains a live-reloading (``auto-refresh'') feature being actively researched in programming-language user-interface communities~\cite{rein2018exploratory,selvaraj2021live}.
By considering (versions of) the existing, relatively standard p5 editor, our formative study aimed to understand which features were important before pursuing more drastic changes within the scope of this work and beyond.

In addition to these more general editors, there are a variety of tools that focus on more limited domains.
For instance, Shadertoy~\cite{shaderToy} provides a browser-based editor for creating and sharing shaders and prominently features procedural and generative visual art.
Like creative coding in general, these editing environments are not limited to the graphical domain.
Tweakable~\cite{Tweakable} and Orca~\cite{Orcas} provide environments for creating programmatically generated music, based on node-and-wire composition and 2D livecoding, respectively.
HappyBrackets~\cite{fraietta2019rapid} more closely reflects our approach to enhancing creativity by augmenting a standard IDE, although it is centered on using IoT devices for musical composition.
MakeCode~\cite{ball2019microsoft} has an editing environment that contains synchronized block and text representations of code with a focus on creating games for microcontroller-based devices.
Most similar to our work, p5.fab~\cite{subbaraman2022p5} modifies the p5 editor to support digital fabrication.

\citet{mitchell2013towards} studied the needs of creative coders through a lab-based study, highlighting the value of visualizing program state, supporting best practices and short iteration cycles, and assisting exploration.
Our findings are closely related to theirs, but located within a classroom and conducted on a longer time-scale---following Frich \etals{}~\cite{frich2018twenty} call for more studies to evaluate extant tools in their \emph{in-vivo} usage context.
This scale and scope informs our different, but complementary, set of themes.

Creative coding IDEs, and other such tools, target users at an intriguing intersection: many are relatively inexperienced but are strongly motivated to use these systems effectively.
Lessons learned from studying users of these systems (\eg{} students in a creative coding classroom) may translate to other venues with non-technical high-engagement users.
Such populations occur widely and include spreadsheet users~\cite{bartram2021untidy}, tinkerers~\cite{Beckwith06Tinkering, Burnett16Finding}, and artists more generally.

\subsection{Advanced Editor Features}
\label{sec:advanced-features}

Many works have experimented with new ways to augment conventional text-based code editors with more interactive capabilities.
Among the plethora of such features, we chose several to consider in this work.
In the first year of our study, students had access to a \textbf{\emph{live-reloading}} feature in the existing p5 editor.
In the second year, we implemented domain-specific \textbf{\emph{graphical widgets}} (namely, color picker and number sliders), and \textbf{\emph{bidirectional shape drawing}}---among many other advanced features being actively researched---because they
are closely aligned with the concerns of creative coding,
were well-received in the first-year survey, and
were feasible to implement given finite resources.
We discuss these features below.
Tanimoto \cite{Tanimoto1990,Tanimoto13live} describes programming affordances on a \emph{liveness} spectrum, relating to the degree of agency that users express in the execution of code. These range from the familiar edit-run cycle to predictive execution, with the always-executing style of live-reloading in the p5 editor falling in the middle.
Immediate feedback evidently has rich educational utility, as in Python Tutor~\cite{Guo13tutor}, and the sprawling number of systems its design informs~\cite{Guo21Million}.
Omnicode~\cite{kang2017omnicode} takes a ``Display all the values'' approach to help novices understand and debug code. They find that the always-on strategy is useful for these purposes, which agrees with Kramer \etals{}~\cite{kramer2014live} findings that live programming helps users fix bugs more quickly than a traditional edit-run cycle.
\citet{ProjectionBoxesInClass} also found that live programming helps students perform some tasks more quickly, but in their study learning outcomes remained unchanged.

Augmenting text with \emph{graphical} representations can provide a more natural way to specify code than textual input. The complexities of these vary from simple inline widgets, such as sliders and color pickers, to more complex designs.
Graphite~\citep{omar2012active} explores a notion of palettes which allow for domain-specific editors, such as for color and regular expressions, surfaced through autocomplete-style menus.
Barista~\citep{ko2006barista} integrates interactive structured visual representations inline with code.
\citet{andersen2020adding} build on this premise by formalizing how GUIs might be integrated directly into Racket code.
Several features we added to the p5 editor follow the hybrid textual-plus-visual approach found in these works, targeting our specific domain and audience, novice creative coding.

Some systems \emph{bidirectionally} synchronize code and GUI manipulations: changes made to either the source text or corresponding graphical output are reflected in the other~\cite{hempel_semi-automated_2016, Utopia, hempel2022maniposynth}.
While this has been most prominently used to create parametric drawings~\cite{hempel_sketch-n-sketch_2019,Glisp}, both ours and previous works suggest potential value for novices as well.
For instance, \citet{hundhausen2009can} find that this type of bidirectional development has educational utility and promotes skill transfer to text-based languages and environments.
Contrastingly, \citet{do2019evaluating} utilize a mixed text-and-direct manipulation approach to teach an Hour of Code course to 5th and 6th graders using a JavaScript-like language. Yet, they did not find as rich an educational benefit, but argue that further development is necessary to situate this UI paradigm in creative-educational contexts.
Our results tentatively suggest that this approach can be useful---in terms of student usage and perception.
However, further study is needed to understand the effect it has on learning.

\subsection{Classroom Studies}
\label{sec:classroom-studies}

Computer science education researchers have studied the potential benefits---regarding gender diversity, retention, and learning outcomes---of emphasizing computing with media in introductory programming courses~\cite{Guzdial2005,MediaCompUCSD,Guzdial2013}.
Our work provides a step toward understanding the role that programming \emph{tools}---as opposed to curricular design---might play in creative-educational settings.

Despite this work taking place in a classroom, however, our aims in this paper are not focused on measuring the pedagogical impact of individual editor features.
Instead, we pursue an understanding of the needs and perceptions of novice creative coders, and our classroom setting allows us to engage with such users on the time scale of an introductory programming course.
As \citet{ToBlockOrNot} argue, ``it is critical that we conduct studies ... analyzing tools not from the perspective of those who have already mastered the content, but instead from the perspective of the learners who the tools is designed for.''
Our work embraces this approach, deriving guidelines for IDE design based on perceived utility of different features ``to better inform educators on how to best utilize them in their classrooms ... [and] provide a roadmap for the improvement of these tools moving forward''~\cite{ToBlockOrNot}.

While we have specifically opted for a text-based environment,
block-based environments are notable for their frequent use with younger learners.
In a study with high school students in a formal classroom setting, \citet{ToBlockOrNot} found that students
(i) considered it easier to read programs as blocks rather than as text,
(ii) liked the visual cues offered by blocks (though this preference diminished over time),
(iii) found blocks easier to compose (via drag-and-drop), and
(iv) liked how the interface organized blocks into related functionality and helped serve as memory aid.
Their study also identified several perceived limitations of blocks: that they are potentially less powerful, slower and more verbose, and inauthentic---in the sense of not ``doing the same kinds of things they will do in `real life' outside of the environment in which learning takes place''~\cite{ThickAuthenticity}.
Notably, even though novice high school students appreciate the pedagogic value of blocks, they still perceive them as inauthentic.
This comports with our findings about skepticism about unfamiliar or advanced features among novices.

A variety of works have employed similar logging studies to ours (often referred to as learning analytics~\cite{ihantola2015educational}).
Some such works use in-course logging studies as a means by which to analyze student progression through assignments, which are described in multiple surveys~\cite{hundhausen2017ide, ihantola2015educational}.
\citet{helminen2013recording} used a similar environment as our own to understand the types of errors students encountered in an introductory Python course.
\citet{vihavainen2014novices} conduct a key-level logging study of novice coding behavior, although they seek to understand student behavior rather than IDE design for novices.
Our work is related to these, but we are less interested in understanding issues like student progress through assignments than the hindrances they encounter in the UI generally.

\section{Course Description}
\label{sec:context}

Our course aimed to teach basic computing skills (\eg{} variables, iteration, and function decomposition) to students with little-to-no-programming experience in the context of creative coding.
Learning to program typically also requires learning many surrounding skills, such as facility with command-line interfaces.
To eliminate such possibly intimidating setup difficulties---and allow tighter integration between our web-based instructional texts and the venue where work was to be done---we decided to centralize all student work within the online p5 editor.

Following common practices in creative coding courses~\cite{levin21Code} and tutorials
(such as from Khan Academy~\cite{khanAcademy} and Happy Coding~\cite{happyCoding}),
we used JavaScript and the p5.js library as the primary learning mediums, although the basics of web programming with HTML and CSS were also introduced.
p5 exposes a variety of drawing and interaction methods as primitive functions (such as \texttt{rect} and \texttt{circle}) which the programmer combines either to make static or dynamic compositions.
While p5 can be used to fully manipulate the native DOM, the majority of coding occurs inside a simplified environment focused on HTML-canvas manipulation.

\begin{figure}[b]
\beginFigureNarrow
\includegraphics[width=\linewidth]{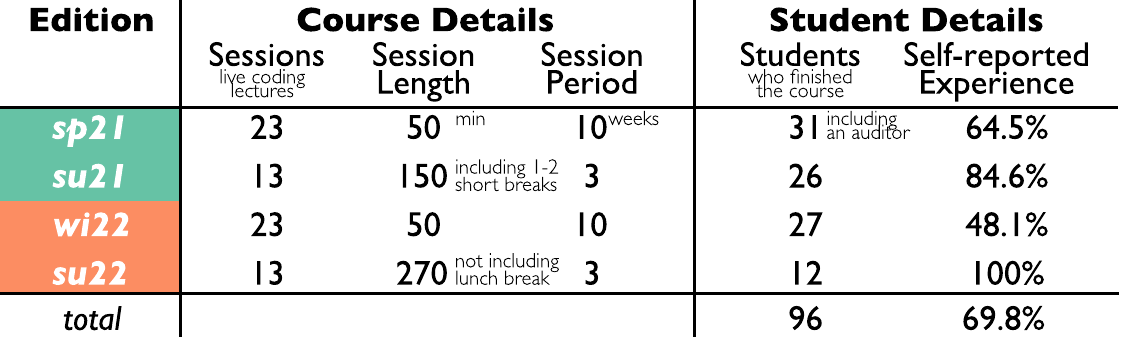}
\caption{
    Course details by edition.
    Experience was found by a pre-course survey that asked \emph{``How much programming experience do you have?''}
    We coded answers into no experience, some (having taken less than a college-level course), or high otherwise. We merge the latter two levels here.
}
\label{fig:participant-count}
\Description{A table showing the course enrollment and details across all three-course versions. It has columns describing the number of sessions, session length, session period, number of students, and percent of students with self-reported experience. Row represent course editions.}
\finishFigureNarrow
\end{figure}
 
We taught four editions of the course, referred to chronologically as \spq{}, \imm{}, \wiq{}, and \sumtwo{} (summarized in \figref{fig:participant-count}).
The \spq{} and \wiq{} editions were ``full'' 10-week college courses, offered from within a computer science department but cross-listed with media arts.
Students had broad academic interests: more than 20 different degree programs were represented by the 58 students (see \theAppendix{} for a breakdown).
The \imm{} and \sumtwo{} editions were intensive 3-week versions taught over the summer to high school students.
A fifth version of the course was taught during the summer of 2021, but was dropped from our analysis because participation was too small to meaningfully analyze.
Required coursework consisted of graded individual homeworks, collected but ungraded exercises, and, in the full editions, an individual self-designed project.
Taking into account differences in assignments and course material, \imm{} and \sumtwo{} were roughly two-thirds of \spq{} or \wiq{}.
Lectures in \spq{} and \imm{} were delivered remotely over Zoom.
The first three weeks of \wiq{} were also taught remotely, with the remaining weeks conducted in a hybrid format (during which students more often joined via Zoom than in person).
The \sumtwo{} edition was taught entirely in person and---with more in-class time (\figref{fig:participant-count})---included more required group work on practice exercises than other editions.
While the course was designed for those with limited experience, we observed high levels of self-reported prior experience in each edition.
These varying levels color some of our observations.
Based on our experience teaching them, the high school students in \imm{} may have been over-confident in their description of their prior experience. However, those in \sumtwo{} did seem to have non-trivial prior experience, perhaps due to a selection bias caused by the course being offered in-person at our university.
Despite higher self-reported prior experience than we expected, in our experience teaching we found that this experience did not necessarily lead to overwhelming mastery of the basic introductory material covered in the course. Therefore, we believe it is fair to view our students, as a group, to be novices.

The progression of assignments was designed to employ fundamental programming concepts (\eg{} variables, function abstraction and decomposition, loops, arrays, and objects)  for various media computation tasks (\eg{} vector graphics drawing, animation, image manipulation, and basic web development).
Aiming to serve the twin goals of teaching programming and fostering its use for creative expression, most assignments were open-ended (as opposed to being prescribed with easily-testable specifications).
For example, one early assignment asked students to make judicious use of variables and arithmetic expressions to implement a symmetric tree drawing of their own design. \figref{fig:tree-collage} shows a sample of student submissions from \wiq{} for this tree assignment.
Additional assignments are described in \theAppendix{}, and the full course materials are available online at \asLink{http://cs111.org/}{cs111.org}.

\section{Methods}
\label{sec:methods}

We now describe the studies that ran alongside each offering of our creative coding course and provide summary statistics.
See appendix for survey instruments, ethics statement, and other  materials.
\subsection{Year 1: Editions \spq{} and \imm{} with \pFiveOne{}}

In the first year of our two-year formative study, we deployed the p5 editor mostly as is.
We added a couple features to support course logistics, but we did not add any new programming affordances.

\begin{figure}[t]
\beginFigureNarrow
\centering
\includegraphics[width=0.9\linewidth]{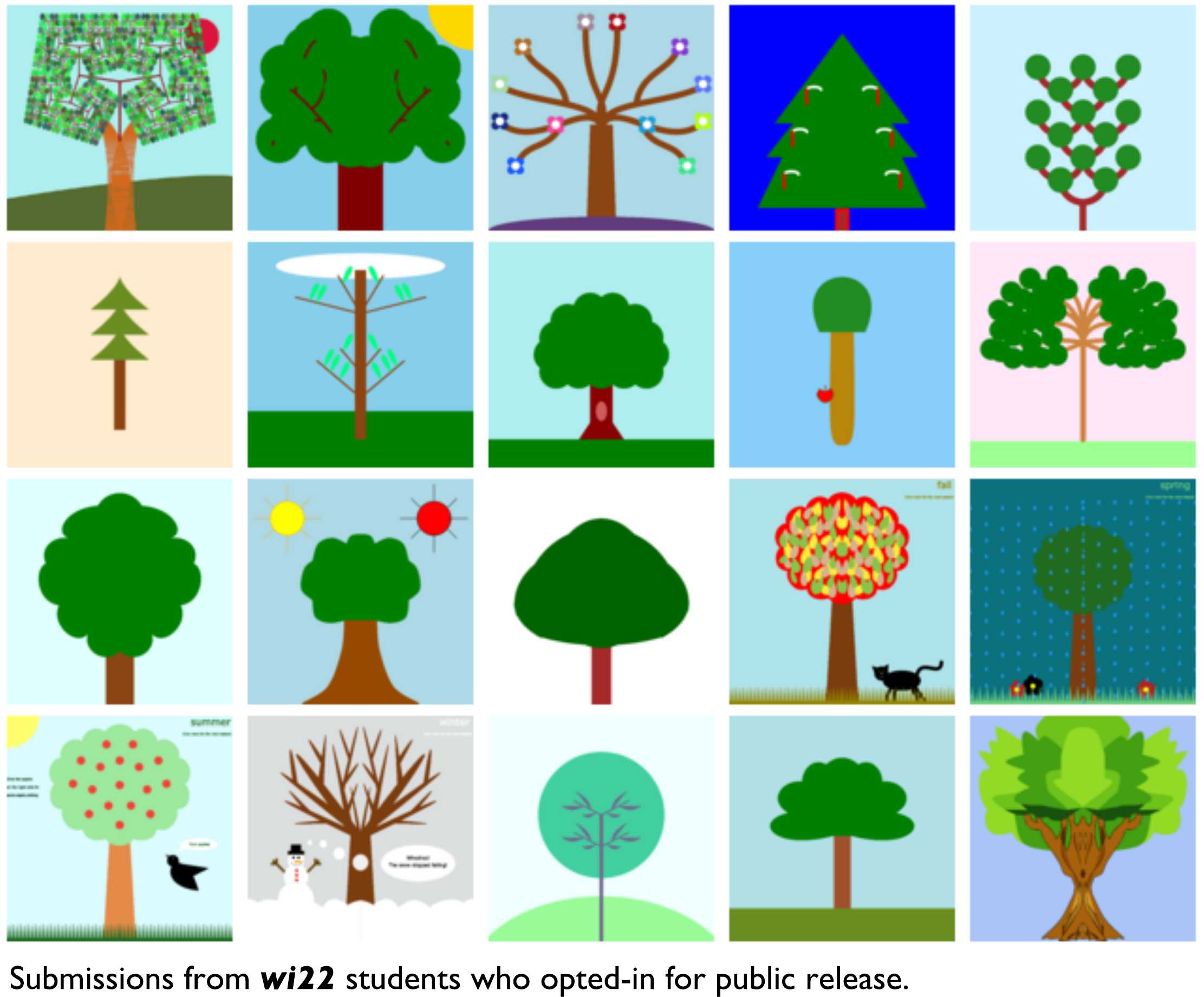}
\caption{
      One assignment in each course involved designing a tree, which exhibits horizontal axial symmetry.
}
\label{fig:tree-collage}
\Description{Images of cartoonish trees that are symmetric across the y-axis. They vary in character, some involving fractals and others being simple lines and triangles.}
\finishFigureNarrow
\end{figure}
 
\subsubsection{Custom Features in \pFiveOne{}}

Our initial fork added two features in support of teaching the class online.
The first enabled students to submit assignments from within the editor to GitHub repositories as pull requests.  Course staff then provided feedback and grading on these pull requests, merging them once complete.
The second mechanism allowed students to click any code example in the online course materials to open the code directly in the editor (without intervening copy-pastes or file-saves).
We removed features which were either not relevant to the class or would have negatively affected the course design (such as project sharing).
\subsubsection{(Per-User) Log Study}

We ran a study in the first two course offerings (\spq{} and \imm{}) to collect information about the coding behavior of students who opted-in to participate.
For these students, we captured the state of each sketch on every execution, save, submission, and structure edit (\eg{} find and replace)
throughout the course.
Logs were sent to a cloud-based server which only recorded events generated by study participants.
This ensured that all students experienced the same level of network traffic regardless of study involvement and thus did not penalize participants.

Student consent (and parental consent for students under 18) was sought prior to the course as part of a pre-course on-boarding process, which was also used to gather GitHub identification for submitting assignments and gauge prior experience levels (\figref{fig:participant-count}).
Students were not compensated for their involvement in the log study as participation did not modify the course experience.
Students were able to retract consent at any time during the course.
Logs were not analyzed during the course.
Although relatively coarse-grained, the logged events capture overall trends and patterns in the use of basic editor features.
In contrast, a key-level log study (as in Vihavainen \etals{}~\cite{vihavainen2014novices} study of novices' first weeks with an IDE) might have enabled more detailed observations at increased cost, both in terms of data collection and analysis, without clearly supporting our research questions about feature usage.

During this study we collected $\sim$0.5 million logged actions spread across $\sim$5500 sessions, which we define as periods of interactivity with $\leq$15 minutes between any two actions. On average sessions lasted $\mu$=$23.3$ minutes with a standard deviation of $\sigma$=$38.9$ minutes (listed as $\mu$$\pm$$\sigma$ hereafter).
    Our analysis of error frequency did not shed light on our research questions, but we provide summary statistics about observed run-time errors in \theAppendix{}.

    \subsubsection{(Long-Format) Feature Survey}\label{sec:feature-survey}

    After both \spq{} and \imm{}, students were invited to take part in an online survey soliciting their experiences using \pFiveOne{} and opinions about various features.
    They were asked about a series of features (\figref{fig:useful-features}), each presented as a static image with a paragraph of descriptive text.
    The feature progression was bookended by free-text questions on more general topics, such as debugging and code organization.

    A total of 25 \spq{} and \imm{} students participated in the survey.
    Participation in the log study was not a prerequisite.
    Our survey tool did not report the working time, but based on piloting, we believe that the survey took 20-40 minutes to complete.
    Participants were paid \$30 for completing the survey.
    Demographic data was not collected beyond a self-reported experience level.

    For each feature, the survey asked both free-text questions and Likert-item style rating questions (how ``Useful'' is the feature, and how ``Often'' would they use it).
    The number of surveyed features (17) was rather high, and we did not randomize their order.
    So, to help calibrate ratings, at the end of the progression a table summarizing the features asked for additional Likert-item style numerical ratings (how ``Interested'' they were in each feature).
    We found slightly negative correlations (Spearman's r) with presentation order and rating:
    Useful: $r$=-0.117,         %
    Interested: $r$=-0.118, and %
    Often: $r$=-0.133           %
    with $p$$\leq$0.015.
        These metrics exhibited good agreement: $r$=0.817 (Useful/Often), $r$=0.635 (Useful/Interested), $r$=0.655 (Often/Interested) with $p$<0.001.
        Thus calibrated, we focus only on perceived Usefulness.
        This selection is further informed by the technology acceptance model~\cite{davis1989perceived}, which suggests that users' perceived usefulness is indicative of subsequent usage, and is thus a more helpful quality than estimated Interest or frequency of use.
        We found that only in-context docs was statistically significantly ($r$=-0.308, $p$<0.01) correlated with self-reported experience. This slightly negative correlation is also reflected in the qualitative comments about that feature (see \figref{fig:skeptics} and \secref{sec:skeptics}).

        We included a mixture of features for general computation (such as ``Interactive Value Inspector'' and ``Linked Copy-and-Paste'') as well as creative coding (such as ``Coding by Drawing Tools'' and ``Canvas Ruler'') that would be understandable based on their experience in the course.
        The complete list of surveyed features is shown in \figref{fig:useful-features} and described in the appendix.
        Other features, such as notebook-style programming or multi-canvas editors (such as Stamper~\cite{burgess2020stamper}) were considered, but not included---we believed that textual descriptions or static renderings would be unlikely to give effective motivation for their utility, and students may incorrectly forecast their experience of such unfamiliar features.
        A key limitation of our survey is the simple static presentation of features.
        Participant perceptions might have differed if they had watched video demonstrations or been able to experiment with the features.

\begin{figure}[t]
\beginFigureNarrow
\centering
\includegraphics[width=\linewidth]{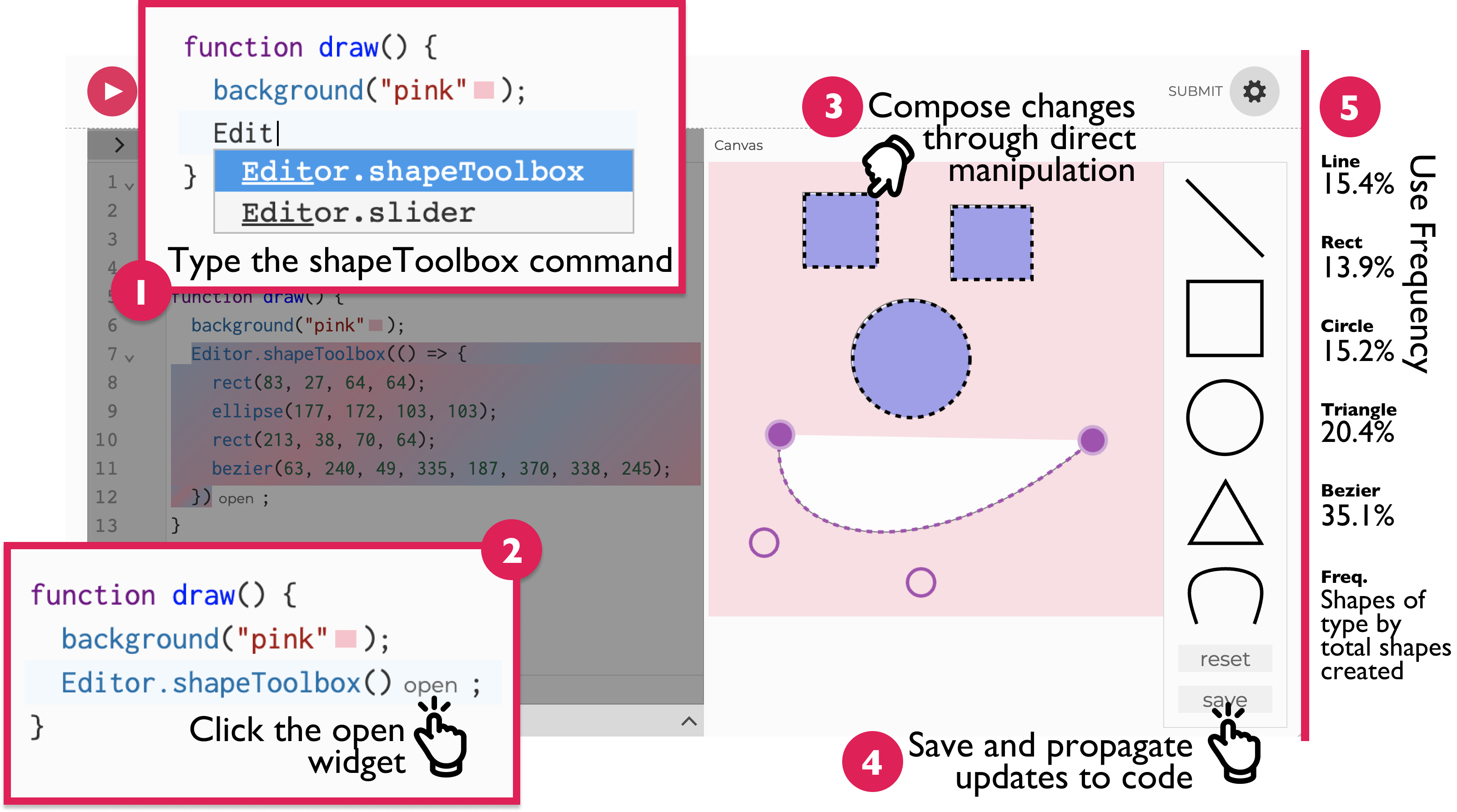}
\caption{
      The Shape Toolbox is used to create simple compositions of shape-drawing code through simple GUI actions.
      (1)-(4): Interaction workflow.
      (5): Frequency of shape usage.
}
\label{fig:shape-toolbox}
\Description{An annotated screenshot of the Shape Toolbox. It consists of three overlapping comic-style panels. In the first the user types Editor.ShapeToolbox into the editor. In the second they click the open button next to the typed words. In the third they have apparently drawn a smiley-face using the shape toolbox using boxes for eyes, a circle for a nose, and a curve for a mouth. To the right of the figure is a list of usage frequency by tool type, bezier curves are highest at 44 percent, while the others are each close to 13 percent.}
\finishFigureNarrow
\end{figure}

        \subsection{Year 2: Editions \wiq{} and \sumtwo{} with \pFiveTwo{}}

        After gleaning student predilections in our formative Year 1 studies, we modified our editor to investigate these stated preferences.
        We used \pFiveTwo{} in \wiq{} and \sumtwo{}, during which we ran two more studies.
        Whereas we customized \pFiveOne{} to improve course logistics, our changes in \pFiveTwo{} were motivated by the first year results.

        \subsubsection{Custom Features in \pFiveTwo{}}
        \label{sec:y2-custom-features}

        We implemented the top-3 unimplemented features from the survey (see \figref{fig:useful-features} or \figref{fig:often-useful} in the appendix): Color Pickers, Autocomplete, and Shape Toolbox (which was a synthesis of Coding by Drawing Tools and Directly Manipulate Shapes).
        Given limited resources, we forwent the Canvas Ruler (the next most-preferred feature) because there is a simple workaround for identifying positions (\eg{} console logging the \texttt{mouseX} and \texttt{mouseY} on mouse movement), although we intend to address it in future editions.
        Several highly-rated features (\eg{} Time Travel Slider, p5 State Displays, and Interactive Value Inspector) were more speculative and thus deemed beyond the scope of this work.
        We made two further modifications based on observations, namely, adjusting Auto-refresh and adding Number Sliders
        (which are common in interactive documents~\cite{ExplorableExplanations}).
        Two of these new features are accessed by calling special functions.
        \verb+Editor.slider(min, max, value)+ renders a Number Slider (\inlineFig{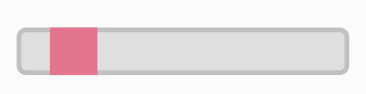}) for \verb+value+ in the range from \texttt{min} to \texttt{max} (see \figref{fig:annotated-ui}), with an optional fourth \verb+step+ argument to override the default continuous-dragging behavior.
        One alternative design is to store the metadata (range and step) in special comments, for example, as in Juxtapose~\cite{hartmann2008design}.
        Such an approach warrants comparison to our chosen design in future work. However, we elected to use the function call API to mimic a common p5 function for creating dynamic sliders,
        which students already learned (\verb+createSlider+~\cite{p5Slider}).

        The most novel feature introduced in \pFiveTwo{}, the Shape Toolbox, is also accessed through a function call-based workflow.
        The user calls \verb+Editor.shapeToolbox()+ and clicks a button to open the shape-drawing GUI.
        The text area is then disabled and a simple drawing toolbox is overlaid atop the output window.
        Using the Toolbox, users create compositions in the output pane
        by adding, translating, rotating, and scaling shapes (including primitive shapes and Bezier segments) with direct manipulation.
        Once satisfied, users click ``save'' to update the code---shape commands are called in the body of a function passed to \verb+Editor.shapeToolbox+.
        \figref{fig:shape-toolbox} depicts this workflow.
        Translation back to the code is achieved through a simple template matching method allowed by a one-to-one mapping between drawn elements and lines of code.
        We decided against always displaying the shape-drawing GUI for two reasons.
        First, not all calls to shape drawing functions can have GUIs---this is the subject of research on bidirectional editing.
        At one of the end the spectrum is a very simple approach that maintains a top-level ``scratchpad'' function (where all new shape-drawing calls would be added), and at the other end are heavyweight and expressive techniques in prior work~\cite{hempel_semi-automated_2016, hempel_sketch-n-sketch_2019, Glisp}---the former would be more restrictive than our chosen approach, and the latter beyond the scope of this work.
        Second, shape-drawing is only one aspect of our creative coding tasks; our approach displays the GUI only when the user explicitly opts to use it.
\begin{figure*}[t]
      \centering
      \includegraphics[width=\linewidth]{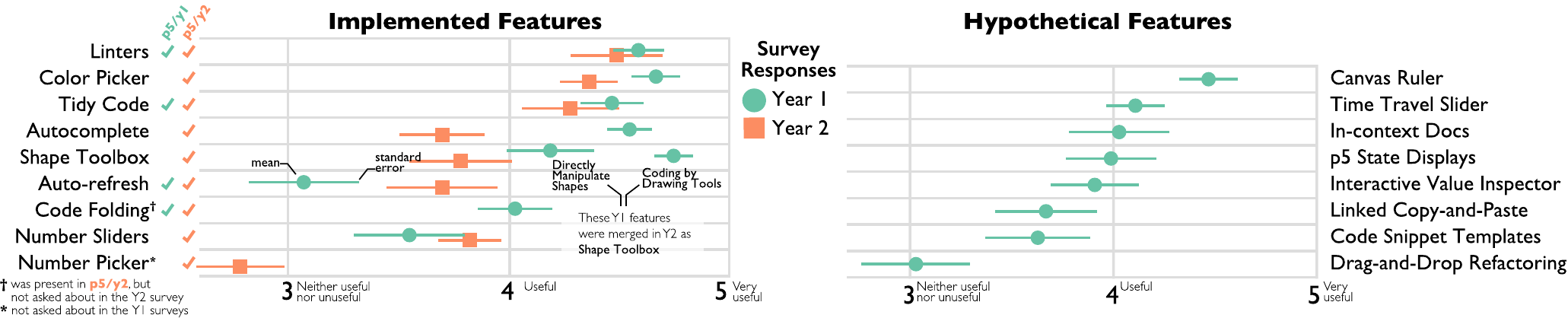}
      \caption{
            The features surveyed across both years and how ``Useful'' they were deemed to be on a 5-point Likert scale.
      }
      \label{fig:useful-features}
      \Description{Two sets of box plots. The left is labeled "implemented features'' and shows box plots for both the Year 1 survey and the Year 2 survey. The right is labeled "Hypothetical features'' and shows boxplots for just Year 1. A data table describing this figure can be found in Figure 19.}
\end{figure*}

        \subsubsection{(Aggregate-Use) Log Study}

        We collected anonymized usage of \pFiveTwo{} features during \wiq{} and \sumtwo{}.
        Logs were collected through a customized version of Umami~\cite{umami}, a self-hosted privacy-minded tracker.
        We elected not to capture full-sketch snapshots in this study because our planned analyses
        for the second year
        did not require them.
        We believed lighter weight instrumentation would allow us to capture more fine-grained usage patterns, such as the length of sessions. %
        Finally, this reconfiguration to full anonymity gave us leeway to collect data on all course participants, rather than just those who opted-in.

        Through this process we collected $\sim$1.2 million events across $\sim$6730 sessions (defined as before).
        Due to a configuration error (present only in \wiq{}), events were not collected with unique session identifiers, although we were able to reconstruct 75.4\% of the sessions---the remainder are excluded from analyses requiring specific session information.
        While the incomplete data is unfortunate, it still provides a more detailed picture of activity than in Year 1, which saw 68\% log study participation across \spq{} and \imm{}.
        Within this reduced sample, sessions lasted $\mu$$=$$24.1$$\pm$$38.0$ minutes. %

    \subsubsection{(Short-Format) Feature Survey}

    Near the end of \wiq{}, students were invited to take an abbreviated version of the Year 1 survey (\secref{sec:feature-survey}), containing only features that were added or improved upon in \pFiveTwo{}.
    Following the structure of the previous survey, we asked about frequency of use and Usefulness for features one at a time, followed by a summary table asking about Interest and a suite of reflection questions.
    Participants were compensated with extra credit roughly equivalent to 1\% of the final course grade.

    A total of \numYearTwoSurvey{} students participated in the survey.
    Our survey provider did not measure time taken to respond, but based on piloting we believe that the survey took 10-15 minutes to complete.
    Presentation order was not correlated with any of our metrics ($p$=0.779-0.939).
    Again, the ratings exhibited reasonable agreement: $r$=0.727 (Useful/Often), $r$=0.607 (Useful/Interested), $r$=0.693 (Often/Interested) with $p$<0.001.
    As with Year 1, we focus only on Usefulness in the body of the text (see \theAppendix{} for the others).
    We found prior experience to be statistically significantly ($p$<0.01) correlated with only a single feature, auto-refresh, for which there was a somewhat negative correlation ($r$=-0.487).
\section{Analysis}
\label{sec:results}

We now reflect on the features, connecting them to the themes summarized in \secref{sec:intro}, denoted\takeawayStatic{} through\takeawaySkeptic{}. We consider features implemented in both \pFiveOne{} and \pFiveTwo{} (\secref{sec:both-features}), followed by those added in \pFiveTwo{} (\secref{sec:only-two}), and then introduce concerns that cut across multiple features (\secref{sec:skeptics} and \secref{sec:creativity}).
Our analysis draws on data from the survey studies (summarized in \figref{fig:useful-features}) and the log studies as appropriate.
Participants from the  \spq{}, \imm{}, \wiq{}, and \sumtwo{}
surveys are referred to as \pA{1-16}, \pB{1-9}, \pC{1-19}, and \pD{1-4}, respectively, and are colored by year.

\subsection{Features in Both \pFiveOne{} and \pFiveTwo{}}\label{sec:both-features}

We begin by considering features present in both editor versions.

\subsubsection{Linting}

This static analysis tool eagerly executes after small code edits, checking simple syntactic assertions akin to spell check for code. It was well received in both years and was mostly seen as helpful, although sometimes impolite.

Students found linting to be
\pQuote{very helpful}~(\pA{5,7,17,20}, \pC{10,16}, \pD{1, 2}) and \pQuote{very useful} (\pA{19}, \pC{2,4,6,12,18}, \pD{4}), because it
\pQuote{saves time and energy}~(\pA{4}) and
shows \pQuote{where I needed to go to fix simple bugs}~(\pA{16}).
\pC{10} believed that debugging \pQuote{would be way more annoying without it} because \pQuote{it's not always obvious what you did wrong}~(\pD{4}).
Whereas 86.0\% of executions in Year 1 passed lint, in Year 2 (where we had visibility into all lint runs) code passed 13.5\% of lint runs (which happened after most small text edits). This may indicate that students address lint errors before running code as a simple integrity check, or that the analyses are executed too early; however, student comments seem to indicate the former.
Unlike other features, students were incentivized to attend to it, as the absence of lint errors was a small part of homework grades
(98.7\% and 97.2\% of submissions in Years 1 and 2, respectively, passed lint).

Beyond code style, linting can provide opportunities to expose novices to other best practices.
For example, CSSLint~\cite{cssLint} (used in \pFiveTwo{}) explained that the \texttt{*} selector is considered bad practice because it is inefficient.
Indeed, \pC{3} felt that linting \pQuote{trained me to think and type in a certain way}, and
\pA{5} observed that it could be \pQuote{a nice way to point out when I am making stylistic errors (instead of [Tidy Code] just magically fixing all of them for me).}
Utilizing this well-received channel for introducing programming features and practices is an opportunity for future IDE design.%
\takeawayStatic{}

Participants also offered ideas to improve the feature.
Because the editor eagerly ran the linter, \pQuote{the yellow line warning[s] often exist all the time. It annoys me}~(\pB{4}).
Instead, some students would have preferred not to see lint errors \pQuote{until I finish typing}~(\pA{13}) or \pQuote{before finishing a line of code}~(\pB{5})---the mechanics of exactly when and how to display errors for incomplete code will require careful design (as considered, \eg{} in Hazel~\cite{Hazelnut,HazelnutLive}).
Others expressed a desire for more nuance---\pQuote{acknowledging the difference between `This Must Be Changed To Have Nice Code™' and `hey, maybe consider changing this thing!'}~(\pA{5})---and control---being able to \pQuote{ignore/exit out of a warning}~(\pA{3}).
Poorly-received default choices and persistent errors can repel users.
As an extreme example, one \wiq{} student decided to use Replit~\cite{replit}, rather than \pFiveTwo{}, for their final project because too many (CSSLint) errors seemed irrelevant or unclear how to fix.
Linters integrated with editors in this way do not offer mechanisms to override general advice or to indicate that the user knows what they are doing.
This is impolite computing~\cite{whitworth2005polite}: it forgoes user agency and generally is perceived as a pest.
Avoiding these pitfalls is important to leverage the instructive opportunities offered by the well-received, static analysis-informed tools.%
\takeawayStatic{}

\subsubsection{Tidy Code}

Auto-formatters provide on-demand code restyling without semantic modification, and are common in professional coding workflows~\cite{RautiainenTools}.
We often encouraged the use of this tool---called Tidy Code in the p5 editor---in lectures, but we did not incentivize its usage in grading.
It was invoked manually (from the top menu bar or keyboard shortcut) rather than being executed on every save.
Like linting, this feature was generally well received.
Students found auto-formatting to be
\pQuote{super useful}~(\pA{15}) and
\pQuote{very satisfying}~(\pA{2}).
The formatting choices were not always appreciated, however.
Whereas \pC{15} \pQuote{only rarely preferred my own organization},
\pA{12} felt the results \pQuote{appeared less organized, such as having irregular line breaks} and
\pA{10} \pQuote{worried it would mess up my organization.}
We observed that students in Year 1 often (needlessly) invoked auto-formatting twice in a row. In particular there was a probability of 16.15\% and 8.65\% (in \spq{} and \imm{}) of auto-formatted code being auto-formatted again right away---with similar behavior observed for saves (see appendix for details).
This suggests that providing clear code-state signals (analogous to linting's visual indicators) may reduce needless anxiety-motivated saves and tidyings.
The presence of this behavior in Year 1 suggests it was likely repeated in Year 2;
however, the aforementioned configuration error prevented us from collecting auto-formatting usage.
While simple indicators may seem to be trivial UI modifications, we suggest that it will impact the perception and understanding of such features.

Several students would have liked the feature to be customizable, rather than enforcing a fixed set of
\pQuote{preferences that should not be forced by tidy code}~(\pC{16}).
Indeed, some students would have liked auto-formatting better
\pQuote{if it was a little configurable}~(\pA{16}); for example,
\pQuote{if there [were] multiple common/standard rulesets there could be a way to choose which you want to follow}~(\pA{5}).
Furthermore, it \pQuote{would be helpful to be able to specif[y] which block of code to tidy}~(\pA{13}).
Thus, extending well-chosen defaults with ways to selectively customize style preferences---a notion which has been referred to as ``code style sheets''~\cite{LubinPLATEAU}---could further increase the politeness of this feature and thus its utility.%
\takeawayStatic{}

Like the teachable moments offered by linting, \pB{8} felt they \pQuote{Learnt a lot about code organization using this feature!}
As implemented, however, the results of auto-formatting are updated in the code box without explanation.
Better would be for the editor to \pQuote{show you what you are doing `incorrectly'}~(\pC{19}),
for example, using visual highlights and annotations to explain the differences---which could also serve as scaffolding to introduce version control tools.

\subsubsection{Auto-refresh}
\label{sec:auto-refresh}

This feature re-executes code upon text edits---a workflow demonstrating ``level-3 liveness''~\cite{Tanimoto1990,Tanimoto13live}.
Auto-refresh was present in both \pFiveOne{} (inherited from the original editor) and in \pFiveTwo{} (where it was modified).
In principle, live feedback would seem particularly helpful in a creative coding context as programs are often updated with small graphical adjustments, and thus well matched with a short edit-run cycle.
It was also well matched with our setting: the Normalized Programming State model~\cite{carter2015normalized} suggests that spending longer periods of time in syntactically unknown states (such as when the code has not been executed in a while) is negatively correlated with program success.
This needs to be balanced with the cognitive load~\cite{hundhausen2017ide} caused by repeated executions.

\begin{figure}[t]
\beginFigureNarrow
\centering
\includegraphics[width=\linewidth]{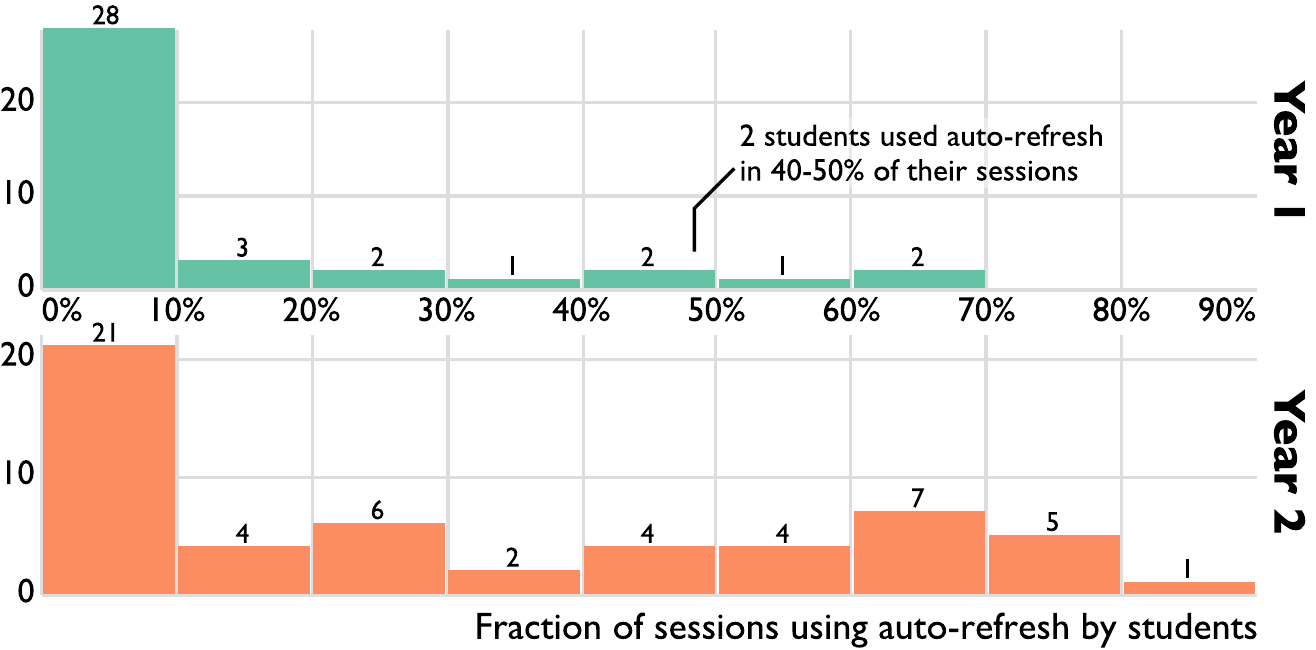}
\caption{
      Histograms of the fraction of sessions where a student used auto-refresh any amount.
      58.7\% and 5.1\% of students never used auto-refresh in Years 1 and 2 respectively.
}
\label{fig:live-frac}
\Description{A pair of vertically stacked bar charts. The top one is labeled Year 1 and the lower is Labeled Year 2. Each value is labeled with a number corresponding to its value. In both charts the 0-10 percent bar is much bigger than the others.}
\finishFigureNarrow
\end{figure}

Auto-refresh in \pFiveOne{} did not achieve a fruitful balance.
As indicated in \figref{fig:live-frac}, only a handful of participants regularly used it and most students used it rarely, if at all.
The survey responses color this imbalance.
Whereas \pA{9} \pQuote{used this all the time and loved it}, finding it %
\pQuote{way easier than clicking the `play' button all the time},
others felt that the keyboard hotkey was sufficient~(\pA{16}, \pB{1,8}, \pD{1}).

More important than convenience were differing views on the fundamental interaction model itself.
\pB{7} appreciated the ability \pQuote{to see what I was creating as I coded}, finding it useful even though \pQuote{error messages that kept popping up got in the way a little},
while others found the errors \pQuote{very distracting}~(\pA{5,6}).
Participants felt the feature was
\pQuote{running incomplete code unintentionally}~(\pB{6}) and
\pQuote{when you don't want it to}~(\pB{1}).
Instead, some students felt robbed of their agency over their code, desiring \pQuote{to be the boss of when my code reran}~(\pA{5}) and
in \pQuote{control my own pace}~(\pB{2}), only running the code when
\pQuote{I know I have something that I want to see}~(\pA{13}).
This suggests that, while spending too much time in syntactically invalid states may be detrimental~\cite{carter2015normalized}, spending too \emph{little} time is also problematic.
Developing a careful understanding of the tradeoffs is an important avenue for future live programming work.%
\takeawayLive{}

For the purposes of this work, we made only simple changes to auto-refresh in \pFiveTwo{} based on our observations from the first year.
We increased the refresh delay from 400ms to 1s, and, more importantly, in the event that executed code had lint errors---a proxy for run-time errors---the editor did \emph{not} refresh the canvas, instead indicating that it was ``stale.''
Thus, in the (many) cases when edits are incomplete or erroneous, the canvas remains visually stable.

The modified auto-refresh was modestly better received, with its Usefulness increasing from $\mu$$=$$3.1$ to $\mu$$=$$3.7$.
        In addition, per \figref{fig:live-frac}, it was used more often---although we note that auto-refresh was demonstrated more at the beginning of \wiq{} and \sumtwo{} than in prior editions.
        A one-sided t-test indicates that students in Year 2 used auto-refresh significantly  ($p$<0.001) more often.
        Yet, the overall balance remained far from perfect.
        Some participants were \pQuote{stressed} at
        \pQuote{all the errors that pop up as I implement new things}~(\pC{15}) and
        \pQuote{before I got to fix them}~(\pC{4}).
        These negative views seemed more likely to come from those with prior experience ($r$=-0.487, $p$<0.001), which may suggest that expectations are set by experience with tools exhibiting a different execution cadence.
        Others, however, found it
        \pQuote{very useful for certain exercises that needed lots of small adjustments}~(\pC{3}) and
        \pQuote{very helpful when using trial and error}~(\pC{16}).
        Overall, we observed no significant changes in user behavior after modifying auto-refresh, despite the improved perception of the feature. This again underscores that designing UIs to be polite (or at least not irritating) is critical to their usage.

        \subsection{Features Only in \pFiveTwo{}}\label{sec:only-two}

        Next, we consider the features that were added in \pFiveTwo{}.
        While Year 2 survey responses are based on hands-on experience with the features in \pFiveTwo{},
        Year 1 responses are based on descriptions in the survey and experience with other tools.
        Feature use in \wiq{} is shown in \figref{fig:feature-use-in-year-2}.

        \subsubsection{Shape Toolbox}
        \label{sec:shapetoolbox}

        The most significant addition to \pFiveTwo{} was the Shape Toolbox feature that allowed GUI-based specification of  primitive shapes using direct manipulation which generated matching code (\figref{fig:shape-toolbox}).
        The constituent parts of this feature were highly perceived in Year 1:
    $\mu$$=$$4.8$$\pm$$0.44$ for Coding by Drawing Tools, and
    $\mu$$=$$4.2$$\pm$$1.0$ for Directly Manipulate Shapes.
        Some students believed it would be \pQuote{very beginner friendly}~(\pA{3}) and would make work \pQuote{a lot easier and faster}~(\pB{7}).
        Others believed it would also reduce errors (\pA{9}) and help with debugging  ~(\pB{6}).

        Help programming curved shapes---such as the trees in \figref{fig:tree-collage}---was particularly enticing:
        \pQuote{for bezier curves, changing the input values rarely produced an expected result}~(\pA{12}), highlighting a gulf of execution~\cite{norman2013design}.
        The process usually involved \pQuote{lots of trial and error}~(\pA{3}), sometimes resulting in student disengagement: \pQuote{Coding the bezier curves manually turned me off of them, and I did not attempt them in my work}~(\pA{14}).
        However, that same student noted \pQuote{If I had had a tool like this, I certainly would have used them.}

        Several students in Year 2 embraced the feature.
        For example,
        \pC{18} found it \pQuote{EXTREMELY helpful, especially when it came to drawing Bezier curves. Every time I had to draw a curve, I used the shape toolbox. I probably would have cried without it.}
        \pC{10} mentioned that it \pQuote{Was very nice to use it to get approximate coordinates then fine tune them after.}

\begin{figure}[t]
\beginFigureNarrow
\centering
\includegraphics[width=\linewidth]{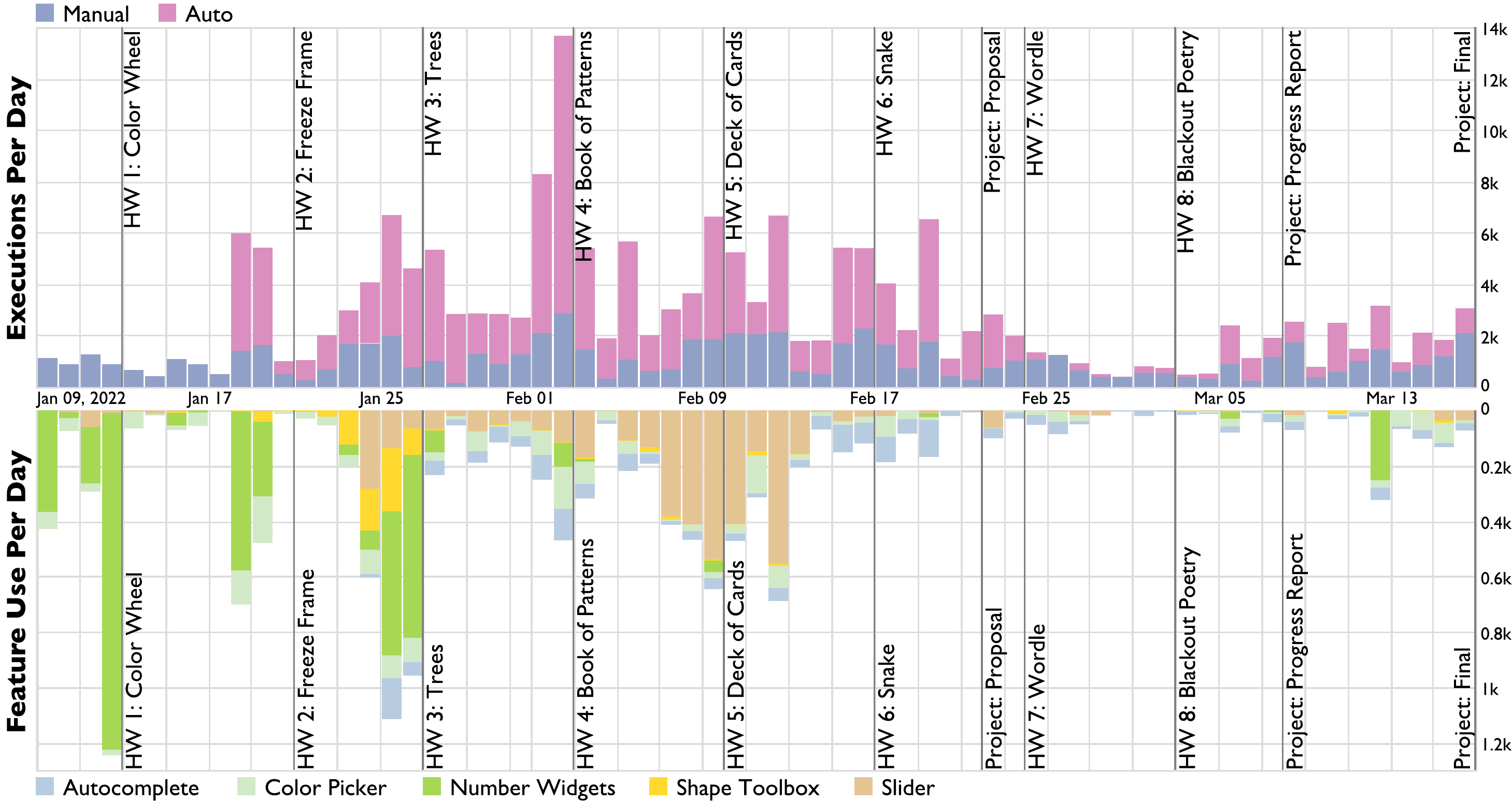}
\caption{
      Feature use in \wiq{} was guided by course content.
      For instance, autocomplete was demonstrated prior to HW3 and sliders were included in the starter code for HW5.
}
\label{fig:feature-use-in-year-2}
\Description{A pair of stacked time series bar charts. The shared x-axis is the day since the start of the winter quarter, the y-axis in the top bar chart is executions per day, and the lower is feature use per day.}
\finishFigureNarrow
\end{figure}

        Although the feature was
        \pQuote{very useful for beginner projects} (\pC{2}),
        several students, including \pC{6}, \pQuote{used them less as time progressed.}
        Shape Toolbox was used often for the tree homework (see HW3 in \figref{fig:feature-use-in-year-2}), and use per execution by week was minimal after that assignment,
        being used in only 2.12\% of all (available) sessions.
        Perhaps because the feature did not have a stable visual presence (as with the auto-refresh button), some students \pQuote{completely forgot this existed, but I think it would have been really really useful if I had remembered} (\pC{4}).
        In addition, although we expected the feature to be used extensively for HW 2, in \wiq{} \verb+Editor.shapeToolbox+ was announced but not demonstrated in class until after the assignment was released.
        Bezier curves accounted for the majority of invocations (see \figref{fig:shape-toolbox}.5).
        Toolbox sessions (from open to save) lasted $\mu$$=$$22$$\pm$$30$ seconds, indicating that it may have been used relatively often to make small graphical adjustments, as opposed to building larger compositions.

        Students may have continued to use them later in the course \pQuote{if it allowed for some of the shapes that are more complicated}~(\pC{16}).
        Further limiting the utility of the feature, within an invocation shape drawing functions allowed only literals---once a student wanted to use variables and arithmetic expressions, the Toolbox would no longer open.
        \pQuote{[C]reating an object without this feature would be better because of the precision}~(\pB{5}) afforded by variables, expressions, and so on.
        Thus, the feature ultimately fell short of what students imagined: $\mu$$=$$3.8$$\pm$$1.2$.

        Bidirectional updates are being explored in a growing number of systems (as in \sns{}~\cite{hempel_sketch-n-sketch_2019}), but there remain significant technical and UI design challenges to explore, before even considering their value to novices.
        As predicted by a couple students, more feature-rich bidirectional synchronization would need to reconcile
        ambiguous graphical interactions (\pQuote{There are many parameters and it would be hard to make it so it manipulates them one at a time}~(\pB{5})) and
        their effect on other parts of the program (\pQuote{My only but major concern would be that it doesn't confuse the other lines of code, and that it may not run the way the programmer wants to use it}~(\pB{6})).

        Nevertheless, the experience suggests that even a simple implementation of this very desirable feature was promising.
        However, as we discuss in \secref{sec:skeptics}, many students were skeptical about the effect of this feature on learning.
        \takeawaySkeptic{}

        \subsubsection{Autocomplete}

        We enabled a simple autocomplete menu~\cite{codeMirrorSix} and populated it with p5-specific \emph{identifiers} (variables and function names), \emph{syntax templates} (common patterns,
        like \verb+for+-loops with holes),
        and commands for invoking the Shape Toolbox and Number Sliders.
        Passively supporting learning in this way would seem to be a natural fit for our setting, but some students were leery of it.

        Year 1 survey respondents anticipated autocomplete positively ($\mu$$=$$4.6$$\pm$$0.5$),  believing it would help in several ways. For example, to
        \pQuote{increase speed and productivity when coding}~(\pB{9}) and
        \pQuote{make it faster to get debugging done}~(\pB{6}).
        In addition, \pA{13} believed autocomplete would encourage better code style:
        \pQuote{not having dynamic autocomplete incentivizes me to write non-descriptive function names and variables for the sake of efficiency.}
        Participants also believed autocomplete would help
        \pQuote{discover new features}~(\pB{5}),
        \pQuote{expose us to new things we didn't know existed}~(\pB{7}), and
        provide \pQuote{an idea of what to write or what could be written}~(\pB{4}).
        These beliefs are in line with how professional programmers use autocomplete to debug and explore APIs~\cite{empiricalcodecompletion2015}.

        However, the experienced reality of \pFiveTwo{} fell short ($\mu$$=$$3.7$$\pm$$1.0$) of anticipation.
        Autocomplete was used in only 12\% of sessions (with 33.6\% selections being templates), although it was used progressively less as \wiq{} and \sumtwo{} went on.
        While this relative infrequency of use may be related to
        the simple implementation (which did not include embedded documentation or other common guidance features) or
        the emphasis later in the course on web programming (the DOM was not thoroughly reflected in the autocomplete suggestions),
        this trend appears to agree with how \citet{vihavainen2014novices} observed novice usage of autocomplete. They note that 27.3\% of novices initially used autocomplete to create a particular command (Java's system print), which decreased to 1.64\% after a week of use.

        This appears to suggest that autocomplete can serve as a vehicle for teaching: it is \pQuote{a useful guide until I was able to type certain things in by memory} (\pC{13}).
        Some perceived the ability to
        \pQuote{stop memorizing certain code}~(\pA{9}) as a benefit, while others thought
        \pQuote{it's a give and take}~(\pC{7}) and might hinder \pQuote{programmers' knowledge about commands and their forms in the long run}~(\pB{8}).
        We return to this concern about the effect on learning in \secref{sec:skeptics}.%
        \takeawaySkeptic{}
        Beyond these hesitancies, it is unclear why more students did not engage with the feature, although some noted that it can be  \pQuote{annoying when you already know what you want}~(\pC{15})---which suggests that the clutter\takeawayClutter{} or cognitive noise\takeawayLive{} may be a factor.
        Given this diversity of opinion, we suggest that configurability is important to designing such features politely, as some students (such as \pD{4}) wanted to be able to turn off autocomplete (to limit its disturbances).

        \subsubsection{Color Pickers}

        Integrating a color picker into an editor for creative coding was perceived as very useful in the Year 1 surveys ($\mu$$=$$4.7$$\pm$$0.56$).
        Whereas \pA{4}, probably like many students, did not pick colors as much in the second half of the course,
        \pA{6} said \pQuote{I had a color picker tab open for every single assignment.}
        Based on this enthusiasm, we implemented a modal color picker dialog box in \pFiveTwo{}~(\figref{fig:annotated-ui}), following a sentiment from \pA{13} that such a design \pQuote{would probably be more helpful than in the code to prevent clutter.}
        \takeawayClutter{}
        Note that a similar color picker was recently added to the p5 editor, highlighting the value of this feature. However, ours supports more color formats, namely, web color names---driven by a participant suggestion (\pB{9})---and RGB values as numeric lists.

        After experiencing color pickers, students viewed them positively ($\mu$$=$$4.4$$\pm$$0.65$),
with the \wiq{} students using them frequently in the first half of the course, as depicted in \figref{fig:feature-use-in-year-2}.
They helped make the process \pQuote{more efficient}~(\pA{9}),
\pQuote{[taking] out the hassle}~(\pC{6}) of
\pQuote{open[ing] up another program}~(\pA{2}).
Color pickers may foster creativity, as they could \pQuote{let me pick some irregular colors}~(\pB{2}).

Several participants also voiced support for the idea, suggested in the survey prompt, for an eyedropper tool.
Others suggested additional features inspired by drawing programs like Illustrator, such as
grid~(\pD{1}), zoom~(\pC{15}), \pQuote{better proportions}~(\pA{3}), or a way to
\pQuote{group lines and shapes and move them all at once}~(\pB{7}).
Such lightweight and familiar tools from creativity domains are natural enhancements---as long as they are not impolitely imposed\takeawayStatic{}---that we intend to investigate in the future.

\subsubsection{Number Pickers and Sliders}

``Scrubbers''~\cite{ScrubbingCalculator}, which allow direct manipulation of numeric values by dragging, are often touted in live programming systems and interactive documents~\cite{ExplorableExplanations}  as being representative of the value of those environments.
Despite the overlap between live programming's close connection to the visual domain and the interests of creative coding, the clutter\takeawayClutter{} and lack of control\takeawayLive{} brought on by these features impeded adoption.

Some Year 1 students were positive about hover-based Number Sliders, believing they would allow them to
\pQuote{experiment with the code more quickly}~(\pB{6}) and
\pQuote{more efficiently}~(\pB{9}).
However, some worried that \pQuote{it could make the editor look more crowded}~(\pA{3}), while \pA{5} noted \pQuote{I would rather just do it myself.}

Nevertheless, we added Number Sliders to \pFiveTwo{}, which appear (per \secref{sec:y2-custom-features}) inline via \verb+Editor.slider(min, max, value)+, as well as Number Pickers (\figref{fig:annotated-ui}), which are buttons surrounding each number literal that allow it to be  incremented and decremented (\inlineFig{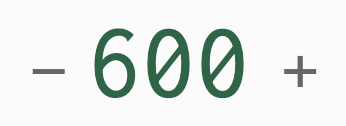}).
(Such small modifications explain the large absolute number of Number Picker events in \figref{fig:feature-use-in-year-2}.)
Students found these additions could be a \pQuote{quick, helpful way to make sure my assignments didn't break at a larger scale}~(\pC{6}), as was the case for HW5 (cf. \figref{fig:feature-use-in-year-2}).
Although scrubbers were perhaps most useful toward the latter stages of a task,
\pQuote{when I'm playing around with my final result}~(\pC{17}),
\pC{13} felt they \pQuote{allowed me to tap into my creativity.}

Yet, per \figref{fig:useful-features}, the feature was not so highly rated.
A recurring theme is that scrubbers---in various configurations---felt
\pQuote{messy}~(\pB{4}, \pC{9}),
\pQuote{disrupted the look of the code}~(\pC{4}) \takeawayClutter{}
or were just generally unnecessary~(\pB{10}).
\pA{14} felt that the transitory changes would be confusing, and hard to maintain a model of different parameter configurations.\takeawayLive{}
Others wanted more refinement in the numeric type, such as limiting it to numbers \pQuote{divisible by five}~(\pA{5}).
While these features are typically well-used in graphical applications like Figma, it seems that this type of feature is \pQuote{trying to solve or better a process that needs no help}~(\pB{10}).
While there is evident overlap between our domain with other artistic settings,
not every translated feature will match the interests of learners.

\subsection{On Skepticism}
\label{sec:skeptics}

\begin{figure}[t]
\beginFigureNarrow
\includegraphics[width=\linewidth]{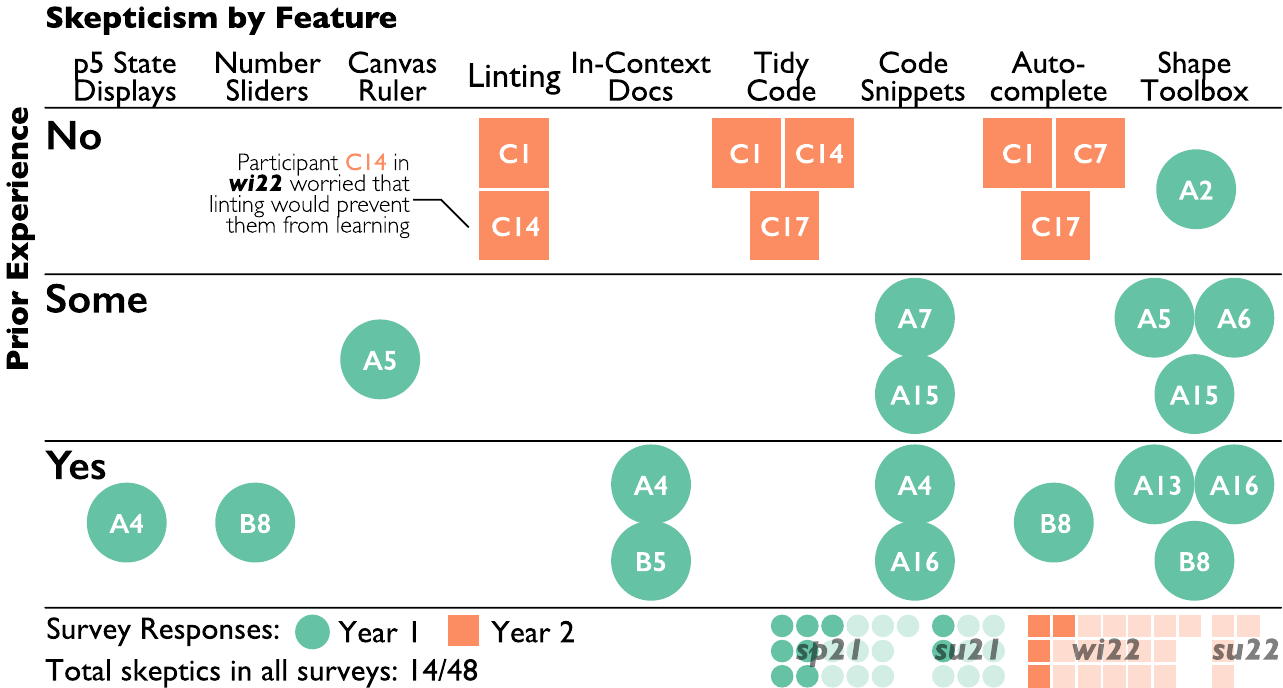}
\vspace{-2em}
\caption{
    Some skeptical survey respondents worried that some features would deleteriously affect learning.
}
\label{fig:skeptics}
\Description{A isotype plot showing student skepticism. Three rows are divided by experience (No / Some / Yes). There are nine columns, each showing a different feature. There are circles and rectangles labeled with a participant ID, indicating that the student was skeptical about that feature. The columns have been sorted based on the number of skeptical students. In the lower right-hand corner are waffle plots for each course edition showing the ratio of skeptics to survey respondents in that edition.}
\finishFigureNarrow
\end{figure}

Next, we grapple with the perception that some tools take away learning opportunities that may be needed to \pQuote{become a good programmer.}
\takeawaySkeptic{}
Several features were perceived as making things too easy for novices.
\figref{fig:skeptics} summarizes how ``skeptics'' worried about different features.
These perceptions are valuable: as the technology acceptance model~\cite{davis1989perceived} and related theories highlight, \emph{perceived usefulness is a central part of whether a system is ultimately used}.

Several students worried how syntax templates (see appendix) and autocomplete balanced the tradeoff between augmenting their abilities and enfeebling their development of skills.
Whereas \pA{5} was \pQuote{not sure if it actually matters} to practice memorizing names and function signatures,
\pC{17} weighed the tradeoff according to the goals of the student:
\pQuote{I wouldn't consider it a horrible thing for those who don't want to go into coding professionally/too much}---implying that a more serious programmer might indeed miss out on practicing an important skill.
\pA{16} thought such features \pQuote{might reduce some of the learning by doing that you get when coding, so I'm not sure if it's great for a class. I learn through my coding mistakes and this would reduce the number of mistakes, so a mixed bag.}

There were similar concerns about in-editor documentation (also discussed in the appendix).
For example, \pA{15} said that
\pQuote{new coders need to learn the process of going into the manual.}
\pA{4} reconciled the aforementioned tradeoff as follows:
\pQuote{However, depending on the specific goal of this course, if it is to focus more on the creative coding aspect and not necessarily `become a good programmer' then in-context docs would be awesssommeee.}

Some of these concerns might be ameliorated by introducing a notion of documentation or autocomplete levels (in a similar style as DrRacket's language levels~\cite{marceau2011mind}), which gradually adjust what information is available as new concepts are introduced. However, without sufficient signaling, students might construct mental models of the information present in the feature and then dismiss all subsequent configurations.

Conflicting views over the Shape Toolbox in Year 1 were most striking.
\pQuote{This feature would be so useful and allow for more creative opportunities especially for beginner coders}~(\pA{10}).
It would also be a \pQuote{useful learning tool}~(\pA{3}) by allowing students to
\pQuote{see how the code changes in order to learn how certain parts of the code are working}~(\pB{9}).
But many students were skeptical.
This \pQuote{feels like cheating!}~(\pA{6}) and
\pQuote{saves way too much work for the new learners}~(\pA{15}).
\pQuote{It's way too useful but can hinder with the learning process of basics of coding. As a student, I won't want this but as a programmer who knows the basics, it's a nice feature}~(\pB{8}).
\pQuote{I think some of these features while helpful would have discouraged learning. Some of the most rewarding parts was sweating through inconvenient parts}~(\pA{2}).
As shown in \figref{fig:skeptics}, some of this hesitation was self-censorship by students with little or no prior experience.

However, among Year 2 participants (all of whom had access to the feature), there were \emph{no} skeptics of the feature.
Perhaps the idea of others having improved tools is jarring, while students who are given improved tools simply worry about the plenty of challenging learning left to do.
We view this situation as akin to giving students calculators in a math class:
they help with specific classes of tasks that, once simplified, enable learning about richer topics.

Students seem to construct a naive model of what makes a good programmer, suggested above as being someone who has memorized the entire language and does not depend on digital assistants or developer-experience tools, thereby dismissing behaviors besides this as being inauthentic.
We suggest that reorganizing and reforming this model is part of the value that classroom-based computing education offers, as it can help to offer a thicker model of what is authentic~\cite{ThickAuthenticity}. Enhancements to novice-oriented IDEs such may also help to dispel these notions if they are \emph{perceived} as realistic tools rather than as something akin to training wheels.

\subsection{On Creativity}
\label{sec:creativity}

Finally, we consider the role of creativity in our editor.
While there exists little agreement on what creativity means in HCI research~\cite{frich2018twenty},
we found that students espoused two clear views on how tools might help them creatively: automating tasks that impede of creativity and helping explore unknown functionality.
For students, creativity often appeared to be something which typical coding tasks stood in the way of; obstacles that some technical interventions could ameliorate.
Shape Toolbox was emblematic of this style of reduction. For instance, \pA{10} believed that such a tool would \pQuote{allow for more creative opportunities especially for beginner coders,} and \pB{9} believed that it \pQuote{could help with planning ideas for art projects and increase creativity.} As noted in \secref{sec:shapetoolbox} students in \pFiveTwo{} embraced this feature and appeared to use it to reduce the tedium required to precisely locate shapes, thereby making greater room for artistic expression.
Others highlighted the value that tools that reduced tedious tasks, such as picking individual coordinates through a ruler (\eg{} \pA{3} and \pA{9}) or identifying which lines corresponded to which components of the image (\pB{9}).
Summarizing this view, \pC{16} observed that \pQuote{I think what this editor did well as an art tool was streamlining certain common processes.}

Beyond reducing tedium were opportunities for exploration, which were manifested both as moments of play (\pA{8}) or fun (\pA{4}, \pC{6}), as well as discovering new functionality.
For instance \pA{12} believed that \pQuote{comprehensive documentation would have allowed me to be both more creative,} a view which was confirmed by \pC{17}, who believed that such features help do \pQuote{things I don't yet know how to do by myself} and thereby \pQuote{help me be more creative.}
\pA{9} believed that surfacing program state might encourage reflection and discovery, such as by seeing that \pQuote{circle has a round stroke cap, so it might make me wonder what other shapes the stroke cap could have.}
\pA{14} believed that an assistant that made artistic suggestions might be well received, \pQuote{[f]or instance, if I'm editing text, and I was given suggestions for font, color, etc.} %
Similarly, \pA{4} noted that it would be useful receive suggestions to help inspire their designs, for example, through \pQuote{videos on youtube, images, articles.}
Some features, such as number sliders, were highlighted as being only valuable \pQuote{when I'm playing around with my final result} (\pC{17}), but that they \pQuote{encourage a lot of experimentation and creativity} (\pA{4}).
These observations align with prior work, which highlighted the value of providing assistance in exploring the space of possible designs in creative coding contexts~\cite{mitchell2013towards} and in creativity support tools generally~\cite{shneiderman2007creativity}.
Whether a student's primary purpose was closer to coding or to making art was an additional source of skepticism to those in \secref{sec:skeptics}.
Again, regarding Shape Toolbox:
\pQuote{This would be great, but would reduce the amount of time figuring out the code. This would make it more an explicit art tool, and less a `make art with code' tool}~(\pA{16}).
\pA{5} similarly noted that \pQuote{feels a little too much like draw-ing for my taste} and took the class
\pQuote{with the primary goal of getting better at programming so I'd want to do things the code-y way}.
Similarly, \pC{17} felt that it \pQuote{hinder[ed] the process of creative discovery-- including trial and error}, however this was mostly not an issue as they sought to be \pQuote{to be more accurate than creative} in this course.
While it is natural to want tools to be familiar, we believe that new authoring paradigms
(\eg{} bidirectional programming)
should be viewed as complementary rather than antagonistic.

\section{Discussion}\label{sec:discussion}

This paper explored the observed behaviors and surveyed perceptions of novice programmers in a creative coding course. To wrap up, we recap the main themes, reflect on the connection between our work and other domains, describe limitations of our studies, and offer avenues for future work.

\subsection{Recap: Themes}

In our analysis, we chose four recurring themes to highlight.

\begin{description}[leftmargin=0cm, rightmargin=0cm, itemsep=1em]
      \item[\takeawayStatic{}\iconSep{} Static Analyses.]
            We observed that simple static analyses were seen as supportive of a variety of types of work---notable given that error messages sometimes are obstacles in introductory settings~\cite{becker2019errormessages}.
            Polite lightweight assistants that respect user agency, like those expressed through linting or auto-formatting, can be a helpful platform on which to learn and test new skills with confidence.
            On the other hand, \pA{5} noted that they \pQuote{would also probably prefer to do things by hand} rather than use advanced features because there lacked visual indicators of a particular action's effect%
            ---highlighting the importance of clear effect-forecasting for feature trust.

      \item[\takeawayLive{}\iconSep{} Liveness.]
            We saw that overeager evaluation can overwhelm and stress users through distracting updates that are unsynchronized with their expected edit-run cadence. Live programming offers enticing benefits for novice and creative contexts (\eg{} feedback immediacy or a closeness of mapping between code and graphics). Yet, these interaction challenges for non-expert settings are not yet thoroughly understood, leaving open questions about how to blend user control with system eagerness in a profitable way that maintains an experience level-attuned sense of agency.

      \item[\takeawayClutter{}\iconSep{} Clutter.]
            We noted that amateurs are mindful of how the editing space can become overwhelming if too much visual noise or unfamiliar forms of interaction are introduced.
            For instance, students are aware that individual features (\eg{} Number Scrubbers, lint errors, and autocomplete menus) can break their flow.
            We highlight the difficulty and importance of developing design guidelines that can aid the development of novel features within these constraints.

      \item[\takeawaySkeptic{}\iconSep{} Skepticism.]
            Finally, we discussed how user perceptions of a feature can inspire skepticism about its propriety in learning environments.
            Year 1 students believed that the Shape Toolbox would impede learning; however, those who used it in Year 2 did not share that concern, instead viewing it as a convenience.
            Year 2 students also saw knowledge assistants such as autocomplete as detrimental to their development as programmers.
            We believe it is valuable future work to better understand what types of features and knowledge assistants are likely to be viewed as detrimental.

\end{description}

\subsection{Connections to Other Domains}

Next we reflect on how our findings may apply more broadly.
\subsubsection{Programming Pedagogy}
Our work is merely situated within a classroom; we do not seek to make claims about the learning effects of the features we studied---this is an important, separable direction for future work.
Yet, some of our themes may carry over to pedagogically-minded editors in more general learning contexts.

We suggest that skepticism\takeawaySkeptic{} about features perceived as being too useful, such as autocomplete, may continue to be prevalent in learning contexts. Such concerns might be circumvented by emphasizing tools that help correct, rather than help complete, such as how linting~\takeawayStatic{} can identify an error while also providing justification and explanation for that error.
We also note the value of having a programming environment that is perceived as being approachable (\pA{3,5}, \pB{1}).
Furthermore, tools having not \pQuote{got in the way}~(\pC{16}) or otherwise cluttering~\takeawayClutter{} the display in unhelpful ways seem intuitively valuable, particularly in learning contexts.
Similarly, live execution\takeawayLive{} may be beneficial in non-visual contexts as it promotes immediate feedback, such as by rerunning a test suite dynamically, as in Jest's watch mode~\cite{JestWatchMode} or Huang \etals{} use of projection boxes~\cite{ProjectionBoxesInClass} in a classroom  to expose live values.

Finally, like others before us~\cite{greenberg2007processing, Malita20From, wood2016computational}, we found that a curriculum centered around media-art topics%
---as opposed to more abstract content often found in intro CS courses---%
invited a broad range of students who might not otherwise study CS in a formal setting (\theAppendix{} lists majors represented in the courses).
\subsubsection{Other Domains}
Editors specialized to a given domain can make adaptations that aid that context. In this work we focused on creative coding and designed affordances specific to this domain, however our findings might be applied in related contexts.
We highlight the value of bidirectional editing, linting, and designing editors with their effects on creativity in mind.

Bidirectional synchronization of code and effect (such as in our Shape Toolbox) seems to be an especially valuable approach in domains that  have a prominent visual component. This has been explored by \citet{asai2020integrated} as a mechanism to clean and synthesize data for statistical modeling, as well by \citet{deline2021glinda} and \citet{wu2020B2} for data science tasks such as modeling and analysis.
We suggest such synchronization might be usefully applied to other visualization contexts (like preparing charts for presentation), as well as other creative coding contexts.
Such interfaces may potentially reduce tedium in certain tasks and, more fundamentally, may provide opportunities for learning about the domain, for example, demonstrating how to achieve a particular effect using code.

Next, we highlight that linters (or other static analysis tools~\takeawayStatic{}) can provide a straightforward channel for introducing newcomers to basic principles and best practices of a particular domain.
While they have already been explored in some contexts---such as for spreadsheets~\cite{barowy2018excelint} and visualizations~\cite{mcnutt2020surfacing, hopkins2020visualint}---additional fields such as data science~\cite{mcnutt2023MetaCells} and music editors might integrate these concepts as well in order to surface best practices, such as highlighting statistical fallacies, helping guide usage with unusual tools
(\eg{} Orca~\cite{Orcas}),
or surfacing accidental discordance or inaudible components in music editors.
As discussed, such ambient assistants should be designed in a polite manner (e.g., through granular dismissal of advice) to avoid being irritating and then dismissed.

While most technical tasks require some amount of creativity, we argue that features in editors in creativity centered-domains should be constructed in order to align specifically with goals of either reducing tedium or aiding in exploration.
\citet{barke2022grounded} observe a similar pattern of exploration vs. acceleration in use of the AI-powered code assistant Copilot for traditional, non-creative software development tasks, suggesting overlap in editor features that support creative coding and coding more generally.
\citet{compton2021conversation} argues for IDEs with features that are valuable unto themselves---for example, for being playful or thought-provoking---rather than their use as a means to end. %
Non-productivity focused techniques may be useful in creative coding contexts more generally, perhaps as a design advisor as \pA{4} suggested.
These additions may drive unexpected patterns of usage, leading to new types of discovery through play---which might even valuable in technical domains like data science or visualization~\cite{2020-data-crafting}. %
At the same time, such interventions may inspire skepticism \takeawaySkeptic{} about their authenticity if they are perceived as too whimsical or unrealistic.

\subsection{Limitations and Future Work}

As described throughout, our study had a number of limitations. These included data collection errors (such as the configuration error in \wiq{}) and the relative simplicity of the survey.
For instance, our use of static images---as opposed to videos or interactive prototypes---limited our ability to accurately explore reactions to proposed features. However, the use of non-interactive stimuli (following~\citet{kery2020mage}) allowed respondents to project their own beliefs about the features and ignore potentially distracting low-level bugs or stylistic issues.
Further, we only implemented a subset all designs we identified, so we cannot make inferences about what features would be most valuable in general. Instead, we focus only on the observed themes and interactions with implemented features.
This approach was a coarse and inexpensive way to identify and explore some potentially fruitful features, however not all such features were necessarily identified nor considered.
Future work could implement more of the identified features---and also augment our observations with lab studies---to better understand the effects of particular features.
Furthermore, whereas our work investigated how novices \emph{perceive} the utility of various editor features, subsequent work should also investigate their \emph{pedagogical} effects on learning outcomes---%
one notable point for comparison is that Oviatt \etal{} \cite{oviatt2006quiet} found novel interfaces can hinder learning.

The biases of our particular student populations may not be reflective of a more general student population,
however the views of the college-aged (\spq{}, \wiq{}) students seem aligned with those of the high-school students (\sumtwo{}). In addition, they are in agreement with those of \imm{}  students who, because of pandemic era-distancing, attended from around the world and thus drawn from a substantially different population.
Our own biases were likely projected onto the students in teaching this material, and different instructors may have inspired different responses in students.
To this end, student perceptions are likely reflective of the context and content of the work they were asked to do.
For instance, the open-ended nature of many assignments likely shaped student opinions of the features we asked about, which may have been different under more structured programming tasks.
In future work we would like to reexamine our findings by teaching the course to and soliciting feedback from students from other institutions, age groups, and backgrounds.
Students were generally positive about the editor being online and the way in which our feedback and submission systems were integrated (\pA{3,5}), with \pB{1} noting that they were especially beginner-friendly.
Nevertheless, the choice to use a web tool had limitations.
Students with inconsistent internet connections struggled with the online environment (prompting \pB{6} to suggest an offline mode), while others had computers that were unable to handle the computational weight of a larger web application (which made some students hesitant to explore some editor features).
For instance, \pA{2} noted that they hesitated to use auto-refresh because \pQuote{my computer was already very slow and I didn't want my code to crash while it was running.}
These concerns were particularly prominent during the fully online \spq{} and \imm{} editions.
While in-person teaching has resumed (as in \wiq{} and \sumtwo{}), that consideration of how to build novice-oriented tools that support those with limited internet connectivity or less powerful computers should not cease.
While our target population in this work was students,
in future work we wish to understand what features instructors see as valuable or concerning in such a setting.
Similarly, it would be useful to consider whether these user interface patterns are applicable to professional artists working in creative coding spaces---questions which are closely connected to Li \etals{}~\cite{li2021we} study of the tools that artists make for themselves.
Of particular relevance are artist-designed custom coding environments used for teaching and artistic practice (such as Field~\cite{FieldDownie}).

In sum,
creative coding has been, and continues to be, fertile soil for HCI research.
We believe that studying the problems users in these creative domains face is valuable unto itself, and is ever more relevant as creative coding becomes an increasingly common way to introduce computing and to make art.
\begin{acks}
    We are grateful to those who made our courses possible, including the course staff (Brian Hempel, Angela Liu, and Bhakti Shah), Kevin Workman for allowing us to incorporate his Happy Coding tutorials, and the p5 community and developers for building such useful tools.
    We thank
    Lilian Huang,
    Shriram Krishnamurthi,
    Elsie Lee-Robbins,
    Justin Lubin,
    and the anonymous reviewers
    for their helpful commentary.
    Finally, we thank our students, without whom this work could not have taken place.
    This work was supported in part by the University of Chicago College Innovation Fund.
\end{acks}

\bibliographystyle{ACM-Reference-Format}
\bibliography{./main}


\begin{thebibliography}{108}


\ifx \showCODEN    \undefined \def \showCODEN     #1{\unskip}     \fi
\ifx \showDOI      \undefined \def \showDOI       #1{#1}\fi
\ifx \showISBNx    \undefined \def \showISBNx     #1{\unskip}     \fi
\ifx \showISBNxiii \undefined \def \showISBNxiii  #1{\unskip}     \fi
\ifx \showISSN     \undefined \def \showISSN      #1{\unskip}     \fi
\ifx \showLCCN     \undefined \def \showLCCN      #1{\unskip}     \fi
\ifx \shownote     \undefined \def \shownote      #1{#1}          \fi
\ifx \showarticletitle \undefined \def \showarticletitle #1{#1}   \fi
\ifx \showURL      \undefined \def \showURL       {\relax}        \fi
\providecommand\bibfield[2]{#2}
\providecommand\bibinfo[2]{#2}
\providecommand\natexlab[1]{#1}
\providecommand\showeprint[2][]{arXiv:#2}

\bibitem[p5(2021)]%
        {p5}
 \bibinfo{year}{{2021}}\natexlab{}.
\newblock \bibinfo{title}{p5.js}.
\newblock \bibinfo{howpublished}{\url{https://p5js.org/}}.
\newblock
\newblock
\shownote{Accessed 9/21/21}.


\bibitem[p5S(2021)]%
        {p5Slider}
 \bibinfo{year}{{2021}}\natexlab{}.
\newblock \bibinfo{title}{p5.js: createSlider}.
\newblock
  \bibinfo{howpublished}{\url{https://p5js.org/reference/\#/p5/createSlider}}.
\newblock
\newblock
\shownote{Accessed 9/17/21}.


\bibitem[p5e(2021)]%
        {p5editor}
 \bibinfo{year}{{2021}}\natexlab{}.
\newblock \bibinfo{title}{p5.js editor}.
\newblock
  \bibinfo{howpublished}{\url{https://github.com/processing/p5.js-web-editor}}.
\newblock
\newblock
\shownote{Accessed 9/17/21}.


\bibitem[Uto(2021)]%
        {Utopia}
 \bibinfo{year}{2021}\natexlab{}.
\newblock \bibinfo{title}{Utopia}.
\newblock
  \bibinfo{howpublished}{\url{https://github.com/concrete-utopia/utopia}}.
\newblock


\bibitem[Twe(2022)]%
        {Tweakable}
 \bibinfo{year}{2022}\natexlab{}.
\newblock \bibinfo{title}{Tweakable: an online programming environment for
  audio and video}.
\newblock \bibinfo{howpublished}{\url{https://tweakable.org/}}.
\newblock
\newblock
\shownote{Accessed 8/25/22}.


\bibitem[Academy(2021)]%
        {khanAcademy}
\bibfield{author}{\bibinfo{person}{Khan Academy}.}
  \bibinfo{year}{{2021}}\natexlab{}.
\newblock \bibinfo{title}{Computer Programming}.
\newblock
  \bibinfo{howpublished}{\url{https://www.khanacademy.org/computing/computer-programming}}.
\newblock
\newblock
\shownote{Accessed 4/3/2022}.


\bibitem[Alaboudi and LaToza(2021)]%
        {alaboudi2021edit}
\bibfield{author}{\bibinfo{person}{Abdulaziz Alaboudi} {and}
  \bibinfo{person}{Thomas~D LaToza}.} \bibinfo{year}{2021}\natexlab{}.
\newblock \showarticletitle{Edit-Run Behavior in Programming and Debugging}. In
  \bibinfo{booktitle}{\emph{Symposium on Visual Languages and Human-Centric
  Computing (VL/HCC)}}. \bibinfo{publisher}{{IEEE}}, \bibinfo{pages}{1--10}.
\newblock
\urldef\tempurl%
\url{https://doi.org/10.1109/VL/HCC51201.2021.9576170}
\showDOI{\tempurl}


\bibitem[Ambrose et~al\mbox{.}(2010)]%
        {ambrose2010learning}
\bibfield{author}{\bibinfo{person}{Susan~A Ambrose}, \bibinfo{person}{Michael~W
  Bridges}, \bibinfo{person}{Michele DiPietro}, \bibinfo{person}{Marsha~C
  Lovett}, {and} \bibinfo{person}{Marie~K Norman}.}
  \bibinfo{year}{2010}\natexlab{}.
\newblock \bibinfo{booktitle}{\emph{How Learning Works: Seven Research-based
  Principles for Smart Teaching}}.
\newblock \bibinfo{publisher}{John Wiley \& Sons}, \bibinfo{address}{New York}.
\newblock


\bibitem[Andersen et~al\mbox{.}(2020)]%
        {andersen2020adding}
\bibfield{author}{\bibinfo{person}{Leif Andersen}, \bibinfo{person}{Michael
  Ballantyne}, {and} \bibinfo{person}{Matthias Felleisen}.}
  \bibinfo{year}{2020}\natexlab{}.
\newblock \showarticletitle{Adding interactive visual syntax to textual code}.
\newblock \bibinfo{journal}{\emph{Proceedings of the ACM on Programming
  Languages (OOPSLA)}}  \bibinfo{volume}{4} (\bibinfo{year}{2020}),
  \bibinfo{pages}{1--28}.
\newblock


\bibitem[Asai et~al\mbox{.}(2020)]%
        {asai2020integrated}
\bibfield{author}{\bibinfo{person}{Kentaro Asai}, \bibinfo{person}{Tsukasa
  Fukusato}, {and} \bibinfo{person}{Takeo Igarashi}.}
  \bibinfo{year}{2020}\natexlab{}.
\newblock \showarticletitle{Integrated Development Environment with Interactive
  Scatter Plot for Examining Statistical Modeling}. In
  \bibinfo{booktitle}{\emph{{SIGCHI Conference on Human Factors in Computing
  Systems}}}. \bibinfo{pages}{1--7}.
\newblock


\bibitem[Ball et~al\mbox{.}(2019)]%
        {ball2019microsoft}
\bibfield{author}{\bibinfo{person}{Thomas Ball}, \bibinfo{person}{Abhijith
  Chatra}, \bibinfo{person}{Peli de Halleux}, \bibinfo{person}{Steve Hodges},
  \bibinfo{person}{Michal Moskal}, {and} \bibinfo{person}{Jacqueline Russell}.}
  \bibinfo{year}{2019}\natexlab{}.
\newblock \showarticletitle{Microsoft MakeCode: Embedded Programming for
  Education, in Blocks and TypeScript}. In \bibinfo{booktitle}{\emph{{ACM}
  {SIGPLAN} Workshop on SPLASH-E}}. \bibinfo{publisher}{{ACM}},
  \bibinfo{pages}{7--12}.
\newblock
\urldef\tempurl%
\url{https://doi.org/10.1145/3358711.3361630}
\showDOI{\tempurl}


\bibitem[Barke et~al\mbox{.}(2022)]%
        {barke2022grounded}
\bibfield{author}{\bibinfo{person}{Shraddha Barke}, \bibinfo{person}{Michael~B
  James}, {and} \bibinfo{person}{Nadia Polikarpova}.}
  \bibinfo{year}{2022}\natexlab{}.
\newblock \showarticletitle{Grounded Copilot: How Programmers Interact with
  Code-Generating Models}.
\newblock \bibinfo{journal}{\emph{arXiv preprint arXiv:2206.15000}}
  (\bibinfo{year}{2022}).
\newblock


\bibitem[Barowy et~al\mbox{.}(2018)]%
        {barowy2018excelint}
\bibfield{author}{\bibinfo{person}{Daniel~W Barowy}, \bibinfo{person}{Emery~D
  Berger}, {and} \bibinfo{person}{Benjamin Zorn}.}
  \bibinfo{year}{2018}\natexlab{}.
\newblock \showarticletitle{ExceLint: automatically finding spreadsheet formula
  errors}.
\newblock \bibinfo{journal}{\emph{Proceedings of the ACM on Programming
  Languages (OOPSLA)}}  \bibinfo{volume}{2} (\bibinfo{year}{2018}),
  \bibinfo{pages}{1--26}.
\newblock


\bibitem[Bartram et~al\mbox{.}(2022)]%
        {bartram2021untidy}
\bibfield{author}{\bibinfo{person}{Lyn Bartram}, \bibinfo{person}{Michael
  Correll}, {and} \bibinfo{person}{Melanie Tory}.}
  \bibinfo{year}{2022}\natexlab{}.
\newblock \showarticletitle{Untidy Data: The Unreasonable Effectiveness of
  Tables}.
\newblock \bibinfo{journal}{\emph{{IEEE Transactions on Visualization and
  Computer Graphics}}} \bibinfo{volume}{28}, \bibinfo{number}{1}
  (\bibinfo{year}{2022}), \bibinfo{pages}{686--696}.
\newblock
\urldef\tempurl%
\url{https://doi.org/10.1109/TVCG.2021.3114830}
\showDOI{\tempurl}


\bibitem[beautify web(2021)]%
        {jsBeautify}
\bibfield{author}{\bibinfo{person}{beautify web}.}
  \bibinfo{year}{{2021}}\natexlab{}.
\newblock \bibinfo{title}{JSBeautify}.
\newblock
  \bibinfo{howpublished}{\url{https://github.com/beautify-web/js-beautify}}.
\newblock
\newblock
\shownote{Accessed 4/3/2022}.


\bibitem[Becker et~al\mbox{.}(2019)]%
        {becker2019errormessages}
\bibfield{author}{\bibinfo{person}{Brett~A. Becker}, \bibinfo{person}{Paul
  Denny}, \bibinfo{person}{Raymond Pettit}, \bibinfo{person}{Durell Bouchard},
  \bibinfo{person}{Dennis~J. Bouvier}, \bibinfo{person}{Brian Harrington},
  \bibinfo{person}{Amir Kamil}, \bibinfo{person}{Amey Karkare},
  \bibinfo{person}{Chris McDonald}, \bibinfo{person}{Peter{-}Michael Osera},
  \bibinfo{person}{Janice~L. Pearce}, {and} \bibinfo{person}{James Prather}.}
  \bibinfo{year}{2019}\natexlab{}.
\newblock \showarticletitle{Compiler Error Messages Considered Unhelpful: The
  Landscape of Text-Based Programming Error Message Research}. In
  \bibinfo{booktitle}{\emph{Working Group Reports on Innovation and Technology
  in Computer Science Education, ITiCSE-WGR}}. \bibinfo{publisher}{{ACM}},
  \bibinfo{pages}{177--210}.
\newblock
\urldef\tempurl%
\url{https://doi.org/10.1145/3344429.3372508}
\showDOI{\tempurl}


\bibitem[Beckwith et~al\mbox{.}(2006)]%
        {Beckwith06Tinkering}
\bibfield{author}{\bibinfo{person}{Laura Beckwith}, \bibinfo{person}{Cory
  Kissinger}, \bibinfo{person}{Margaret~M. Burnett}, \bibinfo{person}{Susan
  Wiedenbeck}, \bibinfo{person}{Joseph Lawrance}, \bibinfo{person}{Alan~F.
  Blackwell}, {and} \bibinfo{person}{Curtis~R. Cook}.}
  \bibinfo{year}{2006}\natexlab{}.
\newblock \showarticletitle{Tinkering and gender in end-user programmers'
  debugging}. In \bibinfo{booktitle}{\emph{{SIGCHI Conference on Human Factors
  in Computing Systems}}}. \bibinfo{publisher}{{ACM}},
  \bibinfo{pages}{231--240}.
\newblock
\urldef\tempurl%
\url{https://doi.org/10.1145/1124772.1124808}
\showDOI{\tempurl}


\bibitem[Bragdon et~al\mbox{.}(2010)]%
        {bragdon2010code}
\bibfield{author}{\bibinfo{person}{Andrew Bragdon}, \bibinfo{person}{Steven~P.
  Reiss}, \bibinfo{person}{Robert~C. Zeleznik}, \bibinfo{person}{Suman
  Karumuri}, \bibinfo{person}{William Cheung}, \bibinfo{person}{Joshua Kaplan},
  \bibinfo{person}{Christopher Coleman}, \bibinfo{person}{Ferdi Adeputra},
  {and} \bibinfo{person}{Joseph J.~LaViola Jr.}}
  \bibinfo{year}{2010}\natexlab{}.
\newblock \showarticletitle{Code bubbles: rethinking the user interface
  paradigm of integrated development environments}. In
  \bibinfo{booktitle}{\emph{International Conference on Software Engineering
  {(ICSE)}}}. \bibinfo{publisher}{{ACM}}, \bibinfo{pages}{455--464}.
\newblock
\urldef\tempurl%
\url{https://doi.org/10.1145/1806799.1806866}
\showDOI{\tempurl}


\bibitem[Brandt et~al\mbox{.}(2010)]%
        {brandt2010example}
\bibfield{author}{\bibinfo{person}{Joel Brandt}, \bibinfo{person}{Mira
  Dontcheva}, \bibinfo{person}{Marcos Weskamp}, {and} \bibinfo{person}{Scott~R
  Klemmer}.} \bibinfo{year}{2010}\natexlab{}.
\newblock \showarticletitle{Example-centric programming: integrating web search
  into the development environment}. In \bibinfo{booktitle}{\emph{{SIGCHI
  Conference on Human Factors in Computing Systems}}}.
  \bibinfo{pages}{513--522}.
\newblock


\bibitem[Burgess et~al\mbox{.}(2020)]%
        {burgess2020stamper}
\bibfield{author}{\bibinfo{person}{Cameron Burgess}, \bibinfo{person}{Dan
  Lockton}, \bibinfo{person}{Maayan Albert}, {and} \bibinfo{person}{Daniel
  Cardoso~Llach}.} \bibinfo{year}{2020}\natexlab{}.
\newblock \showarticletitle{Stamper: An Artboard-Oriented Creative Coding
  Environment}. In \bibinfo{booktitle}{\emph{Extended Abstracts of the CHI
  Conference on Human Factors in Computing Systems}}.
  \bibinfo{publisher}{{ACM}}, \bibinfo{pages}{1--9}.
\newblock
\urldef\tempurl%
\url{https://doi.org/10.1145/3334480.3382994}
\showDOI{\tempurl}


\bibitem[Burnett et~al\mbox{.}(2016)]%
        {Burnett16Finding}
\bibfield{author}{\bibinfo{person}{Margaret~M. Burnett},
  \bibinfo{person}{Anicia Peters}, \bibinfo{person}{Charles Hill}, {and}
  \bibinfo{person}{Noha Elarief}.} \bibinfo{year}{2016}\natexlab{}.
\newblock \showarticletitle{Finding Gender-Inclusiveness Software Issues with
  GenderMag: {A} Field Investigation}. In \bibinfo{booktitle}{\emph{{SIGCHI
  Conference on Human Factors in Computing Systems}}}.
  \bibinfo{publisher}{{ACM}}, \bibinfo{pages}{2586--2598}.
\newblock
\urldef\tempurl%
\url{https://doi.org/10.1145/2858036.2858274}
\showDOI{\tempurl}


\bibitem[Cao(2021)]%
        {umami}
\bibfield{author}{\bibinfo{person}{Mike Cao}.}
  \bibinfo{year}{{2021}}\natexlab{}.
\newblock \bibinfo{title}{Umami}.
\newblock \bibinfo{howpublished}{\url{https://umami.is/}}.
\newblock
\newblock
\shownote{Accessed 4/3/2022}.


\bibitem[Carter et~al\mbox{.}(2015)]%
        {carter2015normalized}
\bibfield{author}{\bibinfo{person}{Adam~S Carter},
  \bibinfo{person}{Christopher~D Hundhausen}, {and} \bibinfo{person}{Olusola
  Adesope}.} \bibinfo{year}{2015}\natexlab{}.
\newblock \showarticletitle{The Normalized Programming State Model: Predicting
  Student Performance in Computing Courses Based on Programming Behavior}. In
  \bibinfo{booktitle}{\emph{Proceedings of the eleventh annual International
  Conference on International Computing Education Research}}.
  \bibinfo{publisher}{{ACM}}, \bibinfo{pages}{141--150}.
\newblock
\urldef\tempurl%
\url{https://doi.org/10.1145/2787622.2787710}
\showDOI{\tempurl}


\bibitem[Compton(2021)]%
        {compton2021conversation}
\bibfield{author}{\bibinfo{person}{Kate Compton}.}
  \bibinfo{year}{2021}\natexlab{}.
\newblock \showarticletitle{Conversation Starter: Imagining Autotelic IDEs}. In
  \bibinfo{booktitle}{\emph{CEUR Workshop Proceedings}},
  Vol.~\bibinfo{volume}{3217}. CEUR-WS.
\newblock


\bibitem[Compton et~al\mbox{.}(2015)]%
        {compton2015tracery}
\bibfield{author}{\bibinfo{person}{Kate Compton}, \bibinfo{person}{Ben
  Kybartas}, {and} \bibinfo{person}{Michael Mateas}.}
  \bibinfo{year}{2015}\natexlab{}.
\newblock \showarticletitle{Tracery: An Author-Focused Generative Text Tool}.
  In \bibinfo{booktitle}{\emph{International Conference on Interactive Digital
  Storytelling}}. Springer, \bibinfo{pages}{154--161}.
\newblock
\urldef\tempurl%
\url{https://doi.org/10.1007/978-3-319-27036-4\_14}
\showDOI{\tempurl}


\bibitem[CSSLint(2021)]%
        {cssLint}
\bibfield{author}{\bibinfo{person}{CSSLint}.}
  \bibinfo{year}{{2021}}\natexlab{}.
\newblock \bibinfo{title}{CSSLint}.
\newblock \bibinfo{howpublished}{\url{https://github.com/CSSLint/csslint}}.
\newblock
\newblock
\shownote{Accessed 4/3/2022}.


\bibitem[Davis(1989)]%
        {davis1989perceived}
\bibfield{author}{\bibinfo{person}{Fred~D Davis}.}
  \bibinfo{year}{1989}\natexlab{}.
\newblock \showarticletitle{Perceived Usefulness, Perceived Ease of Use, and
  User Acceptance of Information Technology}.
\newblock \bibinfo{journal}{\emph{MIS Quarterly}} (\bibinfo{year}{1989}),
  \bibinfo{pages}{319--340}.
\newblock


\bibitem[DeLine(2021)]%
        {deline2021glinda}
\bibfield{author}{\bibinfo{person}{Robert~A DeLine}.}
  \bibinfo{year}{2021}\natexlab{}.
\newblock \showarticletitle{Glinda: Supporting data science with live
  programming, GUIs and a Domain-specific Language}. In
  \bibinfo{booktitle}{\emph{{SIGCHI Conference on Human Factors in Computing
  Systems}}}. \bibinfo{pages}{1--11}.
\newblock


\bibitem[Do et~al\mbox{.}(2019)]%
        {do2019evaluating}
\bibfield{author}{\bibinfo{person}{Quan Do}, \bibinfo{person}{Kiersten
  Campbell}, \bibinfo{person}{Emmie Hine}, \bibinfo{person}{Dzung Pham},
  \bibinfo{person}{Alex Taylor}, \bibinfo{person}{Iris Howley}, {and}
  \bibinfo{person}{Daniel~W Barowy}.} \bibinfo{year}{2019}\natexlab{}.
\newblock \showarticletitle{Evaluating ProDirect Manipulation in Hour of Code}.
  In \bibinfo{booktitle}{\emph{ACM SIGPLAN Symposium on SPLASH-E}}.
  \bibinfo{pages}{25--35}.
\newblock
\urldef\tempurl%
\url{https://doi.org/10.1145/3358711.3361623}
\showDOI{\tempurl}


\bibitem[Downie and Kaiser(2021)]%
        {FieldDownie}
\bibfield{author}{\bibinfo{person}{Marc Downie} {and} \bibinfo{person}{Paul
  Kaiser}.} \bibinfo{year}{2021}\natexlab{}.
\newblock \bibinfo{title}{Field}.
\newblock \bibinfo{howpublished}{\url{http://openendedgroup.com/field/}}.
\newblock


\bibitem[Edwards(2005)]%
        {edwards2005subtext}
\bibfield{author}{\bibinfo{person}{Jonathan Edwards}.}
  \bibinfo{year}{2005}\natexlab{}.
\newblock \showarticletitle{Subtext: Uncovering the Simplicity of Programming}.
  In \bibinfo{booktitle}{\emph{{ACM} {SIGPLAN} Conference on Object-Oriented
  Programming, Systems, Languages, and Applications, {OOPSLA}}}.
  \bibinfo{pages}{505--518}.
\newblock
\urldef\tempurl%
\url{https://doi.org/10.1145/1094811.1094851}
\showDOI{\tempurl}


\bibitem[Facebook(2022)]%
        {JestWatchMode}
\bibfield{author}{\bibinfo{person}{Facebook}.}
  \bibinfo{year}{{2022}}\natexlab{}.
\newblock \bibinfo{title}{Jest CLI Options}.
\newblock \bibinfo{howpublished}{\url{https://jestjs.io/docs/cli}}.
\newblock
\newblock
\shownote{Accessed 11/15/20}.


\bibitem[Fowkes et~al\mbox{.}(2017)]%
        {fowkes2017autofolding}
\bibfield{author}{\bibinfo{person}{Jaroslav Fowkes}, \bibinfo{person}{Pankajan
  Chanthirasegaran}, \bibinfo{person}{Razvan Ranca}, \bibinfo{person}{Miltiadis
  Allamanis}, \bibinfo{person}{Mirella Lapata}, {and} \bibinfo{person}{Charles
  Sutton}.} \bibinfo{year}{2017}\natexlab{}.
\newblock \showarticletitle{Autofolding for Source Code Summarization}.
\newblock \bibinfo{journal}{\emph{IEEE Transactions on Software Engineering}}
  \bibinfo{volume}{43}, \bibinfo{number}{12} (\bibinfo{year}{2017}),
  \bibinfo{pages}{1095--1109}.
\newblock
\urldef\tempurl%
\url{https://doi.org/10.1109/TSE.2017.2664836}
\showDOI{\tempurl}


\bibitem[Fraietta et~al\mbox{.}(2019)]%
        {fraietta2019rapid}
\bibfield{author}{\bibinfo{person}{Angelo Fraietta}, \bibinfo{person}{Oliver
  Bown}, \bibinfo{person}{Sam Ferguson}, \bibinfo{person}{Sam Gillespie}, {and}
  \bibinfo{person}{Liam Bray}.} \bibinfo{year}{2019}\natexlab{}.
\newblock \showarticletitle{Rapid composition for networked devices:
  HappyBrackets}.
\newblock \bibinfo{journal}{\emph{Computer Music Journal}}
  \bibinfo{volume}{43}, \bibinfo{number}{2-3} (\bibinfo{year}{2019}),
  \bibinfo{pages}{89--108}.
\newblock


\bibitem[Frich et~al\mbox{.}(2018)]%
        {frich2018twenty}
\bibfield{author}{\bibinfo{person}{Jonas Frich}, \bibinfo{person}{Michael
  Mose~Biskjaer}, {and} \bibinfo{person}{Peter Dalsgaard}.}
  \bibinfo{year}{2018}\natexlab{}.
\newblock \showarticletitle{Twenty years of creativity research in
  human-computer interaction: Current state and future directions}. In
  \bibinfo{booktitle}{\emph{Proceedings of the 2018 Designing Interactive
  Systems Conference}}. \bibinfo{pages}{1235--1257}.
\newblock


\bibitem[Greenberg(2007)]%
        {greenberg2007processing}
\bibfield{author}{\bibinfo{person}{Ira Greenberg}.}
  \bibinfo{year}{2007}\natexlab{}.
\newblock \bibinfo{booktitle}{\emph{Processing: creative coding and
  computational art}}.
\newblock \bibinfo{publisher}{Apress}.
\newblock


\bibitem[Greenberg et~al\mbox{.}(2012)]%
        {greenberg2012creative}
\bibfield{author}{\bibinfo{person}{Ira Greenberg}, \bibinfo{person}{Deepak
  Kumar}, {and} \bibinfo{person}{Dianna Xu}.} \bibinfo{year}{2012}\natexlab{}.
\newblock \showarticletitle{Creative Coding and Visual Portfolios for {CS1}}.
  In \bibinfo{booktitle}{\emph{ACM Technical Symposium on Computer Science
  Education {(SIGCSE)}}}. \bibinfo{pages}{247--252}.
\newblock
\urldef\tempurl%
\url{https://doi.org/10.1145/2157136.2157214}
\showDOI{\tempurl}


\bibitem[Guo(2021)]%
        {Guo21Million}
\bibfield{author}{\bibinfo{person}{Philip Guo}.}
  \bibinfo{year}{2021}\natexlab{}.
\newblock \showarticletitle{{Ten Million Users and Ten Years Later: Python
  Tutor’s Design Guidelines for Building Scalable and Sustainable Research
  Software in Academia}}. In \bibinfo{booktitle}{\emph{{ACM Symposium on User
  Interface Software and Technology (UIST)}}}.
\newblock
\urldef\tempurl%
\url{https://doi.org/10.1145/3472749.3474819}
\showDOI{\tempurl}


\bibitem[Guo(2013)]%
        {Guo13tutor}
\bibfield{author}{\bibinfo{person}{Philip~J. Guo}.}
  \bibinfo{year}{2013}\natexlab{}.
\newblock \showarticletitle{Online Python Tutor: Embeddable Web-based Program
  Visualization for {CS} Education}. In \bibinfo{booktitle}{\emph{ACM Technical
  Symposium on Computer Science Education {(SIGCSE)}}}.
  \bibinfo{publisher}{{ACM}}, \bibinfo{pages}{579--584}.
\newblock
\urldef\tempurl%
\url{https://doi.org/10.1145/2445196.2445368}
\showDOI{\tempurl}


\bibitem[Guzdial(04  )]%
        {MediaCompTeach}
\bibfield{author}{\bibinfo{person}{Mark Guzdial}.}
  \bibinfo{year}{{2004--}}\natexlab{}.
\newblock \bibinfo{title}{{Media Computation Teachers Website}}.
\newblock
  \bibinfo{howpublished}{\url{http://coweb.cc.gatech.edu/mediaComp-teach}}.
\newblock


\bibitem[Guzdial(2013)]%
        {Guzdial2013}
\bibfield{author}{\bibinfo{person}{Mark Guzdial}.}
  \bibinfo{year}{2013}\natexlab{}.
\newblock \showarticletitle{{Exploring Hypotheses about Media Computation}}. In
  \bibinfo{booktitle}{\emph{ACM Conference on International Computing Education
  Research (ICER)}}.
\newblock


\bibitem[Guzdial and Forte(2005)]%
        {Guzdial2005}
\bibfield{author}{\bibinfo{person}{Mark Guzdial} {and} \bibinfo{person}{Andrea
  Forte}.} \bibinfo{year}{2005}\natexlab{}.
\newblock \showarticletitle{Design Process for a Non-Majors Computing Course}.
\newblock \bibinfo{journal}{\emph{ACM SIGCSE Bulletin}} \bibinfo{volume}{37},
  \bibinfo{number}{1} (\bibinfo{year}{2005}), \bibinfo{pages}{361--365}.
\newblock
\urldef\tempurl%
\url{https://doi.org/10.1145/1047344.1047468}
\showDOI{\tempurl}


\bibitem[Hartmann et~al\mbox{.}(2008)]%
        {hartmann2008design}
\bibfield{author}{\bibinfo{person}{Bj{\"o}rn Hartmann}, \bibinfo{person}{Loren
  Yu}, \bibinfo{person}{Abel Allison}, \bibinfo{person}{Yeonsoo Yang}, {and}
  \bibinfo{person}{Scott~R Klemmer}.} \bibinfo{year}{2008}\natexlab{}.
\newblock \showarticletitle{{Design as Exploration: Creating Interface
  Alternatives Through Parallel Authoring and Runtime Tuning}}. In
  \bibinfo{booktitle}{\emph{{ACM Symposium on User Interface Software and
  Technology (UIST)}}}. \bibinfo{pages}{91--100}.
\newblock
\urldef\tempurl%
\url{https://doi.org/10.1145/1449715.1449732}
\showDOI{\tempurl}


\bibitem[Hasimoto(2021)]%
        {Glisp}
\bibfield{author}{\bibinfo{person}{Baku Hasimoto}.}
  \bibinfo{year}{2021}\natexlab{}.
\newblock \bibinfo{title}{Glisp}.
\newblock \bibinfo{howpublished}{\url{https://github.com/baku89/glisp}}.
\newblock


\bibitem[Haverbeke et~al\mbox{.}(2021)]%
        {codeMirrorSix}
\bibfield{author}{\bibinfo{person}{Marijn Haverbeke} {et~al\mbox{.}}}
  \bibinfo{year}{{2021}}\natexlab{}.
\newblock \bibinfo{title}{Code Mirror 6}.
\newblock \bibinfo{howpublished}{\url{https://codemirror.net/6/}}.
\newblock
\newblock
\shownote{Accessed 4/3/22}.


\bibitem[Helminen et~al\mbox{.}(2013)]%
        {helminen2013recording}
\bibfield{author}{\bibinfo{person}{Juha Helminen}, \bibinfo{person}{Petri
  Ihantola}, {and} \bibinfo{person}{Ville Karavirta}.}
  \bibinfo{year}{2013}\natexlab{}.
\newblock \showarticletitle{Recording and Analyzing In-Browser Programming
  Sessions}. In \bibinfo{booktitle}{\emph{Koli Calling International Conference
  on Computing Education Research}}. \bibinfo{pages}{13--22}.
\newblock
\urldef\tempurl%
\url{https://doi.org/10.1145/2526968.2526970}
\showDOI{\tempurl}


\bibitem[Hempel and Chugh(2016)]%
        {hempel_semi-automated_2016}
\bibfield{author}{\bibinfo{person}{Brian Hempel} {and} \bibinfo{person}{Ravi
  Chugh}.} \bibinfo{year}{2016}\natexlab{}.
\newblock \showarticletitle{Semi-{Automated} {SVG} {Programming} via {Direct}
  {Manipulation}}. In \bibinfo{booktitle}{\emph{{ACM Symposium on User
  Interface Software and Technology (UIST)}}}. \bibinfo{pages}{379--390}.
\newblock
\urldef\tempurl%
\url{https://doi.org/10.1145/2984511.2984575}
\showDOI{\tempurl}


\bibitem[Hempel and Chugh(2022)]%
        {hempel2022maniposynth}
\bibfield{author}{\bibinfo{person}{Brian Hempel} {and} \bibinfo{person}{Ravi
  Chugh}.} \bibinfo{year}{2022}\natexlab{}.
\newblock \showarticletitle{Maniposynth: Bimodal Tangible Functional
  Programming}. In \bibinfo{booktitle}{\emph{European Conference on
  Object-Oriented Programming, {ECOOP}}} \emph{(\bibinfo{series}{LIPIcs},
  Vol.~\bibinfo{volume}{222})}. \bibinfo{publisher}{Schloss Dagstuhl -
  Leibniz-Zentrum f{\"{u}}r Informatik}, \bibinfo{pages}{16:1--16:29}.
\newblock
\urldef\tempurl%
\url{https://doi.org/10.4230/LIPIcs.ECOOP.2022.16}
\showDOI{\tempurl}


\bibitem[Hempel et~al\mbox{.}(2019)]%
        {hempel_sketch-n-sketch_2019}
\bibfield{author}{\bibinfo{person}{Brian Hempel}, \bibinfo{person}{Justin
  Lubin}, {and} \bibinfo{person}{Ravi Chugh}.} \bibinfo{year}{2019}\natexlab{}.
\newblock \showarticletitle{Sketch-n-{Sketch}: {Output}-{Directed}
  {Programming} for {SVG}}. In \bibinfo{booktitle}{\emph{ACM Symposium on User
  Interface Software and Technology (UIST)}}. \bibinfo{pages}{281--292}.
\newblock
\urldef\tempurl%
\url{https://doi.org/10.1145/3332165.3347925}
\showDOI{\tempurl}


\bibitem[Hendrix et~al\mbox{.}(1998)]%
        {hendrix1998visual}
\bibfield{author}{\bibinfo{person}{T~Dean Hendrix}, \bibinfo{person}{James~H
  Cross}, \bibinfo{person}{Larry~A Barowski}, {and} \bibinfo{person}{Karl~S
  Mathias}.} \bibinfo{year}{1998}\natexlab{}.
\newblock \showarticletitle{Visual Support for Incremental Abstraction and
  Refinement in Ada 95}.
\newblock \bibinfo{journal}{\emph{SIGAda Annual International Conference on Ada
  Technology}} \bibinfo{volume}{18}, \bibinfo{number}{6}
  (\bibinfo{year}{1998}), \bibinfo{pages}{142--147}.
\newblock
\urldef\tempurl%
\url{https://doi.org/10.1145/289524.289568}
\showDOI{\tempurl}


\bibitem[Hopkins et~al\mbox{.}(2020)]%
        {hopkins2020visualint}
\bibfield{author}{\bibinfo{person}{Aspen~K Hopkins}, \bibinfo{person}{Michael
  Correll}, {and} \bibinfo{person}{Arvind Satyanarayan}.}
  \bibinfo{year}{2020}\natexlab{}.
\newblock \showarticletitle{VisuaLint: Sketchy in situ annotations of chart
  construction errors}. In \bibinfo{booktitle}{\emph{Computer Graphics Forum}},
  Vol.~\bibinfo{volume}{39}. Wiley Online Library, \bibinfo{pages}{219--228}.
\newblock


\bibitem[Huang et~al\mbox{.}(2022)]%
        {ProjectionBoxesInClass}
\bibfield{author}{\bibinfo{person}{Ruanqianqian~(Lisa) Huang},
  \bibinfo{person}{Kasra Ferdowsi}, \bibinfo{person}{Ana Selvaraj},
  \bibinfo{person}{Adalbert~Gerald Soosai~Raj}, {and} \bibinfo{person}{Sorin
  Lerner}.} \bibinfo{year}{2022}\natexlab{}.
\newblock \showarticletitle{Investigating the Impact of Using a Live
  Programming Environment in a CS1 Course}. In \bibinfo{booktitle}{\emph{ACM
  Technical Symposium on Computer Science Education {(SIGCSE)}}}
  \emph{(\bibinfo{series}{SIGCSE 2022})}. \bibinfo{publisher}{Association for
  Computing Machinery}, \bibinfo{pages}{495–501}.
\newblock
\showISBNx{9781450390705}
\urldef\tempurl%
\url{https://doi.org/10.1145/3478431.3499305}
\showDOI{\tempurl}


\bibitem[Hundhausen et~al\mbox{.}(2009)]%
        {hundhausen2009can}
\bibfield{author}{\bibinfo{person}{Christopher~D Hundhausen},
  \bibinfo{person}{Sean~F Farley}, {and} \bibinfo{person}{Jonathan~L Brown}.}
  \bibinfo{year}{2009}\natexlab{}.
\newblock \showarticletitle{Can Direct Manipulation Lower the Barriers to
  Computer Programming and Promote Transfer of Training? An Experimental
  Study}.
\newblock \bibinfo{journal}{\emph{ACM Transactions on Computer-Human
  Interaction (TOCHI)}} \bibinfo{volume}{16}, \bibinfo{number}{3}
  (\bibinfo{year}{2009}), \bibinfo{pages}{1--40}.
\newblock
\urldef\tempurl%
\url{https://doi.org/10.1145/1592440.1592442}
\showDOI{\tempurl}


\bibitem[Hundhausen et~al\mbox{.}(2017)]%
        {hundhausen2017ide}
\bibfield{author}{\bibinfo{person}{Christopher~David Hundhausen},
  \bibinfo{person}{Daniel~M Olivares}, {and} \bibinfo{person}{Adam~S Carter}.}
  \bibinfo{year}{2017}\natexlab{}.
\newblock \showarticletitle{IDE-Based Learning Analytics for Computing
  Education: {A} Process Model, Critical Review, and Research Agenda}.
\newblock \bibinfo{journal}{\emph{ACM Transactions on Computing Education
  (TOCE)}} \bibinfo{volume}{17}, \bibinfo{number}{3} (\bibinfo{year}{2017}),
  \bibinfo{pages}{1--26}.
\newblock
\urldef\tempurl%
\url{https://doi.org/10.1145/3105759}
\showDOI{\tempurl}


\bibitem[hundredrabbits(2021)]%
        {Orcas}
\bibfield{author}{\bibinfo{person}{hundredrabbits}.}
  \bibinfo{year}{{2021}}\natexlab{}.
\newblock \bibinfo{title}{Orca}.
\newblock \bibinfo{howpublished}{\url{https://github.com/hundredrabbits/Orca}}.
\newblock
\newblock
\shownote{Accessed 9/21/21}.


\bibitem[Ihantola et~al\mbox{.}(2015)]%
        {ihantola2015educational}
\bibfield{author}{\bibinfo{person}{Petri Ihantola}, \bibinfo{person}{Arto
  Vihavainen}, \bibinfo{person}{Alireza Ahadi}, \bibinfo{person}{Matthew
  Butler}, \bibinfo{person}{J{\"u}rgen B{\"o}rstler},
  \bibinfo{person}{Stephen~H Edwards}, \bibinfo{person}{Essi Isohanni},
  \bibinfo{person}{Ari Korhonen}, \bibinfo{person}{Andrew Petersen},
  \bibinfo{person}{Kelly Rivers}, {et~al\mbox{.}}}
  \bibinfo{year}{2015}\natexlab{}.
\newblock \showarticletitle{Educational Data Mining and Learning Analytics in
  Programming: Literature Review and Case Studies}.
\newblock \bibinfo{journal}{\emph{Proceedings of the 2015 ITiCSE on Working
  Group Reports}} (\bibinfo{year}{2015}), \bibinfo{pages}{41--63}.
\newblock
\urldef\tempurl%
\url{https://doi.org/10.1145/2858796.2858798}
\showDOI{\tempurl}


\bibitem[jshint(2021)]%
        {jsHint}
\bibfield{author}{\bibinfo{person}{jshint}.} \bibinfo{year}{{2021}}\natexlab{}.
\newblock \bibinfo{title}{JSHint}.
\newblock \bibinfo{howpublished}{\url{https://github.com/jshint/jshint}}.
\newblock
\newblock
\shownote{Accessed 4/3/2022}.


\bibitem[Kang and Guo(2017)]%
        {kang2017omnicode}
\bibfield{author}{\bibinfo{person}{Hyeonsu Kang} {and}
  \bibinfo{person}{Philip~J Guo}.} \bibinfo{year}{2017}\natexlab{}.
\newblock \showarticletitle{Omnicode: {A} Novice-Oriented Live Programming
  Environment with Always-On Run-Time Value Visualizations}. In
  \bibinfo{booktitle}{\emph{{ACM Symposium on User Interface Software and
  Technology (UIST)}}}. \bibinfo{pages}{737--745}.
\newblock
\urldef\tempurl%
\url{https://doi.org/10.1145/3126594.3126632}
\showDOI{\tempurl}


\bibitem[Kery et~al\mbox{.}(2017)]%
        {kery2017variolite}
\bibfield{author}{\bibinfo{person}{Mary~Beth Kery}, \bibinfo{person}{Amber
  Horvath}, {and} \bibinfo{person}{Brad~A Myers}.}
  \bibinfo{year}{2017}\natexlab{}.
\newblock \showarticletitle{Variolite: Supporting Exploratory Programming by
  Data Scientists}. In \bibinfo{booktitle}{\emph{CHI}},
  Vol.~\bibinfo{volume}{10}.
\newblock
\urldef\tempurl%
\url{https://doi.org/10.1145/3025453.3025626}
\showDOI{\tempurl}


\bibitem[Kery et~al\mbox{.}(2020)]%
        {kery2020mage}
\bibfield{author}{\bibinfo{person}{Mary~Beth Kery}, \bibinfo{person}{Donghao
  Ren}, \bibinfo{person}{Fred Hohman}, \bibinfo{person}{Dominik Moritz},
  \bibinfo{person}{Kanit Wongsuphasawat}, {and} \bibinfo{person}{Kayur Patel}.}
  \bibinfo{year}{2020}\natexlab{}.
\newblock \showarticletitle{mage: Fluid moves between code and graphical work
  in computational notebooks}. In \bibinfo{booktitle}{\emph{{ACM Symposium on
  User Interface Software and Technology (UIST)}}}. \bibinfo{pages}{140--151}.
\newblock


\bibitem[Ko and Myers(2006)]%
        {ko2006barista}
\bibfield{author}{\bibinfo{person}{Amy~J Ko} {and} \bibinfo{person}{Brad~A
  Myers}.} \bibinfo{year}{2006}\natexlab{}.
\newblock \showarticletitle{Barista: An implementation framework for enabling
  new tools, interaction techniques and views in code editors}. In
  \bibinfo{booktitle}{\emph{{SIGCHI Conference on Human Factors in Computing
  Systems}}}. \bibinfo{pages}{387--396}.
\newblock


\bibitem[Kobayashi and Igarashi(2007)]%
        {kobayashi2007boomerang}
\bibfield{author}{\bibinfo{person}{Masatomo Kobayashi} {and}
  \bibinfo{person}{Takeo Igarashi}.} \bibinfo{year}{2007}\natexlab{}.
\newblock \showarticletitle{Boomerang: Suspendable Drag-and-Drop Interactions
  Based on a Throw-and-Catch Metaphor}. In \bibinfo{booktitle}{\emph{{ACM
  Symposium on User Interface Software and Technology (UIST)}}}.
  \bibinfo{pages}{187--190}.
\newblock
\urldef\tempurl%
\url{https://doi.org/10.1145/1294211.1294243}
\showDOI{\tempurl}


\bibitem[Kramer et~al\mbox{.}(2014)]%
        {kramer2014live}
\bibfield{author}{\bibinfo{person}{Jan-Peter Kramer}, \bibinfo{person}{Joachim
  Kurz}, \bibinfo{person}{Thorsten Karrer}, {and} \bibinfo{person}{Jan
  Borchers}.} \bibinfo{year}{2014}\natexlab{}.
\newblock \showarticletitle{How Live Coding Affects Developers' Coding
  Behavior}. In \bibinfo{booktitle}{\emph{Symposium on Visual Languages and
  Human-Centric Computing (VL/HCC)}}. IEEE, \bibinfo{pages}{5--8}.
\newblock
\urldef\tempurl%
\url{https://doi.org/10.1109/VLHCC.2014.6883013}
\showDOI{\tempurl}


\bibitem[Lee et~al\mbox{.}(2013)]%
        {lee2013drag}
\bibfield{author}{\bibinfo{person}{Yun~Young Lee}, \bibinfo{person}{Nicholas
  Chen}, {and} \bibinfo{person}{Ralph~E Johnson}.}
  \bibinfo{year}{2013}\natexlab{}.
\newblock \showarticletitle{Drag-and-drop Refactoring: Intuitive and Efficient
  Program Transformation}. In \bibinfo{booktitle}{\emph{International
  Conference on Software Engineering {(ICSE)}}}. IEEE, \bibinfo{pages}{23--32}.
\newblock
\urldef\tempurl%
\url{https://doi.org/10.1109/ICSE.2013.6606548}
\showDOI{\tempurl}


\bibitem[Lerner(2020)]%
        {Lerner20Projection}
\bibfield{author}{\bibinfo{person}{Sorin Lerner}.}
  \bibinfo{year}{2020}\natexlab{}.
\newblock \showarticletitle{Projection Boxes: On-the-fly Reconfigurable
  Visualization for Live Programming}. In \bibinfo{booktitle}{\emph{{SIGCHI
  Conference on Human Factors in Computing Systems}}}.
  \bibinfo{publisher}{{ACM}}, \bibinfo{pages}{1--7}.
\newblock
\urldef\tempurl%
\url{https://doi.org/10.1145/3313831.3376494}
\showDOI{\tempurl}


\bibitem[Levin and Brain(2021)]%
        {levin21Code}
\bibfield{author}{\bibinfo{person}{Golan Levin} {and} \bibinfo{person}{Tega
  Brain}.} \bibinfo{year}{2021}\natexlab{}.
\newblock \bibinfo{booktitle}{\emph{Code as Creative Medium: A Handbook for
  Computational Art and Design}}.
\newblock \bibinfo{publisher}{MIT}.
\newblock


\bibitem[Li et~al\mbox{.}(2021)]%
        {li2021we}
\bibfield{author}{\bibinfo{person}{Jingyi Li}, \bibinfo{person}{Sonia Hashim},
  {and} \bibinfo{person}{Jennifer Jacobs}.} \bibinfo{year}{2021}\natexlab{}.
\newblock \showarticletitle{What We Can Learn From Visual Artists About
  Software Development}. In \bibinfo{booktitle}{\emph{{SIGCHI Conference on
  Human Factors in Computing Systems}}}. \bibinfo{pages}{1--14}.
\newblock
\urldef\tempurl%
\url{https://doi.org/10.1145/3411764.3445682}
\showDOI{\tempurl}


\bibitem[Lieberman(2020)]%
        {openFrameworksForward}
\bibfield{author}{\bibinfo{person}{Zach Lieberman}.}
  \bibinfo{year}{2020}\natexlab{}.
\newblock \bibinfo{title}{openFrameworks}.
\newblock
  \bibinfo{howpublished}{\url{https://openframeworks.cc/ofBook/chapters/foreword.html}}.
\newblock
\newblock
\shownote{Accessed 9/21/21}.


\bibitem[Lubin and Chugh(2020)]%
        {LubinPLATEAU}
\bibfield{author}{\bibinfo{person}{Justin Lubin} {and} \bibinfo{person}{Ravi
  Chugh}.} \bibinfo{year}{2020}\natexlab{}.
\newblock \showarticletitle{Type-Directed Program Transformations for the
  Working Functional Programmer}. In \bibinfo{booktitle}{\emph{Workshop on
  Evaluation and Usability of Programming Languages and Tools (PLATEAU 2019)}}.
  Schloss Dagstuhl-Leibniz-Zentrum f{\"u}r Informatik.
\newblock


\bibitem[Malita and Schuster(2020)]%
        {Malita20From}
\bibfield{author}{\bibinfo{person}{Mihaela Malita} {and} \bibinfo{person}{Ethel
  Schuster}.} \bibinfo{year}{2020}\natexlab{}.
\newblock \showarticletitle{From Drawing to Coding: Teaching Programming with
  Processing}.
\newblock \bibinfo{journal}{\emph{Journal of Computing Sciences in Colleges}}
  \bibinfo{volume}{35}, \bibinfo{number}{8} (\bibinfo{date}{April}
  \bibinfo{year}{2020}), \bibinfo{pages}{245–246}.
\newblock
\showISSN{1937-4771}
\urldef\tempurl%
\url{https://doi.org/10.5555/3417639.3417663}
\showDOI{\tempurl}


\bibitem[M{\u{a}}r{\u{a}}șoiu et~al\mbox{.}(2015)]%
        {empiricalcodecompletion2015}
\bibfield{author}{\bibinfo{person}{Mariana M{\u{a}}r{\u{a}}șoiu},
  \bibinfo{person}{Luke Church}, {and} \bibinfo{person}{Alan Blackwell}.}
  \bibinfo{year}{2015}\natexlab{}.
\newblock \showarticletitle{An empirical investigation of code completion usage
  by professional software developers}. In \bibinfo{booktitle}{\emph{Psychology
  of Programming Interest Group (PPIG 2015)}}. \bibinfo{pages}{59--68}.
\newblock


\bibitem[Marceau et~al\mbox{.}(2011a)]%
        {marceau2011measuring}
\bibfield{author}{\bibinfo{person}{Guillaume Marceau}, \bibinfo{person}{Kathi
  Fisler}, {and} \bibinfo{person}{Shriram Krishnamurthi}.}
  \bibinfo{year}{2011}\natexlab{a}.
\newblock \showarticletitle{Measuring the Effectiveness of Error Messages
  Designed for Novice Programmers}. In \bibinfo{booktitle}{\emph{ACM Technical
  Symposium on Computer Science Education {(SIGCSE)}}}.
  \bibinfo{pages}{499--504}.
\newblock
\urldef\tempurl%
\url{https://doi.org/10.1145/1953163.1953308}
\showDOI{\tempurl}


\bibitem[Marceau et~al\mbox{.}(2011b)]%
        {marceau2011mind}
\bibfield{author}{\bibinfo{person}{Guillaume Marceau}, \bibinfo{person}{Kathi
  Fisler}, {and} \bibinfo{person}{Shriram Krishnamurthi}.}
  \bibinfo{year}{2011}\natexlab{b}.
\newblock \showarticletitle{Mind Your Language: on Novices' Interactions with
  Error Messages}. In \bibinfo{booktitle}{\emph{{ACM} Symposium on New Ideas in
  Programming and Reflections on Software, Onward! 2011, part of {SPLASH}
  '11}}. \bibinfo{pages}{3--18}.
\newblock
\urldef\tempurl%
\url{https://doi.org/10.1145/2048237.2048241}
\showDOI{\tempurl}


\bibitem[McNutt et~al\mbox{.}(2020)]%
        {mcnutt2020surfacing}
\bibfield{author}{\bibinfo{person}{Andrew McNutt}, \bibinfo{person}{Gordon
  Kindlmann}, {and} \bibinfo{person}{Michael Correll}.}
  \bibinfo{year}{2020}\natexlab{}.
\newblock \showarticletitle{{Surfacing Visualization Mirages}}. In
  \bibinfo{booktitle}{\emph{{SIGCHI Conference on Human Factors in Computing
  Systems}}}. \bibinfo{pages}{1--16}.
\newblock


\bibitem[McNutt et~al\mbox{.}(2023)]%
        {mcnutt2023MetaCells}
\bibfield{author}{\bibinfo{person}{Andrew~M. McNutt},
  \bibinfo{person}{Chenglong Wang}, \bibinfo{person}{Rob DeLine}, {and}
  \bibinfo{person}{Steven~M. Drucker}.} \bibinfo{year}{2023}\natexlab{}.
\newblock \showarticletitle{On the Design of AI-powered Code Assistants for
  Notebooks}. In \bibinfo{booktitle}{\emph{{SIGCHI Conference on Human Factors
  in Computing Systems}}}.
\newblock
\newblock
\shownote{To Appear}.


\bibitem[Mikami et~al\mbox{.}(2017)]%
        {mikami2017micro}
\bibfield{author}{\bibinfo{person}{Hiroaki Mikami}, \bibinfo{person}{Daisuke
  Sakamoto}, {and} \bibinfo{person}{Takeo Igarashi}.}
  \bibinfo{year}{2017}\natexlab{}.
\newblock \showarticletitle{Micro-Versioning Tool to Support Experimentation in
  Exploratory Programming}. In \bibinfo{booktitle}{\emph{{SIGCHI Conference on
  Human Factors in Computing Systems}}}. \bibinfo{pages}{6208--6219}.
\newblock
\urldef\tempurl%
\url{https://doi.org/10.1145/3025453.3025597}
\showDOI{\tempurl}


\bibitem[Mitchell and Bown(2013)]%
        {mitchell2013towards}
\bibfield{author}{\bibinfo{person}{Mark~C Mitchell} {and}
  \bibinfo{person}{Oliver Bown}.} \bibinfo{year}{2013}\natexlab{}.
\newblock \showarticletitle{Towards a creativity support tool in processing:
  understanding the needs of creative coders}. In
  \bibinfo{booktitle}{\emph{Australian Computer-Human Interaction Conference:
  Augmentation, Application, Innovation, Collaboration}}.
  \bibinfo{pages}{143--146}.
\newblock


\bibitem[Norman(2013)]%
        {norman2013design}
\bibfield{author}{\bibinfo{person}{Don Norman}.}
  \bibinfo{year}{2013}\natexlab{}.
\newblock \bibinfo{booktitle}{\emph{The design of everyday things: Revised and
  expanded edition}}.
\newblock \bibinfo{publisher}{Basic books}.
\newblock


\bibitem[Omar et~al\mbox{.}(2019)]%
        {HazelnutLive}
\bibfield{author}{\bibinfo{person}{Cyrus Omar}, \bibinfo{person}{Ian Voysey},
  \bibinfo{person}{Ravi Chugh}, {and} \bibinfo{person}{Matthew~A. Hammer}.}
  \bibinfo{year}{2019}\natexlab{}.
\newblock \showarticletitle{Live Functional Programming with Typed Holes}.
\newblock \bibinfo{journal}{\emph{Proceedings of the ACM on Programming
  Languages (POPL)}}  \bibinfo{volume}{3}, Article \bibinfo{articleno}{14}
  (\bibinfo{year}{2019}), \bibinfo{numpages}{32}~pages.
\newblock
\urldef\tempurl%
\url{https://doi.org/10.1145/3290327}
\showDOI{\tempurl}


\bibitem[Omar et~al\mbox{.}(2017)]%
        {Hazelnut}
\bibfield{author}{\bibinfo{person}{Cyrus Omar}, \bibinfo{person}{Ian Voysey},
  \bibinfo{person}{Michael Hilton}, \bibinfo{person}{Jonathan Aldrich}, {and}
  \bibinfo{person}{Matthew~A. Hammer}.} \bibinfo{year}{2017}\natexlab{}.
\newblock \showarticletitle{Hazelnut: A Bidirectionally Typed Structure Editor
  Calculus}. In \bibinfo{booktitle}{\emph{ACM SIGPLAN Symposium on Principles
  of Programming Languages (POPL)}}. \bibinfo{publisher}{Association for
  Computing Machinery}, \bibinfo{pages}{86–99}.
\newblock
\showISBNx{9781450346603}
\urldef\tempurl%
\url{https://doi.org/10.1145/3009837.3009900}
\showDOI{\tempurl}


\bibitem[Omar et~al\mbox{.}(2012)]%
        {omar2012active}
\bibfield{author}{\bibinfo{person}{Cyrus Omar}, \bibinfo{person}{Young~Seok
  Yoon}, \bibinfo{person}{Thomas~D LaToza}, {and} \bibinfo{person}{Brad~A
  Myers}.} \bibinfo{year}{2012}\natexlab{}.
\newblock \showarticletitle{Active Code Completion}. In
  \bibinfo{booktitle}{\emph{International Conference on Software Engineering
  {(ICSE)}}}. IEEE, \bibinfo{pages}{859--869}.
\newblock
\urldef\tempurl%
\url{https://doi.org/10.1109/ICSE.2012.6227133}
\showDOI{\tempurl}


\bibitem[Oviatt et~al\mbox{.}(2006)]%
        {oviatt2006quiet}
\bibfield{author}{\bibinfo{person}{Sharon Oviatt}, \bibinfo{person}{Alex
  Arthur}, {and} \bibinfo{person}{Julia Cohen}.}
  \bibinfo{year}{2006}\natexlab{}.
\newblock \showarticletitle{Quiet interfaces that help students think}. In
  \bibinfo{booktitle}{\emph{{ACM Symposium on User Interface Software and
  Technology (UIST)}}}. \bibinfo{pages}{191--200}.
\newblock


\bibitem[Peppler and Kafai(2009)]%
        {peppler2005creative}
\bibfield{author}{\bibinfo{person}{Kylie Peppler} {and} \bibinfo{person}{Yasmin
  Kafai}.} \bibinfo{year}{2009}\natexlab{}.
\newblock \showarticletitle{Creative Coding: Programming for Personal
  Expression}.
\newblock \bibinfo{journal}{\emph{International Conference on Computer
  Supported Collaborative Learning (CSCL)}}  \bibinfo{volume}{30}
  (\bibinfo{year}{2009}), \bibinfo{pages}{7}.
\newblock


\bibitem[Quilez and Jeremias(2013)]%
        {shaderToy}
\bibfield{author}{\bibinfo{person}{Inigo Quilez} {and} \bibinfo{person}{Pol
  Jeremias}.} \bibinfo{year}{2013}\natexlab{}.
\newblock \bibinfo{title}{ShaderToy}.
\newblock \bibinfo{howpublished}{\url{https://www.shadertoy.com/}}.
\newblock
\newblock
\shownote{Accessed 9/7/2022}.


\bibitem[Rauch et~al\mbox{.}(2019)]%
        {Rauch19Babylonian}
\bibfield{author}{\bibinfo{person}{David Rauch}, \bibinfo{person}{Patrick
  Rein}, \bibinfo{person}{Stefan Ramson}, \bibinfo{person}{Jens Lincke}, {and}
  \bibinfo{person}{Robert Hirschfeld}.} \bibinfo{year}{2019}\natexlab{}.
\newblock \showarticletitle{Babylonian-style Programming - Design and
  Implementation of an Integration of Live Examples Into General-purpose Source
  Code}.
\newblock \bibinfo{journal}{\emph{The Art, Science, and Engineering of
  Programming}} \bibinfo{volume}{3}, \bibinfo{number}{3}
  (\bibinfo{year}{2019}), \bibinfo{pages}{9}.
\newblock
\urldef\tempurl%
\url{https://doi.org/10.22152/programming-journal.org/2019/3/9}
\showDOI{\tempurl}


\bibitem[Rautiainen(2020)]%
        {RautiainenTools}
\bibfield{author}{\bibinfo{person}{Olli Rautiainen}.}
  \bibinfo{year}{2020}\natexlab{}.
\newblock \bibinfo{title}{How to write better code with linting, formatting,
  and analysis tools}.
\newblock
  \bibinfo{howpublished}{\url{https://www.eficode.com/blog/how-to-write-better-code-with-tools}}.
\newblock
\newblock
\shownote{Accessed 4/4/2022}.


\bibitem[Reas and Fry(2007)]%
        {reas2007processing}
\bibfield{author}{\bibinfo{person}{Casey Reas} {and} \bibinfo{person}{Ben
  Fry}.} \bibinfo{year}{2007}\natexlab{}.
\newblock \bibinfo{booktitle}{\emph{Processing: a programming handbook for
  visual designers and artists}}.
\newblock \bibinfo{publisher}{Mit Press}.
\newblock


\bibitem[Rein et~al\mbox{.}(2019)]%
        {rein2018exploratory}
\bibfield{author}{\bibinfo{person}{Patrick Rein}, \bibinfo{person}{Stefan
  Ramson}, \bibinfo{person}{Jens Lincke}, \bibinfo{person}{Robert Hirschfeld},
  {and} \bibinfo{person}{Tobias Pape}.} \bibinfo{year}{2019}\natexlab{}.
\newblock \showarticletitle{Exploratory and Live, Programming and Coding: A
  Literature Study Comparing Perspectives on Liveness}.
\newblock \bibinfo{journal}{\emph{The Art, Science, and Engineering of
  Programming}} \bibinfo{volume}{3}, \bibinfo{number}{1}
  (\bibinfo{year}{2019}).
\newblock
Issue 1.
\urldef\tempurl%
\url{https://doi.org/10.22152/programming-journal.org/2019/3/1}
\showDOI{\tempurl}


\bibitem[Replit(2021)]%
        {replit}
\bibfield{author}{\bibinfo{person}{Replit}.} \bibinfo{year}{{2021}}\natexlab{}.
\newblock \bibinfo{title}{Replit}.
\newblock \bibinfo{howpublished}{\url{https://replit.com/}}.
\newblock
\newblock
\shownote{Accessed 4/3/2022}.


\bibitem[Selvaraj et~al\mbox{.}(2021)]%
        {selvaraj2021live}
\bibfield{author}{\bibinfo{person}{Ana Selvaraj}, \bibinfo{person}{Eda Zhang},
  \bibinfo{person}{Leo Porter}, {and} \bibinfo{person}{Adalbert~Gerald
  Soosai~Raj}.} \bibinfo{year}{2021}\natexlab{}.
\newblock \showarticletitle{Live Coding: {A} Review of the Literature}. In
  \bibinfo{booktitle}{\emph{ACM Conference on Innovation and Technology in
  Computer Science Education}}, Vol.~\bibinfo{volume}{1}.
  \bibinfo{pages}{164--170}.
\newblock
\urldef\tempurl%
\url{https://doi.org/10.1145/3430665.3456382}
\showDOI{\tempurl}


\bibitem[Shaffer and Resnick(1999)]%
        {ThickAuthenticity}
\bibfield{author}{\bibinfo{person}{David~Williamson Shaffer} {and}
  \bibinfo{person}{Mitchel Resnick}.} \bibinfo{year}{1999}\natexlab{}.
\newblock \showarticletitle{{``Thick'' Authenticity: New Media and Authentic
  Learning}}.
\newblock \bibinfo{journal}{\emph{Journal of Interactive Learning Research}}
  \bibinfo{volume}{10}, \bibinfo{number}{2} (\bibinfo{date}{December}
  \bibinfo{year}{1999}), \bibinfo{pages}{195–215}.
\newblock


\bibitem[Shiffman(2021)]%
        {CodingTrain}
\bibfield{author}{\bibinfo{person}{Daniel Shiffman}.}
  \bibinfo{year}{2021}\natexlab{}.
\newblock \bibinfo{title}{Coding Train}.
\newblock \bibinfo{howpublished}{\url{https://thecodingtrain.com/}}.
\newblock


\bibitem[Shneiderman(2007)]%
        {shneiderman2007creativity}
\bibfield{author}{\bibinfo{person}{Ben Shneiderman}.}
  \bibinfo{year}{2007}\natexlab{}.
\newblock \showarticletitle{Creativity support tools: accelerating discovery
  and innovation}.
\newblock \bibinfo{journal}{\emph{Commun. ACM}} \bibinfo{volume}{50},
  \bibinfo{number}{12} (\bibinfo{year}{2007}), \bibinfo{pages}{20--32}.
\newblock


\bibitem[Simon et~al\mbox{.}(2010)]%
        {MediaCompUCSD}
\bibfield{author}{\bibinfo{person}{Beth Simon}, \bibinfo{person}{P\"{a}ivi
  Kinnunen}, \bibinfo{person}{Leo Porter}, {and} \bibinfo{person}{Dov Zazkis}.}
  \bibinfo{year}{2010}\natexlab{}.
\newblock \showarticletitle{{Experience Report: CS1 for Majors with Media
  Computation}}. In \bibinfo{booktitle}{\emph{Conference on Innovation and
  Technology in Computer Science Education (ITiCSE)}}.
\newblock


\bibitem[Subbaraman and Peek(2022)]%
        {subbaraman2022p5}
\bibfield{author}{\bibinfo{person}{Blair Subbaraman} {and}
  \bibinfo{person}{Nadya Peek}.} \bibinfo{year}{2022}\natexlab{}.
\newblock \showarticletitle{p5. fab: Direct Control of Digital Fabrication
  Machines from a Creative Coding Environment}. In
  \bibinfo{booktitle}{\emph{Designing Interactive Systems Conference}}.
  \bibinfo{pages}{1148--1161}.
\newblock


\bibitem[Tanimoto(1990)]%
        {Tanimoto1990}
\bibfield{author}{\bibinfo{person}{Steven~L. Tanimoto}.}
  \bibinfo{year}{1990}\natexlab{}.
\newblock \showarticletitle{{VIVA}: A Visual Language for Image Processing}.
\newblock \bibinfo{journal}{\emph{Journal of Visual Languages and Computing}}
  \bibinfo{volume}{1}, \bibinfo{number}{2} (\bibinfo{date}{June}
  \bibinfo{year}{1990}), \bibinfo{pages}{127–139}.
\newblock
\showISSN{1045-926X}
\urldef\tempurl%
\url{https://doi.org/10.1016/S1045-926X(05)80012-6}
\showDOI{\tempurl}


\bibitem[Tanimoto(2013)]%
        {Tanimoto13live}
\bibfield{author}{\bibinfo{person}{Steven~L. Tanimoto}.}
  \bibinfo{year}{2013}\natexlab{}.
\newblock \showarticletitle{A perspective on the evolution of live
  programming}. In \bibinfo{booktitle}{\emph{Workshop on Live Programming,
  {LIVE}}}. \bibinfo{publisher}{{IEEE}}, \bibinfo{pages}{31--34}.
\newblock
\urldef\tempurl%
\url{https://doi.org/10.1109/LIVE.2013.6617346}
\showDOI{\tempurl}


\bibitem[Toomim et~al\mbox{.}(2004)]%
        {toomim2004managing}
\bibfield{author}{\bibinfo{person}{Michael Toomim}, \bibinfo{person}{Andrew
  Begel}, {and} \bibinfo{person}{Susan~L Graham}.}
  \bibinfo{year}{2004}\natexlab{}.
\newblock \showarticletitle{Managing Duplicated Code with Linked Editing}. In
  \bibinfo{booktitle}{\emph{Symposium on Visual Languages-Human Centric
  Computing (VL/HCC)}}. IEEE, \bibinfo{pages}{173--180}.
\newblock
\urldef\tempurl%
\url{https://doi.org/10.1109/VLHCC.2004.35}
\showDOI{\tempurl}


\bibitem[Victor(2011a)]%
        {ExplorableExplanations}
\bibfield{author}{\bibinfo{person}{Bret Victor}.}
  \bibinfo{year}{2011}\natexlab{a}.
\newblock \bibinfo{title}{Explorable Explanations}.
\newblock
  \bibinfo{howpublished}{\url{http://worrydream.com/ExplorableExplanations/}}.
\newblock


\bibitem[Victor(2011b)]%
        {ScrubbingCalculator}
\bibfield{author}{\bibinfo{person}{Bret Victor}.}
  \bibinfo{year}{2011}\natexlab{b}.
\newblock \bibinfo{title}{Scrubbing Calculator}.
\newblock
  \bibinfo{howpublished}{\url{http://worrydream.com/ScrubbingCalculator/}}.
\newblock


\bibitem[Vihavainen et~al\mbox{.}(2014)]%
        {vihavainen2014novices}
\bibfield{author}{\bibinfo{person}{Arto Vihavainen}, \bibinfo{person}{Juha
  Helminen}, {and} \bibinfo{person}{Petri Ihantola}.}
  \bibinfo{year}{2014}\natexlab{}.
\newblock \showarticletitle{How Novices Tackle their First Lines of Code in an
  {IDE}: Analysis of Programming Session Traces}. In
  \bibinfo{booktitle}{\emph{Koli Calling International Conference on Computing
  Education Research}}. \bibinfo{pages}{109--116}.
\newblock
\urldef\tempurl%
\url{https://doi.org/10.1145/2674683.2674692}
\showDOI{\tempurl}


\bibitem[Vladis et~al\mbox{.}(2020)]%
        {2020-data-crafting}
\bibfield{author}{\bibinfo{person}{Nathalie Vladis}, \bibinfo{person}{Aspen
  Hopkins}, {and} \bibinfo{person}{Arvind Satyanarayan}.}
  \bibinfo{year}{2020}\natexlab{}.
\newblock \bibinfo{title}{{Data Crafting: Exploring Data through Craft and
  Play}}.
\newblock \bibinfo{howpublished}{IEEE VIS Workshop on Data Vis Activities to
  Facilitate Learning, Reflecting, Discussing, and Designing}.
\newblock


\bibitem[Weintrop and Wilensky(2015)]%
        {ToBlockOrNot}
\bibfield{author}{\bibinfo{person}{David Weintrop} {and} \bibinfo{person}{Uri
  Wilensky}.} \bibinfo{year}{2015}\natexlab{}.
\newblock \showarticletitle{{To Block or Not to Block, That is the Question:
  Students' Perceptions of Blocks-Based Programming}}. In
  \bibinfo{booktitle}{\emph{International Conference on Interaction Design and
  Children (IDC)}}.
\newblock


\bibitem[Whitworth(2005)]%
        {whitworth2005polite}
\bibfield{author}{\bibinfo{person}{Brian Whitworth}.}
  \bibinfo{year}{2005}\natexlab{}.
\newblock \showarticletitle{Polite Computing}.
\newblock \bibinfo{journal}{\emph{Behaviour \& Information Technology}}
  \bibinfo{volume}{24}, \bibinfo{number}{5} (\bibinfo{year}{2005}),
  \bibinfo{pages}{353--363}.
\newblock
\urldef\tempurl%
\url{https://doi.org/10.1080/01449290512331333700}
\showDOI{\tempurl}


\bibitem[Wood et~al\mbox{.}(2016)]%
        {wood2016computational}
\bibfield{author}{\bibinfo{person}{Zoe~J Wood}, \bibinfo{person}{Paul Muhl},
  {and} \bibinfo{person}{Katelyn Hicks}.} \bibinfo{year}{2016}\natexlab{}.
\newblock \showarticletitle{Computational Art: Introducing High School Students
  to Computing via Art}. In \bibinfo{booktitle}{\emph{ACM Technical Symposium
  on Computing Science Education}}. \bibinfo{pages}{261--266}.
\newblock
\urldef\tempurl%
\url{https://doi.org/10.1145/2839509.2844614}
\showDOI{\tempurl}


\bibitem[Workman(2021)]%
        {happyCoding}
\bibfield{author}{\bibinfo{person}{Kevin Workman}.}
  \bibinfo{year}{{2021}}\natexlab{}.
\newblock \bibinfo{title}{Happy Coding Tutorials}.
\newblock \bibinfo{howpublished}{\url{https://happycoding.io/}}.
\newblock
\newblock
\shownote{Accessed 4/3/2022}.


\bibitem[Wu et~al\mbox{.}(2020)]%
        {wu2020B2}
\bibfield{author}{\bibinfo{person}{Yifan Wu}, \bibinfo{person}{Joseph~M.
  Hellerstein}, {and} \bibinfo{person}{Arvind Satyanarayan}.}
  \bibinfo{year}{2020}\natexlab{}.
\newblock \showarticletitle{{B2:} Bridging Code and Interactive Visualization
  in Computational Notebooks}. In \bibinfo{booktitle}{\emph{{ACM Symposium on
  User Interface Software and Technology (UIST)}}}. \bibinfo{publisher}{{ACM}},
  \bibinfo{pages}{152--165}.
\newblock
\urldef\tempurl%
\url{https://doi.org/10.1145/3379337.3415851}
\showDOI{\tempurl}


\bibitem[Yoon and Myers(2015)]%
        {yoon2015supporting}
\bibfield{author}{\bibinfo{person}{YoungSeok Yoon} {and}
  \bibinfo{person}{Brad~A Myers}.} \bibinfo{year}{2015}\natexlab{}.
\newblock \showarticletitle{Supporting Selective Undo in a Code Editor}. In
  \bibinfo{booktitle}{\emph{International Conference on Software Engineering
  {(ICSE)}}}, Vol.~\bibinfo{volume}{1}. IEEE, \bibinfo{pages}{223--233}.
\newblock
\urldef\tempurl%
\url{https://doi.org/10.1109/ICSE.2015.43}
\showDOI{\tempurl}


\end{thebibliography}

\appendix
\onecolumn %

\section*{APPENDIX}

In this appendix we provide supplementary material that fell outside the scope of the main content of the paper.

\begin{itemize}
    \item  In \secref{appendix:course} we make several notes about the course design and other ancillary details.
    \item In \secref{appendix:study-details} we provide additional details about several of our studies.
\end{itemize}

\section{The Creative Coding Course}\label{appendix:course}

\newcommand{\ass}[5]{
    \item[#1.] \textbf{#2.} #4 \textit{(#5)}
}

\begin{figure}[h]
    \vspace{2in}
    \includegraphics[width=\linewidth]{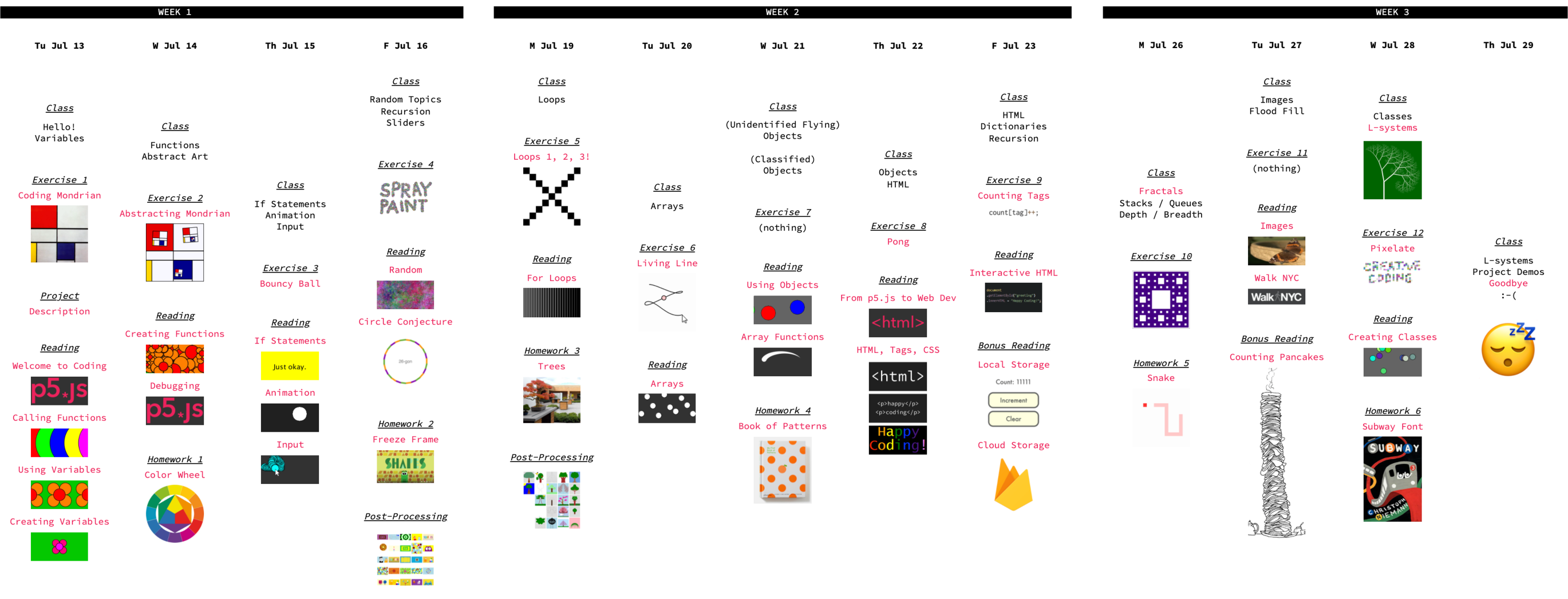}
    \caption{Schedule for \imm{} edition of the course. Readings and their corresponding images were adapted from Workman's HappyCoding tutorials~\cite{happyCoding}.}
    \Description{Three screenshots of the course schedule. Each one shows a single week and is divided into columns for each day. Each day has several pictures relating to activities from that day.}
    \label{fig:imm-course-calendar}
\end{figure}

Here we provide additional context for our course.
In \figref{fig:imm-course-calendar}, we show the schedule for the 3-week \imm{} edition (\asLink{https://www.classes.cs.uchicago.edu/archive/2021/summer/creative-coding/immersion/}{link to course site}).
This edition featured only a sequence of homeworks and exercises, and did not include the self-guided project found in the ``full'' editions of course (\asLink{https://www.classes.cs.uchicago.edu/archive/2021/spring/11111-1/}{sp21}, \asLink{https://www.classes.cs.uchicago.edu/archive/2022/winter/11111-1/}{wi22}).
A similar version of the course was also taught as \asLink{https://www.classes.cs.uchicago.edu/archive/2022/summer/19911-1/}{su22}.
There were six individual homework assignments in \imm{}, the first five of which appeared in all course editions:

\begin{enumerate}

    \ass{1}{Color Wheel}{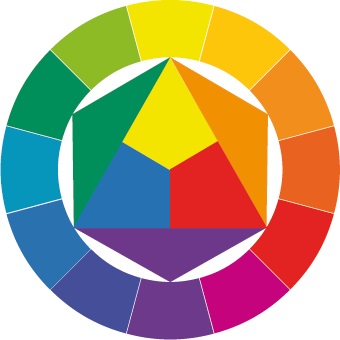}
    {Recreate a given red-yellow-blue color wheel.}
    {function calls, color and shape-drawing APIs, trigonometric expressions}

    \ass{2}{Freeze Frame}{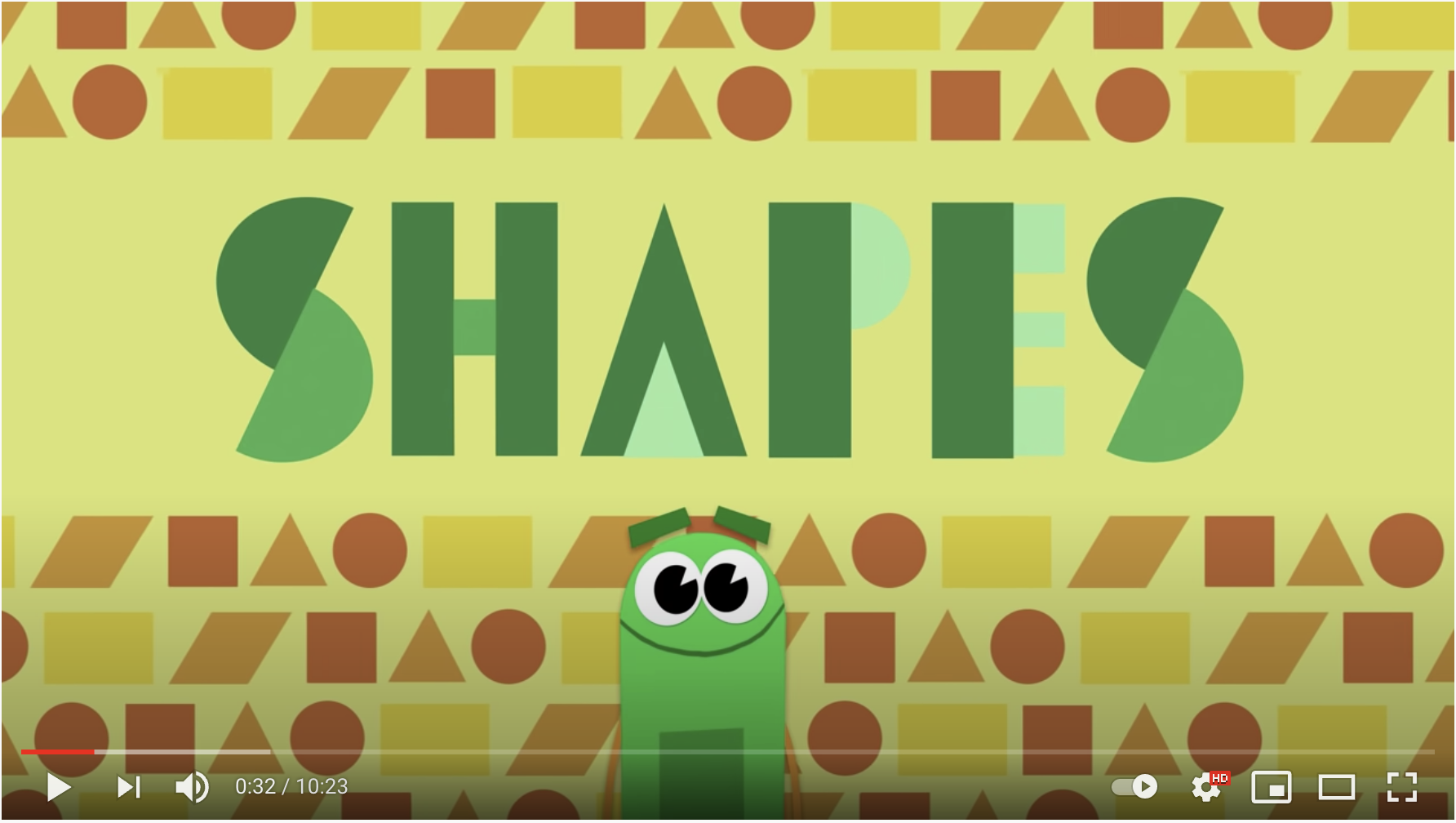}
    {Pick a static frame from the ``StoryBots: Shapes'' video and recreate it.}
    {function calls, color and shape-drawing APIs}

    \ass{3}{Trees}{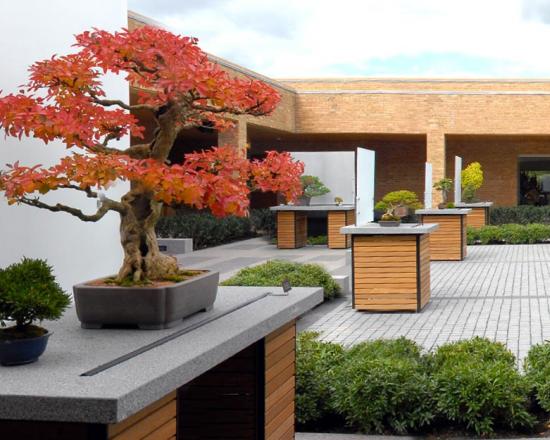}
    {Use variables and arithmetic expressions to implement a symmetric tree drawing.}
    {variables, arithmetic, curves}

    \ass{4}{Book of Patterns}{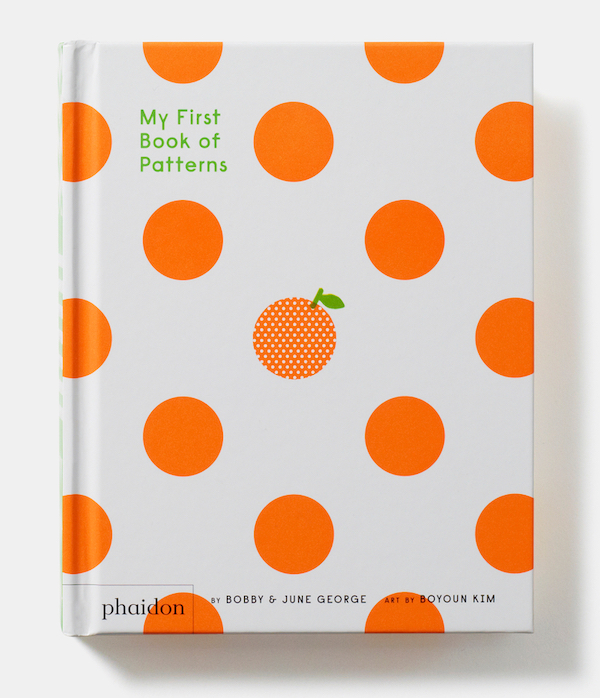}
    {Implement several 2-dimensional grid patterns---stripes, polka dots, checks, plaid, chevron, harlequin, argyle, and honeycomb---inspired by the designs in \textit{My First Book of Patterns}.}
    {nested loops}

    \ass{5}{Snake}{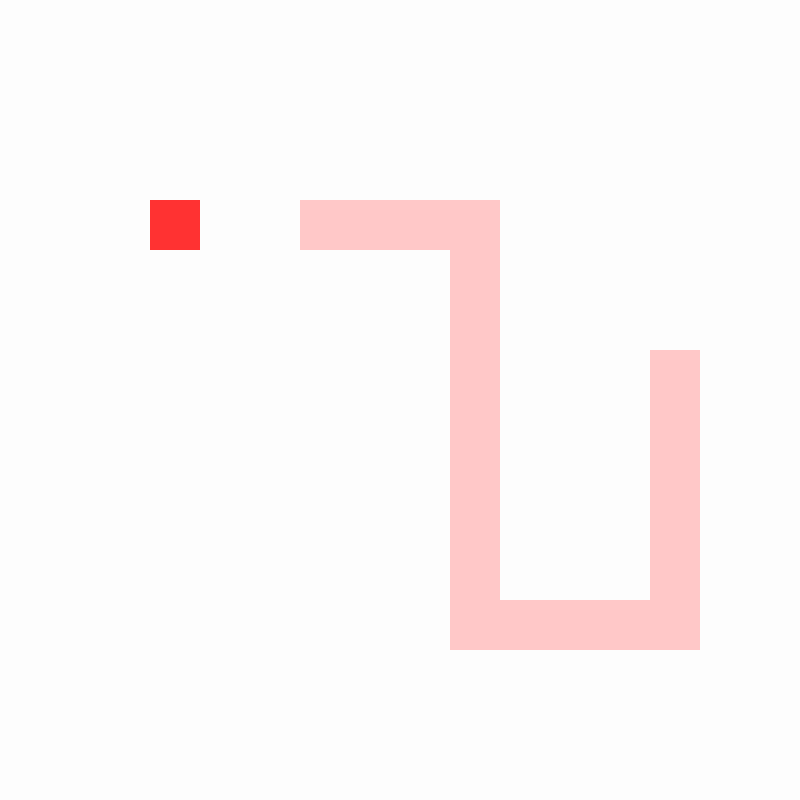}
    {Make a simple version of the classic snake game. Starter code was provided with function stubs for different aspects of a simple model-view-controller architecture}
    {mutable variables, arrays, objects, animation, mouse and keyboard events}

    \ass{6}{Subway Font}{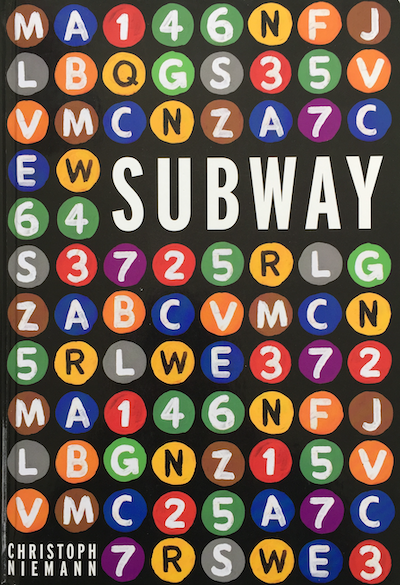} %
    {Rewrite a webpage using a ``font'' that resembles the signage of the New York City subway. As shown on the cover of \textit{Subway}, some letters are rendered white-on-black and others are set atop colored circles.}
    {HTML, CSS, DOM API, dictionaries}

\end{enumerate}

\begin{figure}[h]
    \includegraphics[width=\linewidth]{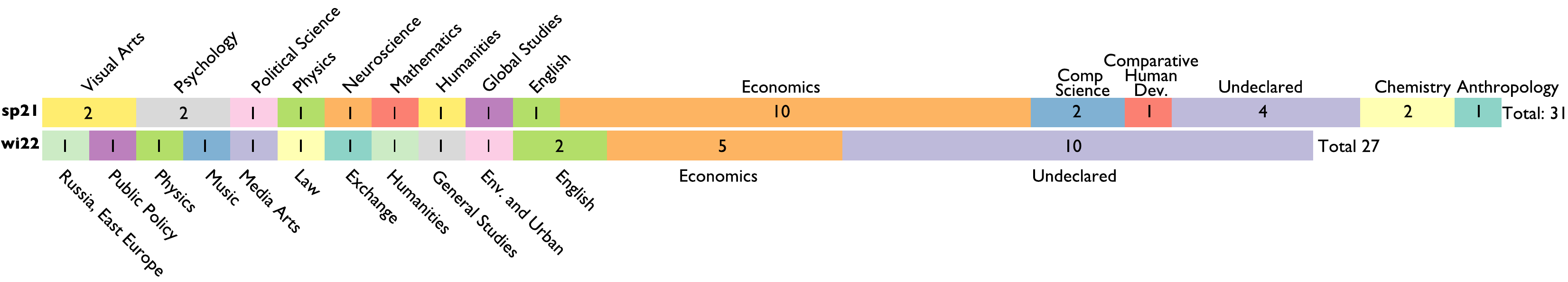}
    \caption{Home departments of students who completed the \spq{} (top) and \wiq{} (bottom) courses.
    }
    \Description{A pair of stacked bar charts. They are colored and labeled based on the home department of each student. The top row shows sp21, and the bottom row shows wi22.}
    \label{fig:departments}
\end{figure}

As highlighted in the main text, we designed our course primarily for college students with little-to-no programming experience who were not planning to major in computer science.
In \spq{}, 4 out of the 31 students were undeclared, and among the remaining 27 students 14 different programs of study were represented.
In \wiq{}, 10 out of the 27 students were undeclared, and among the remaining 17 students 12 different programs of study were represented.
All told, students from 23 different departments participated in the course (\figref{fig:departments}).

Based on both our study and pre-course on-boarding surveys, students self-reported high levels of prior experience (as highlighted in \figref{fig:participant-count}) in each edition of the course.
In \spq{}, students who had previously completed computer science courses at the university---in a couple cases many such courses---were mistakenly allowed to enroll.
This enrollment issue was fixed for \wiq{}, but still nearly half of the students (who completed the course) self-reported prior experience through self-study, courses in high school, and from other university departments or institutions.
In \imm{} and \sumtwo{}, enrollment was unrestricted (the high-school students were not already associated with the university), and a large majority of these students reported prior experience.
In any case, the different levels among our student populations helps color some of the observations in the main text.

\section{Additional Study Details}\label{appendix:study-details}

In this section we present aspects of our studies that did not fit in the main text: the ethics statement for our studies, additional results, followed by descriptions of the hypothetical features from our Year 1 survey that were not implemented in Year 2.

\subsection{Ethics Statement}

All studies were reviewed and determined to be exempt by our university's institutional review board.
We did not collect demographic data, because it was not a core aspect of our investigation.
Although we designed and taught these courses with an eye towards the associated studies, we believe the course materials
we developed and delivered (through lectures, office hours, and online discussions)
were minimally affected by the presence of these studies and our use of custom versions of the p5 editor.

\subsection{Additional Results}

Here we list a series of one-off results that were observed. Then in \secref{sec:code-folding} include an analysis of the code folding feature (original part of the analysis in \secref{sec:both-features}). Finally we provide an additional analysis of the the auto-refresh feature in \secref{sec:live-part-2}.

\begin{figure}[b]
    \vspace{0.75in}
    \centering
    \includegraphics[width=\linewidth]{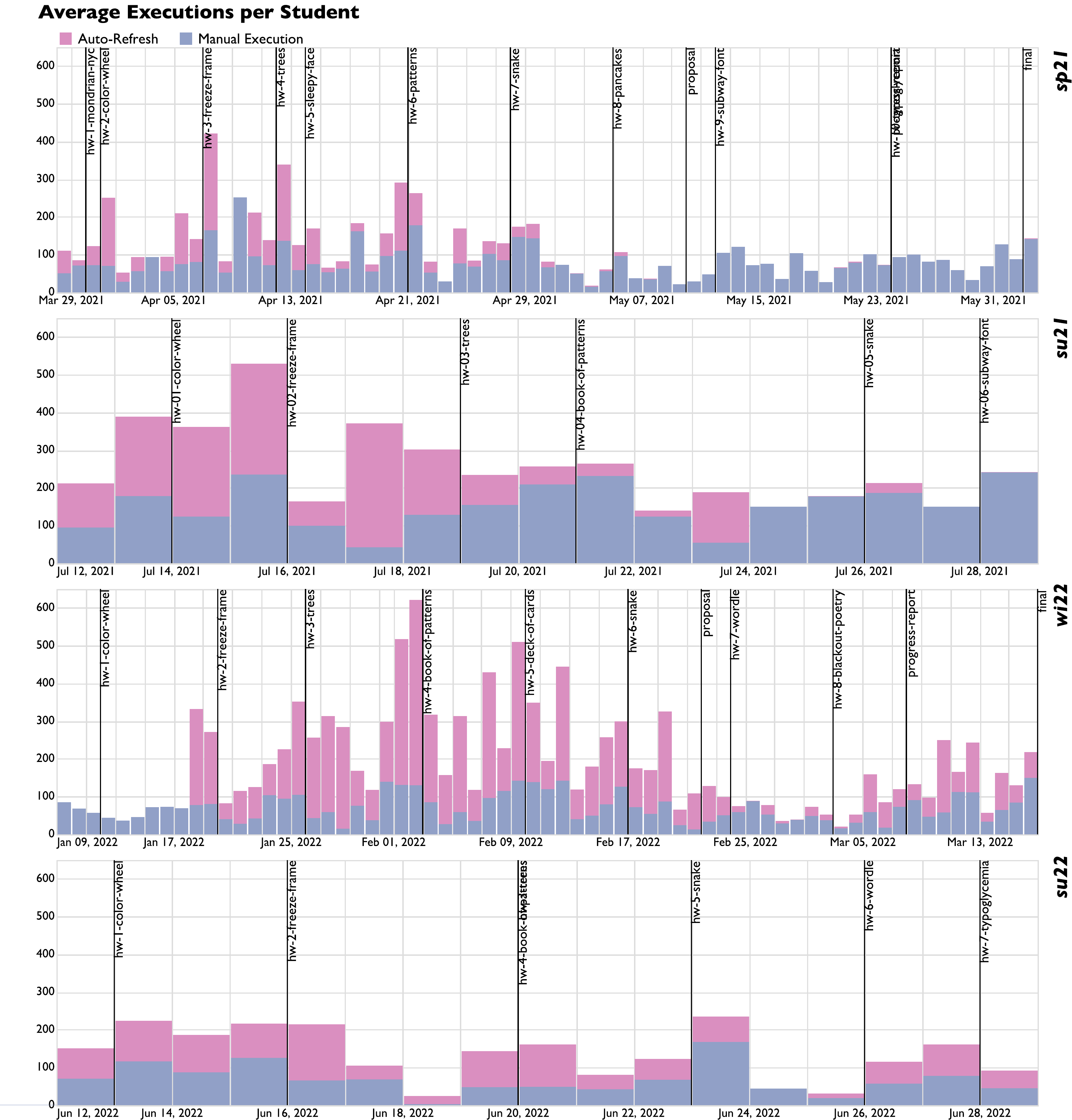}
    \caption{
        A summary of executions for each of the course editions show auto-refresh versus manual execution.
    }
    \label{fig:execution-summary}
    \Description{A trio of stacked bar charts. Each stacked bar chart shows the number of executions per student (y-axis) per day (x-axis) in each of the course editions (sp21, su21, wi21). The bars are divided into auto-refresh executions and manual executions.
    }
\end{figure}

\begin{figure}[b]
    \centering
    \includegraphics[width=\linewidth]{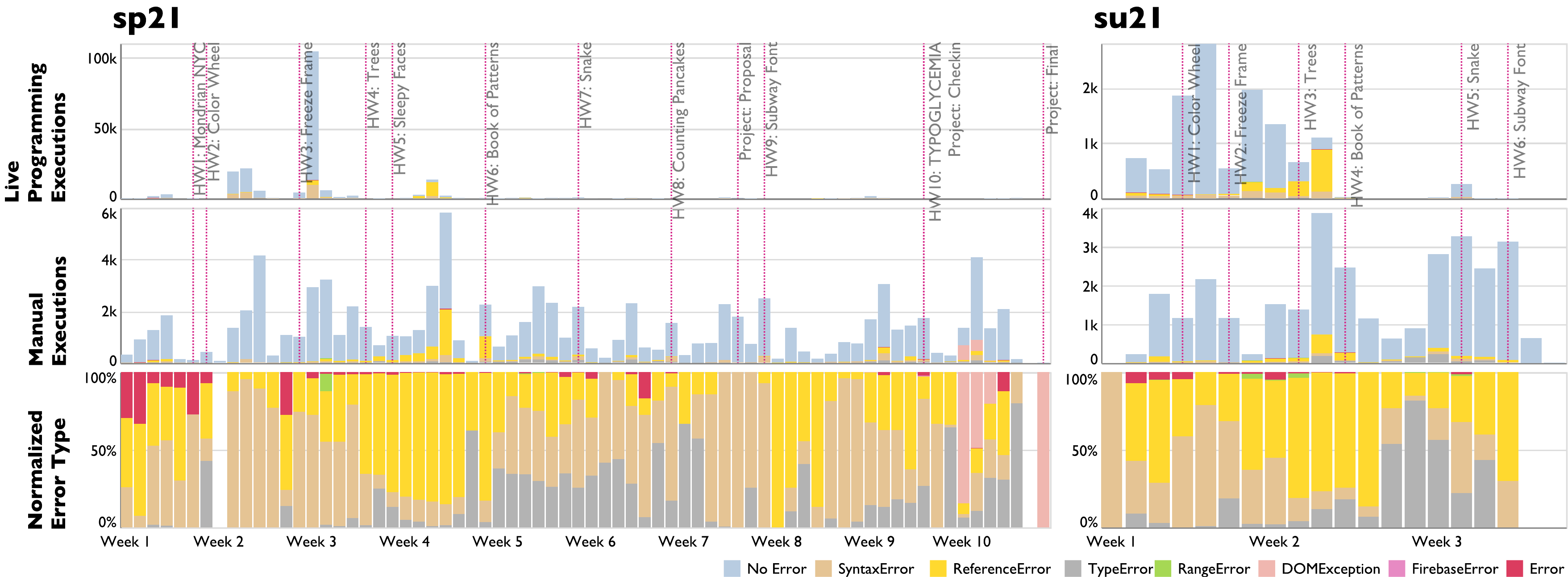}
    \caption{
        Errors over time in the first year of courses. Outliers have not been dropped.
        Our research questions are generally not motivated by analysis of error types, as they were typically driven by course material, rather than by interface design.
    }
    \label{fig:error-dashbaord}
    \Description{ A pair of a trio of stacked bar charts. Each stack consists of a row labeled "Live Programming Executions", "Manual Executions", and "Normalized Error Type". The bars are divided by error type (including a no error type). There is one trio for sp21 and one for su21}
\end{figure}

\begin{figure}[b]
    \centering
    \includegraphics[width=\linewidth]{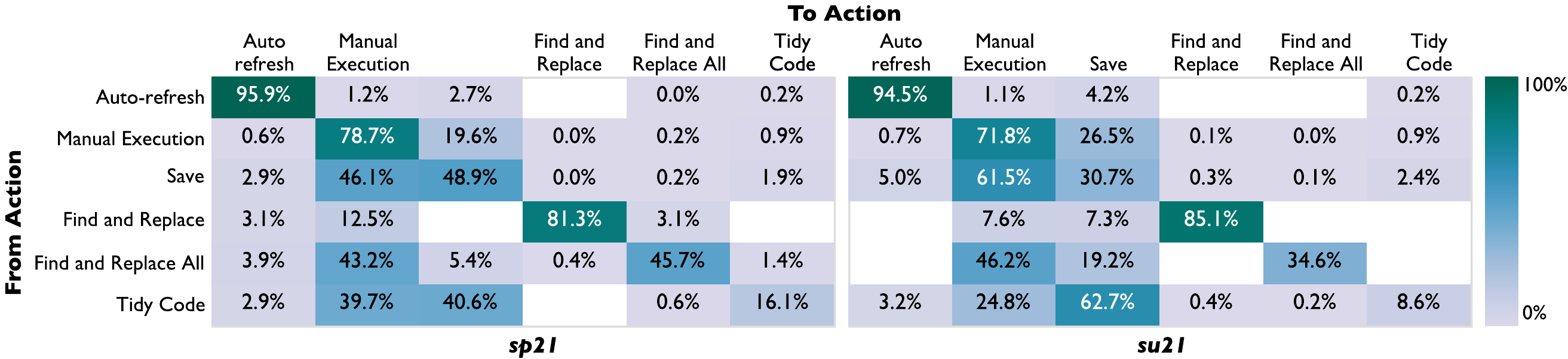}
    \caption{
        The bi-gram action sequence probability in Year 1 shows the rate at which a given action is followed by another particular action.
        We do not show \wiq{} because we mistakenly did not collect Tidy Code executions.
    }
    \label{fig:action-to-action-by-execution}
    \Description{A pair of heat maps. The axes on each are symmetric and consist of a series of simple actions types, including auto-refresh, manual execution, save, find and replace, find and replace all, and tidy code. There is one heatmap for sp21 and one for su21. Cells are labeled as percentages}
\end{figure}

\begin{itemize}

    \item In \figref{fig:often-useful}, we show an alternative depiction of the results from both years of our survey which includes metrics other than the one used in the main text (namely Usefulness).

    \item In \figref{fig:execution-summary} we provide a simple summary of the volume of code executions across all three editions along with assignment due dates, which highlights that execution volume tended to be higher for earlier graphic-only assignments (compared to later assignments which involved interactivity or HTML/CSS).

    \item In \figref{fig:error-dashbaord} we provide a related graphic showing execution history for \spq{} and \imm{}, along with the relative error rates by day.

          In Year 1, our logging scheme did not include a mechanism for explicitly collecting run-time errors, so they were reconstructed post-course by running each logged sketch for 10 seconds and collecting all errors generated during that period.
          This approach may exclude errors students saw, such as those generated through interaction with the sketch or through randomness.
          On average each session had $\mu$$=$$7.27$$\pm$$32.8$ errors, with outliers excluded.
                  Within our reduced sample from Year 2, sessions exhibited $\mu$$=$$30.7$$\pm$$95.8$ errors, again with outliers excluded.
                  A one-sided t-test indicated that there were significantly more mean errors per session in Year 2 ($p$<0.001).
          This increase is likely due to the new collection method, which captured  errors witnessed by the user rather than just reconstructed errors.

    \item In \figref{fig:action-to-action-by-execution} we provide a figure showing the bi-gram action sequence probability of actions in Year 1.

    \item  We then provide tables in \figref{fig:useful-features} for \figref{fig:useful-data-year-1} and \figref{fig:useful-data-year-2}.

\end{itemize}

\subsubsection{Code Folding}
\label{sec:code-folding}

This simple feature, common to most modern editors~\cite{fowkes2017autofolding}, allows functions and other blocks of code to be collapsed and later expanded.
This feature was generally well liked as it made code \pQuote{feel more organized}~(\pA{9}), while helping users avoid \pQuote{being overwhelmed}~(\pA{2}) and making things \pQuote{look neater and less intimidating}~(\pA{1}).
This has the organizational benefit that it is \pQuote{easier to find specific chunks of code}~(\pA{1}), which, as noted by \pA{13} and \pA{16}, reduces the amount of scrolling---these are well-understood benefits of this feature~\cite{hendrix1998visual}.
Being able to organize and navigate code are important concerns for novice creative coders.%
\takeawayClutter{}

However, the feature was not universally appreciated.
Whereas \pB{8} found that code folding \pQuote{Helped a lot while debugging and re-organization!}, \pA{9} asked \pQuote{when debugging, what if the problem is in one of the lines of code that are hidden?}
A number of participants noted that they simply did not use it or did not find it helpful.
Some students only invoked it accidentally (\pA{10}), while others found it confusing (\pA{5,14}) because it did not clearly communicate what code was to be folded.
Code folding, or other interface-based code organization tools, seem especially valuable in this context as most sketches typically involved only a single file
(\eg{} \spq{} and \wiq{} final projects had a median of 1 JavaScript file).
As the file structure abstraction for code organization is underused, there is opportunity for interface-based abstraction.
\begin{figure}[h]
\beginFigureNarrow
\centering
\includegraphics[width=\linewidth]{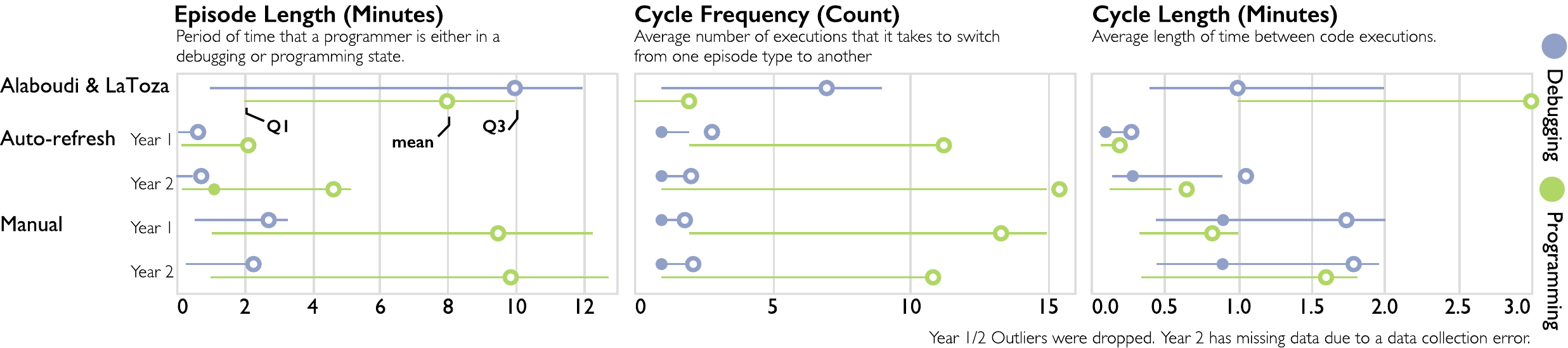}
\caption{
    Comparing how students shifted between debugging and programming states (using different execution styles) against a baseline of professional programmers~\cite{alaboudi2021edit}.
}
\Description{Three stacked box plots showing various metrics. The top is labeled Episode Length (minutes), the middle is labeled Cycle Frequency (Count), and the bottom is labeled Cycle Length (minutes). Each has the same y-axis: Alaboudi and LaToza, Auto-refresh Year 1, Auto-refresh Year 2, Manual Year 1, Manual Year 2.}
\label{fig:a-and-t}
\finishFigureNarrow
\end{figure}
 
\subsubsection{Live Coding}
\label{sec:live-part-2}
In addition to the analysis of the auto-refresh feature considered in the main text (\secref{sec:auto-refresh}), we also sought to understand how student edit-run behavior compared to that of professional programmers.
\figref{fig:a-and-t} shows how auto-refresh usage affected the length, size, and frequency of edit-run cycles with regard to debugging versus programming states adopting the metrics used by Alaboudi and LaToza~\cite{alaboudi2021edit}, who studied how professional programmers shift between debugging and programming states during edit-run cycles in their own work.
A salient difference from the baseline was that the number of executions to transition from a programming to a debugging state (and vice versa) was shorter for our students. This is likely informed by the domain: the professionals were working on projects such as Firefox and Curl, which likely have a different execution cadence than the graphic-oriented work conducted in creative coding.
The programming episode length was similar for the professionals and those using manual execution---although debugging episodes for the latter group were much shorter, suggesting that the errors were much less complex for our students.
However, given the differences in expertise and domain between these groups, it is difficult to identify a primary cause of the changes.

The key observation is that the usage patterns exhibited by our students were not the same as those of professionals, but were not fundamentally dissimilar.
This suggests the potential transferability of our observations about novices to more experienced users.

Using auto-refresh does not appear to have an effect on cycle frequency, although it seems to be associated with shorter episodes and cycle lengths. This coheres with our expectation of auto-refresh, as it triggers executions more quickly than one might with manual execution, but suggests a certain consistency related to task.

\clearpage

 
\clearpage

\definecolor{oftenColor}{HTML}{8DA0CB}
\definecolor{interestedColor}{HTML}{DA8FC0}
\definecolor{usefulColor}{HTML}{A6D854}
\begin{figure}[b]
    \vspace{2in}
    \setlength{\fboxsep}{1pt}
    \centering
    \includegraphics[width=\linewidth]{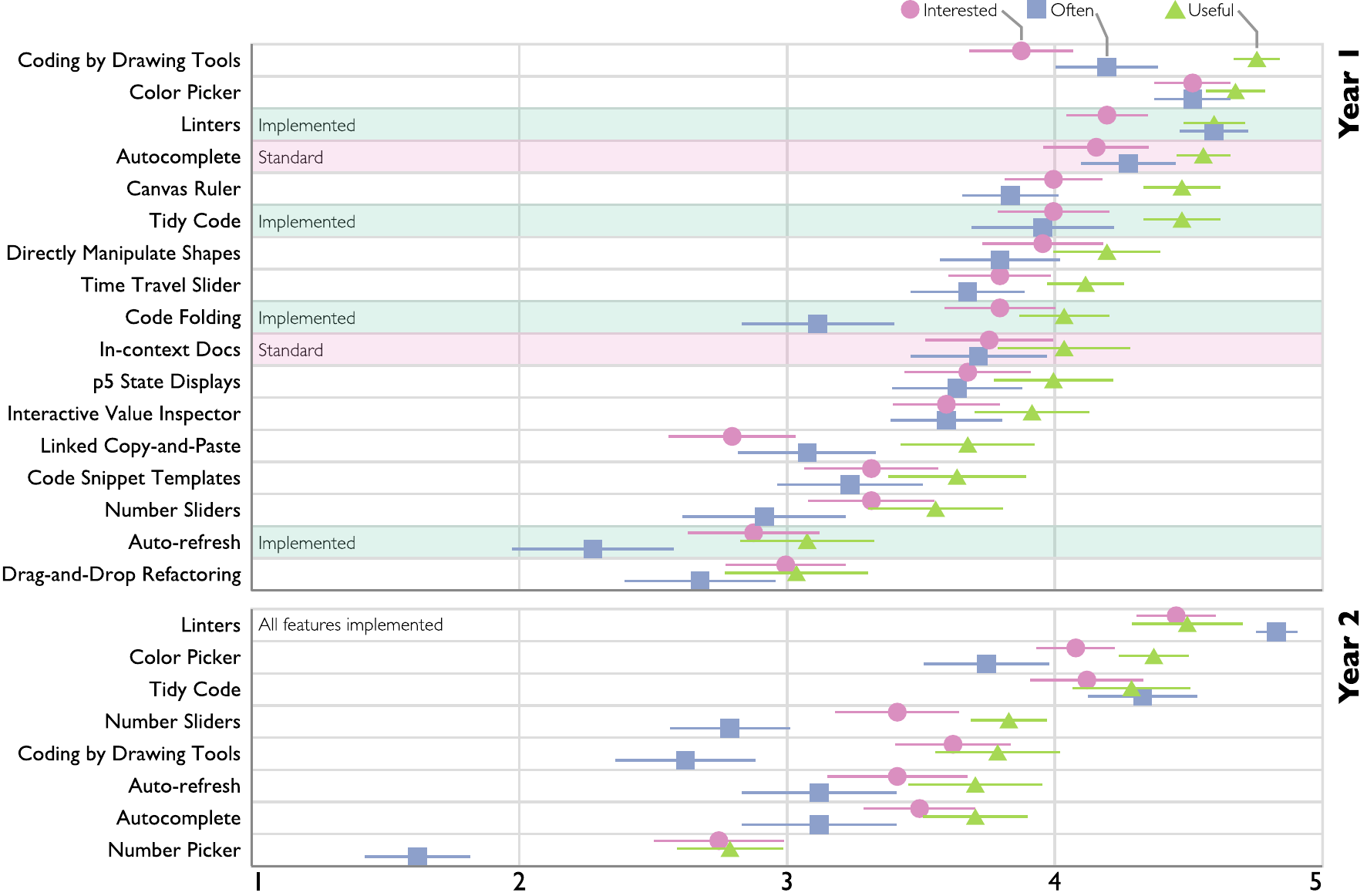}
    \caption{
        Participants in both the long (Year 1) and short (Year 2) survey were asked about a variety of features and rated each of them on how
            {\colorbox{interestedColor}{\hspace{0.01in}\emph{Interested}\hspace{0.02in}}}
        they were in it, how
            {\colorbox{oftenColor}{\hspace{0.01in}\emph{Often}\hspace{0.02in}}}
        they would use it, and how
            {\colorbox{usefulColor}{\hspace{0.01in}\emph{Useful}\hspace{0.02in}}}
        they thought it was.
        Most features considered are non-standard, however several were implemented in our editor or standard (but not implemented).
        It is notable that although some of the most commonly instrumented features are not necessarily predictive of perceived utility.
    }
    \label{fig:often-useful}
    \Description{A pair of box plots showing Year 1 (top) and Year 2 (bottom) survey responses. Each row is a feature asked about in a survey and features a box-plot glyph for each of the metrics (interesting, often, useful). Each plot has been sorted vertically by the Usefulness rating. A table showing the usefulness ratings can be found in Figures 17 and 18.
    }
\end{figure}


\begin{figure}
  \small
  \begin{minipage}{.5\linewidth}
    \begin{tabular}{lp{0.4in}lll}
      \toprule
      Feature Name                & rating mean & $\sigma$ & Q1 & Q3   \\
      \midrule
      Auto-refresh                & 3.08        & 1.26     & 3  & 4.00 \\
      Autocomplete                & 4.56        & 0.51     & 4  & 5.00 \\
      Canvas Ruler                & 4.48        & 0.71     & 4  & 5.00 \\
      Code Folding                & 4.04        & 0.84     & 3  & 5.00 \\
      Code Snippet Templates      & 3.64        & 1.29     & 3  & 5.00 \\
      Coding by Drawing Tools     & 4.76        & 0.44     & 5  & 5.00 \\
      Color Picker                & 4.68        & 0.56     & 4  & 5.00 \\
      Directly Manipulate Shapes  & 4.20        & 1.00     & 4  & 5.00 \\
      Drag-and-Drop Refactoring   & 3.04        & 1.34     & 2  & 4.00 \\
      In-context Docs             & 4.04        & 1.24     & 4  & 5.00 \\
      Interactive Value Inspector & 3.92        & 1.08     & 4  & 4.00 \\
      Linked Copy-and-Paste       & 3.68        & 1.25     & 3  & 5.00 \\
      Linters                     & 4.60        & 0.58     & 4  & 5.00 \\
      Number Sliders              & 3.56        & 1.26     & 3  & 5.00 \\
      Tidy Code                   & 4.48        & 0.71     & 4  & 5.00 \\
      Time Travel Slider          & 4.12        & 0.73     & 4  & 5.00 \\
      p5 State Displays           & 4.00        & 1.12     & 3  & 5.00 \\
      \bottomrule
    \end{tabular}
    \caption{The computed values for \figref{fig:often-useful}, the survey results relating to Usefulness from Year 1.}
    \Description{A table showing the values for Year 1 of Figure 6.}
    \label{fig:useful-data-year-1}
  \end{minipage}%
  \begin{minipage}{.5\linewidth}
    \begin{tabular}{lp{0.4in}lll}
      \toprule
      Feature Name            & rating mean & $\sigma$ & Q1 & Q3   \\
      \midrule
      Auto-refresh            & 3.71        & 1.23     & 3  & 5.00 \\
      Autocomplete            & 3.71        & 0.95     & 3  & 4.00 \\
      Coding by Drawing Tools & 3.79        & 1.14     & 3  & 5.00 \\
      Color Picker            & 4.38        & 0.65     & 4  & 5.00 \\
      Linters                 & 4.50        & 1.02     & 4  & 5.00 \\
      Number Picker           & 2.79        & 0.98     & 2  & 3.25 \\
      Number Sliders          & 3.83        & 0.70     & 3  & 4.00 \\
      Tidy Code               & 4.29        & 1.08     & 4  & 5.00 \\
      \bottomrule
    \end{tabular}
    \caption{The computed values for \figref{fig:often-useful}, the survey results relating to Usefulness from Year 2.}
    \Description{A table showing the values for Year 2 of Figure 6.}
    \label{fig:useful-data-year-2}
  \end{minipage}

\end{figure}

\clearpage
\subsection{Hypothetical Features}

Here we return to the hypothetical features asked about in Year 1 surveys but not implemented in \pFiveTwo{}.
Because responses were based only on a brief description and static image, we limit our discussion of each feature.

\begin{figure}[h!]
    \centering
    \includegraphics[width=0.9\linewidth]{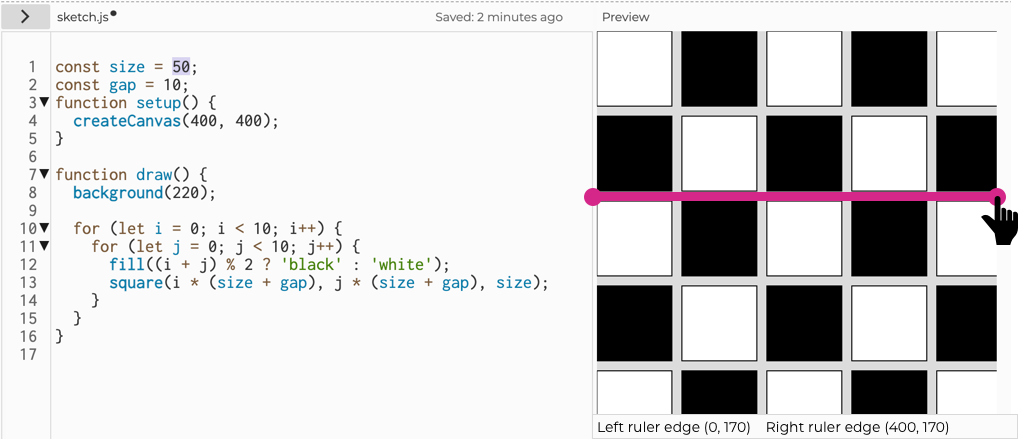}
    \Description{Screenshot showing the proposed canvas ruler IDE enhancement}
\end{figure}

\paragraph{Canvas Ruler} As mentioned earlier, the Canvas Ruler was widely viewed as a useful tool to add for creative coding---however,
\pB{8} felt \pQuote{it would take away the fun of \texttt{mouseX} and \texttt{mouseY}!}
Several additional suggestions were made, such as being able to \pQuote{measure angles, so a ruler and compass.}(\pA{16})
In future editions of the course we intend to return to this feature, as it seems like a natural next step. The primary concern in implementing such an addition would be that it does not clutter the interface\takeawayClutter{}, and perhaps, per commentary on Syntax Templates, be controllable with a keyboard.

\begin{figure}[h!]
    \centering
    \includegraphics[width=0.4\linewidth]{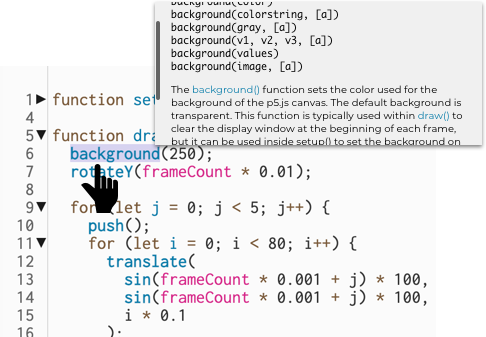}
    \Description{Screenshot showing the proposed in-context docs IDE enhancement}
\end{figure}

\paragraph{In-context Docs}  Many full-featured editors (\eg{}~VS Code) include relevant documentation about language features and user-defined variables as a tooltip.
As with autocomplete, \pB{4} believed In-context Docs would be helpful because \pQuote{gives me an idea of what to [write].}
Several participants echoed this sentiment, believing that it \pQuote{would have drastically widened my skill set}~(\pA{14}).
On the other hand, \pA{4} was \pQuote{actually a little torn by [it] because I think googling and traveling to the reference is really important. It may start off as inconvenient but just becomes more natural with practice}---which is in line with our observations about student skepticism.\takeawaySkeptic{}
Among the quantitative ratings from the surveys in the first year, this feature was the only one that had a statistically significant relationship ($p$<0.01) with self reported experience was in-context docs, in particular exhibiting a negative correlation ($r$=-0.308).
It is possible that an alternative presentation of this feature (perhaps in the search-based style of Blueprint \cite{brandt2010example}) might elicit more positive responses, however, based on these results we believe that users might be similarly skeptical, although exploration of such responses could be usefully explored in future work.

\pagebreak

\begin{figure}[h!]
    \centering
    \includegraphics[width=0.9\linewidth]{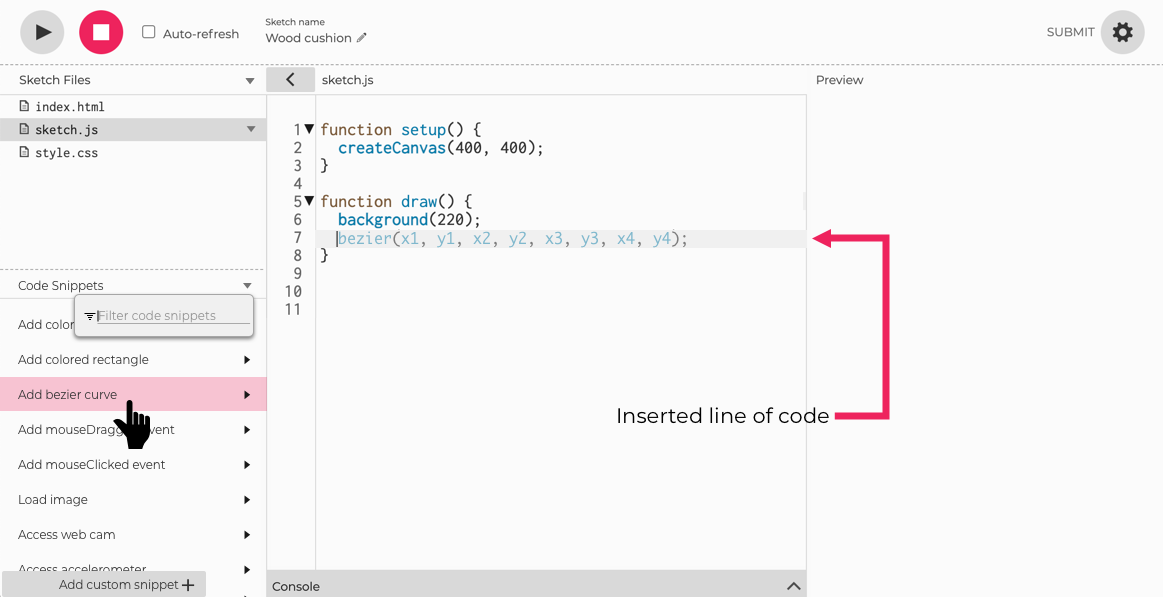}
    \Description{Screenshot showing the proposed syntax templates IDE enhancement}
\end{figure}

\paragraph{Syntax Templates} The syntax templates we implement in our autocomplete were similar to our proposed Code Snippet Templates, however the latter feature was docked  (in the manner of Google Colab's Code Snippet library), and thus required mouse clicks, which may have dampened enthusiasm for the feature:
\pQuote{I think if there were keyboard shortcuts for these then I would use them extensively}~(\pA{13}).
Some thought these features would be an \pQuote{easy way to get students started with no experience}~(\pA{24}), but as discussed in \secref{sec:discussion} others were skeptical.
We still believe this feature would be valuable to implement in the future, possibly integrated into the autocomplete, in order to keep the interface tidy and unencumbered.\takeawayClutter{}

\begin{figure}[h!]
    \centering
    \includegraphics[width=0.9\linewidth]{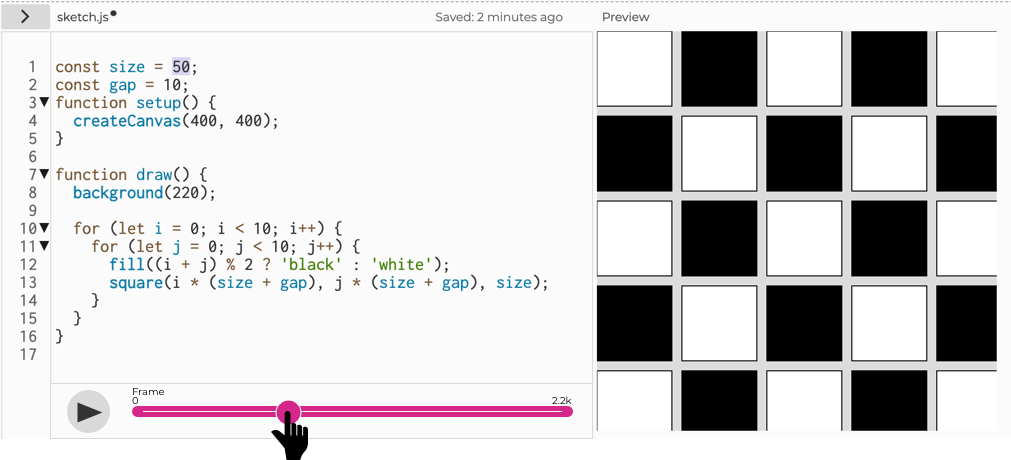}
    \Description{Screenshot showing the proposed time travel slider IDE enhancement}
\end{figure}

\paragraph{Time Travel Slider}
This proposed feature would allow the state of the code execution to be paused and rewound in order to support debugging tasks---which a majority of respondents either understood as a GUI-based shortcut for p5's \texttt{frameRate} setting (which specifies how many times per second the draw loop is called) or as a mechanism for version control, both of which, while interesting, are not the feature we intended. While several students expressed enthusiasm for this latter idea (indicating the potential utility of a Variolite-style~\cite{kery2017variolite} or other selective undos, such as that of \citet{yoon2015supporting} or Mikami \etals{}~\cite{mikami2017micro} Micro-Versioning), this did not yield coherent feedback, beyond confusion about unfamiliar features.

\pagebreak

\begin{figure}[h!]
    \centering
    \includegraphics[width=0.9\linewidth]{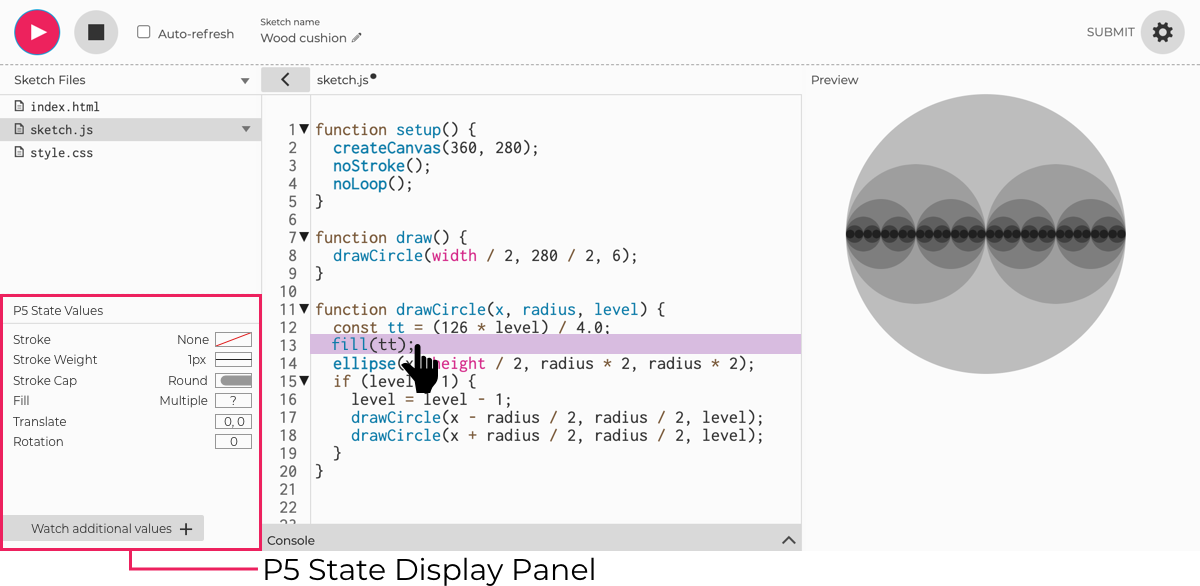}
    \Description{Screenshot showing the proposed p5 state variables IDE enhancement}
\end{figure}

\paragraph{p5 State Displays} A display of current values for common library variables, such as strokeWidth and fill color, at particular lines of code---similar in spirit to object value displays in creativity tools like Illustrator.
Some students were enthusiastic about this feature, noting that it \pQuote{would be extremely useful to be able to see all this information in one place}~(\pB{7}), while others felt it might enrich creativity by showing what options are available (\pA{9}).
Others were less enthusiastic, noting that it would be \pQuote{a little redundant}~(\pA{1}) with running the code, or that it would be tedious (\pA{7}) compared to simply writing code.

\begin{figure}[h!]
    \centering
    \includegraphics[width=0.9\linewidth]{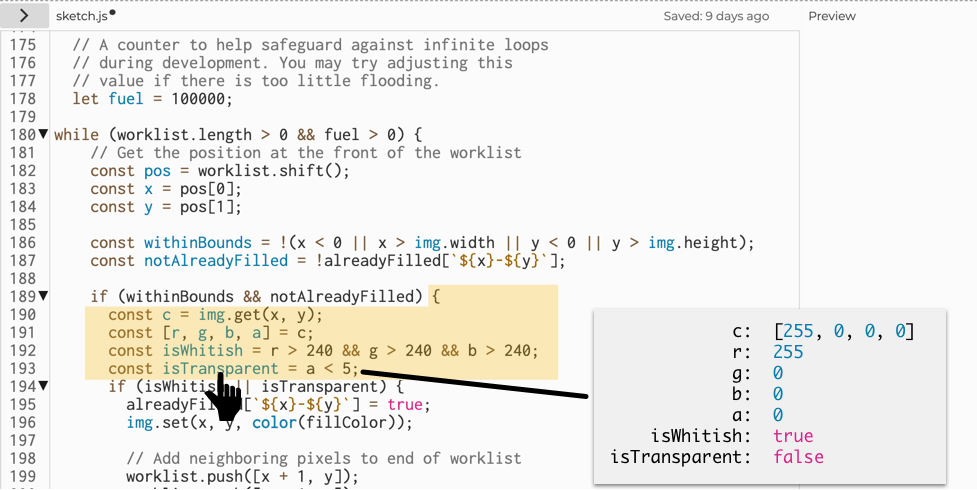}
    \Description{Screenshot showing the proposed interactive value inspector IDE enhancement}
\end{figure}

\paragraph{Interactive Value Inspector}
A growing thread of research allows users to inspect the current value of variables at various lines of code on demand, such as in Lerner's Project Boxes~\cite{Lerner20Projection} or Kang and Guo's~\cite{kang2017omnicode} ``DISPLAY ALL THE VALUES!'' approach to novice coding in Omnicode, as well through live probes~\cite{Rauch19Babylonian}. In this feature we  proposed a Projection Box style feature that included a customizable inspector.

Students were generally enthusiastic about this feature, noting that it would be helpful for beginners (\pB{3,8}) as it would make  \pQuote{loop definitions} (\pA{16}) and debugging (\pA{1,10}). Yet some worried that the implementation might be overwhelming (\pA{5}) or distracting (\pA{6}), or would not substantially improve over \verb+console.log+-based debugging (\pA{14}).\takeawayLive{}
In addition one skeptical student believed that it might \pQuote{make the coder (especially early learner) to be lazy}~(\pA{15}), and prevent them from learning good debugging skills \takeawaySkeptic{}.

\pagebreak

\begin{figure}[h!]
    \centering
    \includegraphics[width=0.7\linewidth]{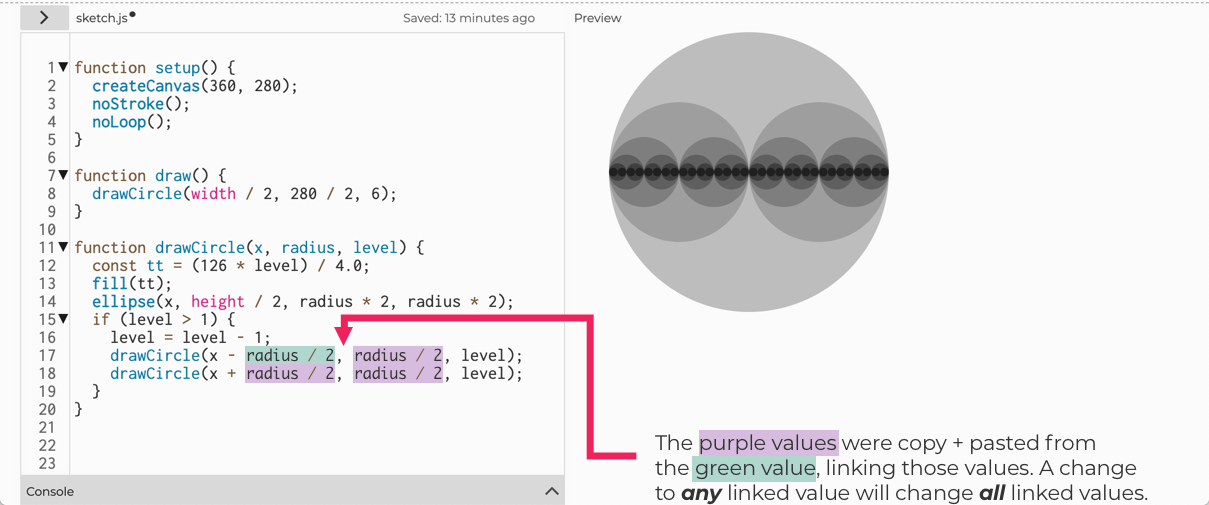}
    \vspace{-0.2in}
    \Description{Screenshot showing the proposed linked copy and paste IDE enhancement}
\end{figure}

\paragraph{Linked Copy-and-Paste}

Vihavainen\etal{}~\cite{vihavainen2014novices} note that novices tend to make heavy use of copy-paste, so a natural point of enhancement then would be to embed variable-style abstraction into copy-paste itself. This idea has been discussed in research works previously~\cite{toomim2004managing, edwards2005subtext}, however is not typically seen in this style of editor.
Some students thought this would be helpful, by \pQuote{facilitat[ing] better organizational practices}~(\pA{12}) or in niche situations (\pA{7,13}).
However, most others were apprehensive about the feature's value. Some noted that it seemed to be a more oblique version of creating a variable (\pA{1,16}).
Some thought that what was already in the editor sufficiently addressed any tasks linked copy-and-paste might accomplish, through regular copy-paste~(\pA{9}) and Find \& Replace~(\pB{8}).
\pA{13} argued that a Sublime-style multi-cursor selection would be more flexible and preferable. We note that multi-cursor support was enabled in our editor (as part of CodeMirror), although students were not explicitly made aware of this functionality.
Others still simply thought it would not be useful, and would \pQuote{creat[e] mess for me}~(\pB{4}) or otherwise be confusing (\pA{10}, \pB{5}).\takeawayClutter{}

\begin{figure}[h!]
    \centering
    \includegraphics[width=0.7\linewidth]{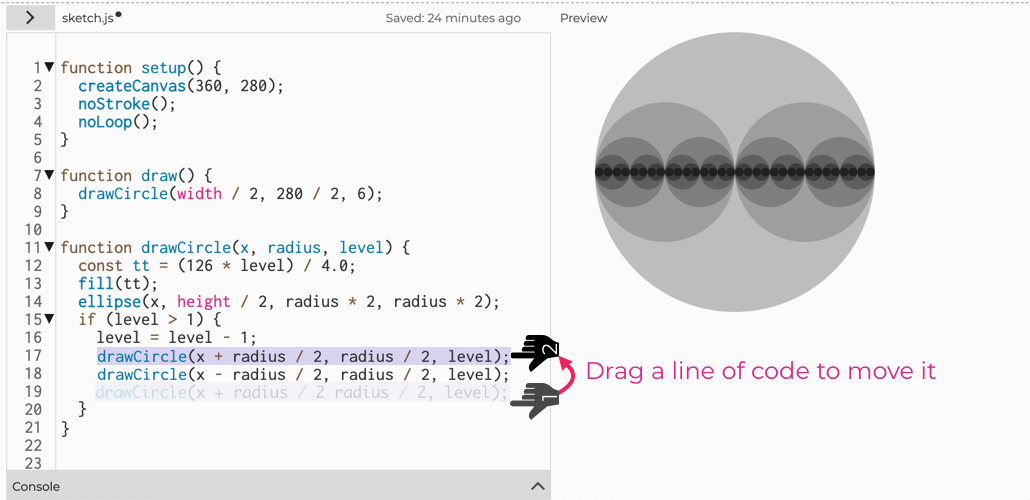}
    \vspace{-0.2in}
    \Description{Screenshot showing the proposed drag and drop IDE enhancement}
\end{figure}

\paragraph{Drag-and-Drop Refactoring}
Clicking and dragging values to create arguments, variables, and other functions in a technique that has been previously explored to useful effect~\cite{lee2013drag}. In this feature we proposed a simple version of this feature, however our presentation lacked the nuance of the presentations used by \citet{lee2013drag}, which may have led feature being rated lowest.
Although a few respondents were intrigued, some said they would prefer copy and paste (\pA{12}, \pB{1,5,6,8}), most were disinterested. For example:
\pQuote{I personally don't like dragging and dropping things because there is room for dragging and dropping into the wrong section especially if your computer is slow. I don't think copy paste was too time consuming and encourages greater accuracy}~(\pA{2}); several others shared these views about efficiency and accuracy.
In addition, there were concerns about the usability of the feature:
\pQuote{Clicking and dragging is not an ergonomic motion on a laptop touchpad}~(\pA{6}).
We highlight this as an especially valuable concern, as dragging may not be an accessible motion for some users, although something like Kobayashi and Igarashi's suspendable drag-and-drop interactions~\cite{kobayashi2007boomerang} may usefully address these concerns.

\paragraph{Other Suggested Features}

Beyond the hypothetical features we presented, some respondents suggested ideas like scratchpads or selective execution contexts similar to some of the ideas expressed in Code Bubbles~\cite{bragdon2010code} or Jupyter notebooks.
Others suggested course-specific affordances, such as hints relevant to the assignment or integration of the assignment directly into the editor. This has a similar flavor as DrRacket's language levels, and \citet{marceau2011measuring} briefly sketched out learner-attuned error messaging levels.
This is similar to Interactive Tutor Systems \cite{hundhausen2009can} which integrate curriculum and course work into a single environment. While this level of integration can be helpful, it may undermine the utility of an in-class instruction model because such interfaces are naturally self- rather than group-paced, although that should be investigated in future work.

\setlength{\parskip}{2pt}
\clearpage
\section{Survey Instrument for Initial Survey --- Year 1 (\protect\spq{}, \protect\imm{})}\label{appendix:survey-1}

\subsection{Page 1: Consent to Participate in a Research Study}

Research Project Title: Post-Course Survey of Students in Creative Coding (2021)\\
Principal Investigator: Ravi Chugh\\
Graduate Student: Andrew McNutt\\
IRB Protocol: IRB21-1062

This form is designed for students younger than 18 years of age who took the Creative Coding Pre-College Immersion class in Summer 2021 and their parents, respectively referred to as “you” and “your child” below. You (or your child) is being asked to take part in a research study. This form has important information about the reason for doing this study, what we will ask you (or your child) to do, and the way we would like to use information about you (or your child) if you choose to allow yourself (or your child) to be in the study.

\textbf{Purpose of Research Study}:
You are (or your child is) being asked to participate in a research study regarding the usability of editors for creative coding. In our recently completed course we used an in-browser editor that was slightly modified from the publicly available p5 editor. We are interested in understanding what editor features might be useful to someone learning to code (particularly in the context of a creative coding course) or otherwise making digital art works. Ultimately, this research may be published and presented at scientific conferences to improve the community’s knowledge about editors for creative coding, and may be used to improve the editor used in future iterations of our course.

\textbf{Participation Procedures and Activities}:
The full extent of the procedure will involve completing this survey. We anticipate that completion of this survey will take up to 60 minutes. Due to the difficulty of determining credit for partial completion, no compensation will be provided for partial completion. At the end of the form you (or your child) will provide a student id and preferred email address, and you (or your child) will receive a \$30 Amazon gift card for participating.

\textbf{Consent and Assent Process}:
If you are (or your child is) 18 years or older, you (or your child) can provide the consent required to opt-in to the study. If you are (or your child is) under 18 years of age, you can give your assent (or you can give your parental consent) to join the study. For students under 18 years of age, participation in this study requires both consent from a parent as well as assent from the student.

\textbf{Risks/Discomforts of Being in this Study}:
The risks to your participation in the survey are those associated with basic computer tasks, including boredom, fatigue, or mild stress. Benefits of Being in this Study The only benefit to you (or your child) is the learning experience from participating in a research study. The benefit to society is the contribution to scientific knowledge.

\textbf{Confidentiality of Data and Limits to Confidentiality}:
Any reports and presentations about the findings from this study will not include your (or your child’s) name or any other identifying information.

\textbf{Use of Your Research Data}:
We will never share the data beyond the University of Chicago research team. However, an analysis of the data may be analyzed and published in scientific conference proceedings or journal articles. The free-text responses provided to any portion of this survey may be quoted in part or in whole in this publication. We will remove any information from the analysis that could identify you (or your child) before providing the analysis for publication.

\textbf{Voluntary Participation and Right to Refuse or Withdraw}:
Participation in this study is voluntary. The decision to participate in this study is entirely up to you and your child. You (or your child) may refuse to take part in the study at any time without prejudice or penalties and will not result in any loss of benefits to which you (or your child) are otherwise entitled.

\textbf{Mandatory Reporting of Child Abuse or Neglect}:
The research study staff are mandated reporters and are required to report suspected child abuse or neglect to the Illinois Department of Child and Family Services. For more information, please see the University policy: https://tinyurl.com/mr26uazn

\textbf{Contact Information for Research Questions and Participation}:
If you have questions or concerns about the study, you can contact the researchers at:

\noindent{} Principal Investigator\\
Ravi Chugh,\\
Associate Professor\\
John Crerar Library\\
University of Chicago\\
5730 S Ellis Ave\\
Chicago, IL 60637\\
Email: rchugh@uchicago.edu

\noindent{} Graduate Student\\
Andrew McNutt,\\
PhD student\\
John Crerar Library\\
University of Chicago\\
5730 S Ellis Ave\\
Chicago, IL 60637\\
Email: mcnutt@uchicago.edu

If you have any questions about your rights as a participant in this research, feel you have been harmed, or wish to discuss other study-related concerns with someone who is not part of the research team,
you can contact the University of Chicago Social and Behavioral Sciences Institutional Review Board (IRB) Office by phone at (773) 702-2915, or by email at sbs-irb@uchicago.edu.

\textbf{Parental Consent}
\vspace{-0.5em}

\begin{enumerate}

    \item Parent full name (Last, First)

    \item Parent Email address

    \item I have read and understood this consent form. Yes $\bigcirc$ no $\bigcirc$

    \item I am a parent and give consent for my child, under 18 years of age, to participate in this study.
          Yes $\bigcirc$ no $\bigcirc$

\end{enumerate}

\textbf{Student Assent}
\vspace{-0.5em}

\begin{enumerate}

    \item Student full name (Last, First)

    \item Student Email address

    \item Student GitHub username (same as used for homework submission in this class)

    \item Student CNetID (the username before your @uchicago.edu email address)

    \item I have read and understood this consent form. Yes $\bigcirc$ no $\bigcirc$

    \item I am a student, under 18 years of age, and give assent to participate in this study. Yes $\bigcirc$ no $\bigcirc$

\end{enumerate}

\newpage

\subsection{Page 2: Introduction and Reflection}

In this section, we’ll ask you some questions about your programming background, and to reflect on your experience during the course.

\begin{enumerate}[topsep=7pt,itemsep=7pt]
    \item \textbf{Pre-Course Experience}. How much programming experience did you have prior to taking the course?

    \item \textbf{Post-Course Confidence}. How confident do you feel in your programming skills after taking this course? Have they improved?

    \item \textbf{Challenges}. What aspect of coding or learning to program gave you the most trouble? As a way to help organize your thinking, consider the assignment that you had the most difficulty with. Could the editor have done anything to help you with that?

    \item \textbf{Debugging}. Think about the experience of debugging. How did you go about doing it? If you used console.log to help debug, did you find it helpful? Did you use any other strategies? Is there anything about it or the debugging process that you wish could have been different?

    \item \textbf{Error Messages}. Think about the error messages you encountered (inline in the code box, in the console area under the code box, in the browser console, or elsewhere). Were they useful? How did you deal with them? Do you wish they were presented differently?

    \item \textbf{Code Organization}. How did you go about organizing your code? For instance, how did you decide where to place variables, create functions? Was there ever a point when your organizational scheme ran into problems, if so how did you handle it? Is there anything the editor could have done to help you during these organizational tasks?

    \item \textbf{Freeze Frame Homework}. Think about the freeze frame assignment (or any other time during the course when you needed to repeatedly edit and re-run the code in order to get particular positions or other values to achieve a desired effect). Is there anything the editor could have done to help you get your image to be just right?

    \item \textbf{External Tools}. It’s natural to use other tools as part of the programming process, such as color eye droppers or p5’s online documentation. Do you think it would be useful to integrate these tools as part of the editor? What other tools can you imagine wanting to be part of your in-editor coding workflow?

    \item \textbf{Desired Features}. What sorts of editor features might have allowed you to be more effective in your coding? What sorts of editor features might have allowed you to be more creative?

\end{enumerate}

\newpage

\subsection{Pages 3-18: Editor Features}

In this section, we’ll ask for your thoughts and opinions about some features that appeared in the editor as well as some hypothetical features that we may implement for future iterations of the course.

\begin{enumerate}[topsep=7pt,itemsep=7pt]
    \item \textbf{Autocomplete} Imagine an editor feature which provides autocomplete suggestions as you type. This would be akin to the predictive text feature found in many messaging applications, but would be sensitive to variables you’ve created and functions available from imported libraries. While this feature appears in some other editors it did not appear in our editor.

          \begin{figure}[h!]
              \centering
              \includegraphics[width=0.7\linewidth]{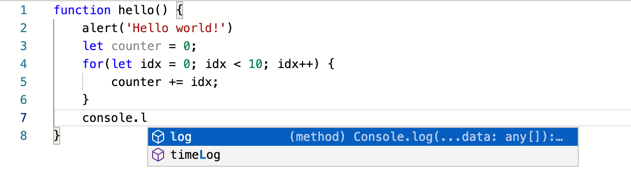}
          \end{figure}

          \begin{enumerate}[topsep=0pt,itemsep=0pt]
              \item Do you think this would be useful? (1) Not Very Useful $\bigcirc$, (2) $\bigcirc$, (3) Neither useful nor unuseful $\bigcirc$, (4) $\bigcirc$, Very Useful $\bigcirc$
              \item How often do you think you would use this feature? (1) Never $\bigcirc$, (2) $\bigcirc$, (3)Occasionally $\bigcirc$, (4) $\bigcirc$, All the time $\bigcirc$
              \item Why or why not? Is there any way you would like to modify this feature to make it more useful?
          \end{enumerate}

    \item \textbf{Linters} Our editor featured a tool called a “linter” that surfaced stylistic or coding errors through on-screen alerts, as in the image below. This tool is analogous to spell- and grammar-checkers in standard word processors. The particular linter used in our editor, called JSHint, tends not to give many warnings for stylistic errors. Other available linters give many more warnings for stylistic errors.

          \begin{figure}[h!]
              \centering
              \includegraphics[width=0.7\linewidth]{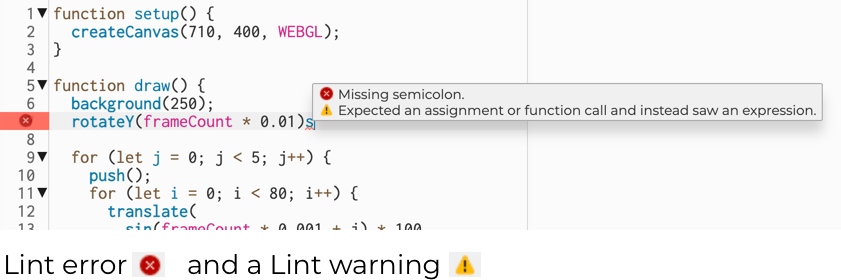}
          \end{figure}

          \begin{enumerate}[topsep=0pt,itemsep=0pt]
              \item Do you think this is useful? (1) Not Very Useful $\bigcirc$, (2) $\bigcirc$, (3) Neither useful nor unuseful $\bigcirc$, (4) $\bigcirc$, Very Useful $\bigcirc$
              \item How often do you think you used this feature? (1) Never $\bigcirc$, (2) $\bigcirc$, (3)Occasionally $\bigcirc$, (4) $\bigcirc$, All the time $\bigcirc$
              \item Why or why not? Is there any way you would like to modify this feature to make it more useful?
          \end{enumerate}

    \item \textbf{Tidy Code} Our editor featured a button called “Tidy Code” which automatically reorganized your code. This feature is sometimes seen in other editors and is more commonly known as an “auto-formatter”. These can typically be configured to enforce a particular coding style.

          \begin{figure}[h!]
              \centering
              \includegraphics[width=0.6\linewidth]{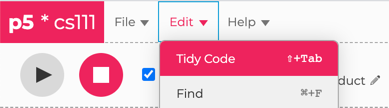}
          \end{figure}

          \begin{enumerate}[topsep=0pt,itemsep=0pt]
              \item Do you think this is useful? (1) Not Very Useful $\bigcirc$, (2) $\bigcirc$, (3) Neither useful nor unuseful $\bigcirc$, (4) $\bigcirc$, Very Useful $\bigcirc$
              \item How often do you think you used this feature? (1) Never $\bigcirc$, (2) $\bigcirc$, (3)Occasionally $\bigcirc$, (4) $\bigcirc$, All the time $\bigcirc$
              \item Why or why not? Is there any way you would like to modify this feature to make it more useful?
          \end{enumerate}

          \vfill \pagebreak
    \item \textbf{Auto-refresh} There is a feature in our editor called “Auto-refresh.” When selected, it re-runs your code every time you finish typing (or sometimes before). This enables small update cycles as you code.

          \begin{figure}[h!]
              \centering
              \includegraphics[width=0.5\linewidth]{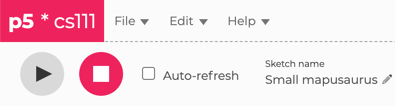}
          \end{figure}

          \begin{enumerate}[topsep=0pt,itemsep=0pt]
              \item Do you think this is useful? (1) Not Very Useful $\bigcirc$, (2) $\bigcirc$, (3) Neither useful nor unuseful $\bigcirc$, (4) $\bigcirc$, Very Useful $\bigcirc$
              \item How often do you think you used this feature? (1) Never $\bigcirc$, (2) $\bigcirc$, (3)Occasionally $\bigcirc$, (4) $\bigcirc$, All the time $\bigcirc$
              \item Why or why not? Is there any way you would like to modify this feature to make it more useful?
          \end{enumerate}

    \item \textbf{Code Folding} There is a feature in our editor, and many other editors, called “code folding” as in the screenshot below. This allows you to collapse certain sections of code, such as functions and loops. The “folded” code is still there and can be referenced from other places, but it’s temporarily hidden and replaced with “...”.

          \begin{figure}[h!]
              \centering
              \includegraphics[width=0.4\linewidth]{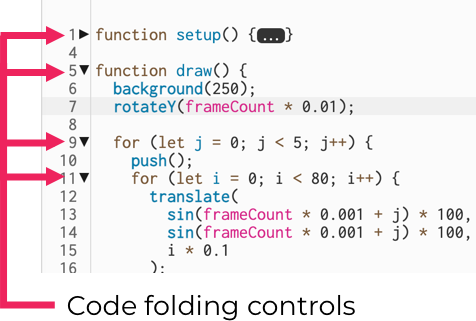}
          \end{figure}

          \begin{enumerate}[topsep=0pt,itemsep=0pt]
              \item Do you think this is useful? (1) Not Very Useful $\bigcirc$, (2) $\bigcirc$, (3) Neither useful nor unuseful $\bigcirc$, (4) $\bigcirc$, Very Useful $\bigcirc$
              \item How often do you think you used this feature? (1) Never $\bigcirc$, (2) $\bigcirc$, (3)Occasionally $\bigcirc$, (4) $\bigcirc$, All the time $\bigcirc$
              \item Why or why not? Is there any way you would like to modify this feature to make it more useful?
          \end{enumerate}

    \item \textbf{Canvas Ruler} Imagine a feature which allows you to place a draggable ruler into the drawing side of the editor. You can use it as a way to visually identify screen coordinates. This feature might involve a way to display the current direction and placement of the coordinate origin, especially with regard to translation and rotation functions.

          \begin{figure}[h!]
              \centering
              \includegraphics[width=0.5\linewidth]{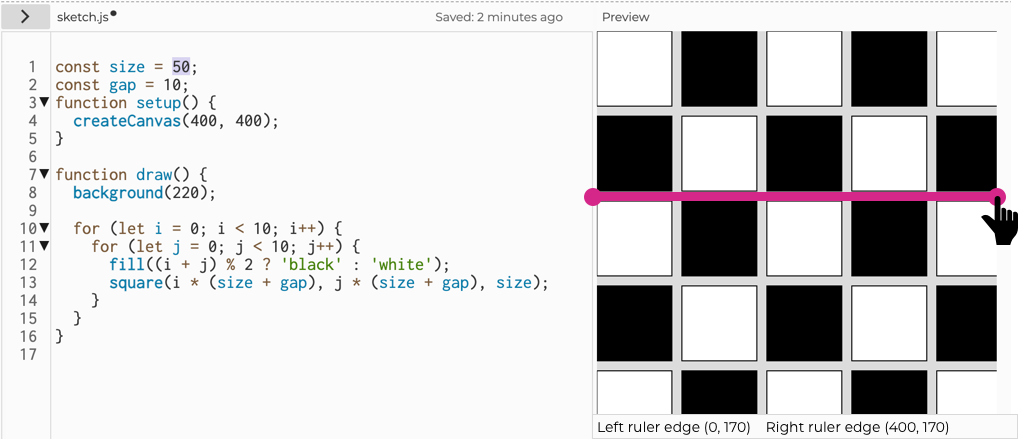}
          \end{figure}

          \begin{enumerate}[topsep=0pt,itemsep=0pt]
              \item Do you think this would be useful? (1) Not Very Useful $\bigcirc$, (2) $\bigcirc$, (3) Neither useful nor unuseful $\bigcirc$, (4) $\bigcirc$, Very Useful $\bigcirc$
              \item How often do you think you would use this feature? (1) Never $\bigcirc$, (2) $\bigcirc$, (3)Occasionally $\bigcirc$, (4) $\bigcirc$, All the time $\bigcirc$
              \item Why or why not? Is there any way you would like to modify this feature to make it more useful?
          \end{enumerate}

          \vfill \pagebreak

    \item \textbf{Number Sliders} Imagine a feature which allows you to modify the numeric values in the code without typing or re-running the program (such as in the image below). With this feature, you click a value of interest and then drag a slider that appears above it to change it. The canvas is continuously re-rendered as you drag the slider. This would be similar to using p5’s slider function, but, rather than just changing the value in the running code, it would also modify the text of the code.

          \begin{figure}[h!]
              \centering
              \includegraphics[width=0.7\linewidth]{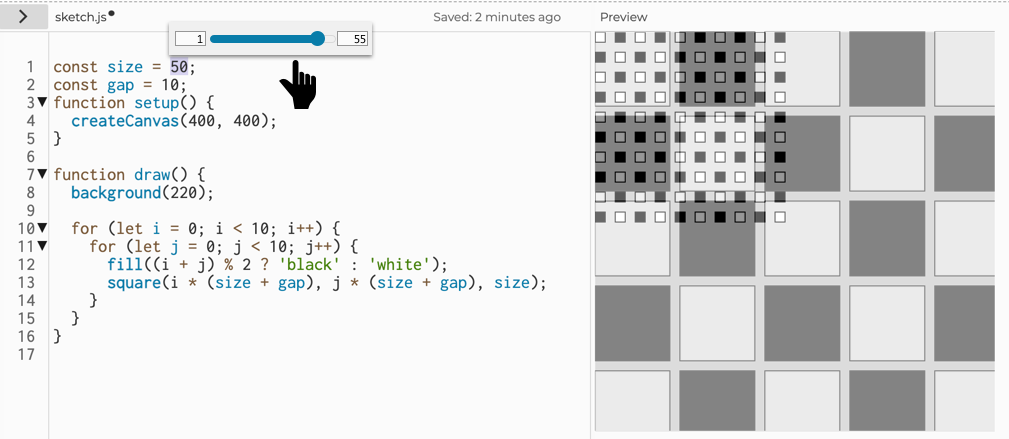}
          \end{figure}

          \begin{enumerate}[topsep=0pt,itemsep=0pt]
              \item Do you think this would be useful? (1) Not Very Useful $\bigcirc$, (2) $\bigcirc$, (3) Neither useful nor unuseful $\bigcirc$, (4) $\bigcirc$, Very Useful $\bigcirc$
              \item How often do you think you would use this feature? (1) Never $\bigcirc$, (2) $\bigcirc$, (3)Occasionally $\bigcirc$, (4) $\bigcirc$, All the time $\bigcirc$
              \item Why or why not? Is there any way you would like to modify this feature to make it more useful?
          \end{enumerate}

    \item \textbf{Color Picker} Imagine having a color picker integrated into the editor. When selecting color values in the code, the color picker could appear on hover (as in the image below) to modify the value, or the tool could be docked into the bottom of the editor (allowing it to be always on). This could include pre-configured or document-based palettes, as in Illustrator or Photoshop.

          \begin{figure}[h!]
              \centering
              \includegraphics[width=0.7\linewidth]{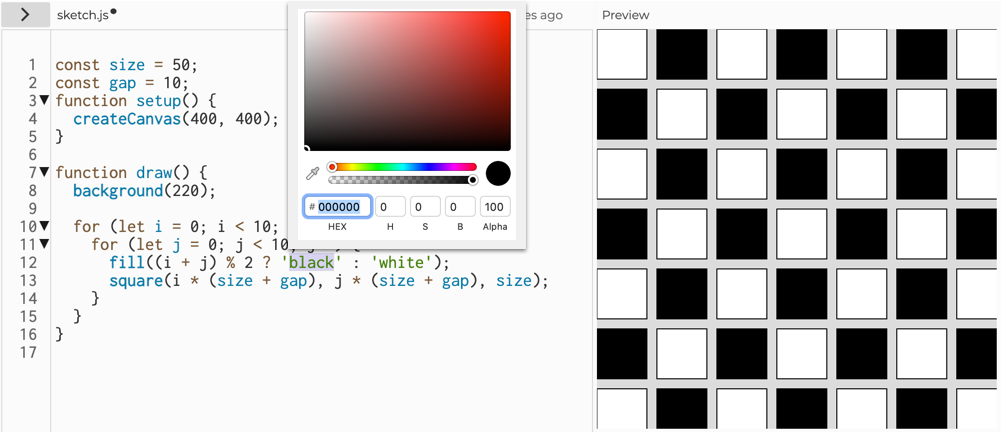}
          \end{figure}

          \begin{enumerate}[topsep=0pt,itemsep=0pt]
              \item Do you think this would be useful? (1) Not Very Useful $\bigcirc$, (2) $\bigcirc$, (3) Neither useful nor unuseful $\bigcirc$, (4) $\bigcirc$, Very Useful $\bigcirc$
              \item How often do you think you would use this feature? (1) Never $\bigcirc$, (2) $\bigcirc$, (3)Occasionally $\bigcirc$, (4) $\bigcirc$, All the time $\bigcirc$
              \item Why or why not? Is there any way you would like to modify this feature to make it more useful?
          \end{enumerate}

          \vfill \pagebreak

    \item \textbf{p5 State Displays} In p5 it is common to set values for variables such as strokeWidth (which describes the width of subsequently drawn lines), or fill (which describes the interior color of subsequently drawn shape). These are examples of ""state variables"". There are a variety of such variables in p5, however (in contrast with digital drawing tools like Photoshop), these variables are not displayed anywhere in the editor. <br/><br/> Imagine a feature where all of the relevant state values are shown, such that when you move the text cursor to a line in your code, the display shows the state values at that point in time. This would allow you to evaluate if your drawing tools are configured as you want them to be.

          \begin{figure}[h!]
              \centering
              \vspace{-0.1in}
              \includegraphics[width=0.7\linewidth]{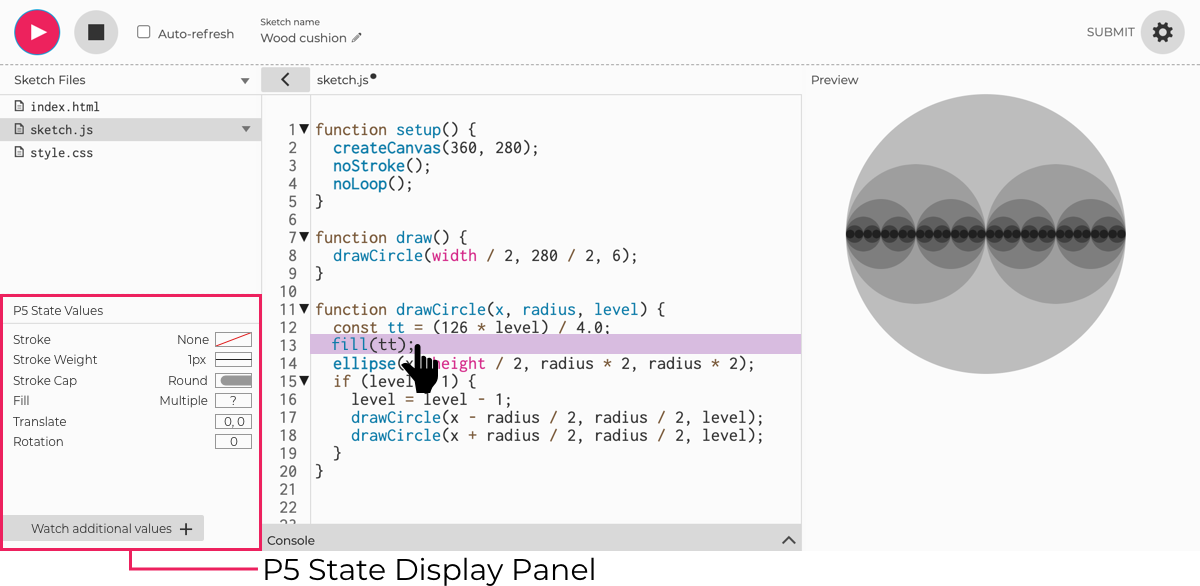}
              \vspace{-0.1in}
          \end{figure}

          \begin{enumerate}[topsep=0pt,itemsep=0pt]
              \item Do you think this would be useful? (1) Not Very Useful $\bigcirc$, (2) $\bigcirc$, (3) Neither useful nor unuseful $\bigcirc$, (4) $\bigcirc$, Very Useful $\bigcirc$
              \item How often do you think you would use this feature? (1) Never $\bigcirc$, (2) $\bigcirc$, (3)Occasionally $\bigcirc$, (4) $\bigcirc$, All the time $\bigcirc$
              \item Why or why not? Is there any way you would like to modify this feature to make it more useful?
          \end{enumerate}

    \item \textbf{Interactive Value Inspector} Imagine a feature which allows you to see the value of the current program execution by hovering over chunks of the code. It would provide similar information as when inserting console.log statements into your code, but instead you would extract that same information through hovering. In contrast with ""p5 State Displays"" (which only shows p5 state variables like fill and strokeWidth) this feature would allow you to see both state variables as well as the value of all variables, including ones you've defined. This information could be presented through a tooltip (as in the below image) or through a docked panel.

          \begin{figure}[h!]
              \centering
              \includegraphics[width=0.6\linewidth]{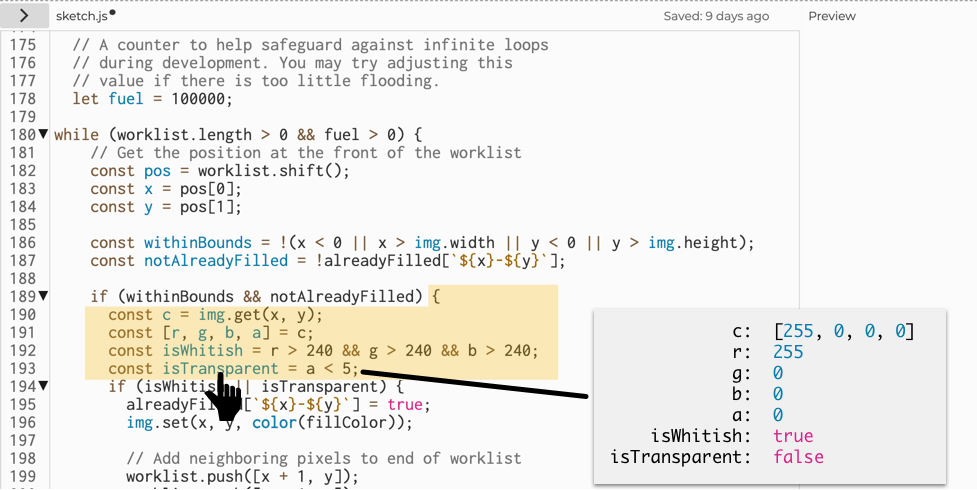}
          \end{figure}

          \begin{enumerate}[topsep=0pt,itemsep=0pt]
              \item Do you think this would be useful? (1) Not Very Useful $\bigcirc$, (2) $\bigcirc$, (3) Neither useful nor unuseful $\bigcirc$, (4) $\bigcirc$, Very Useful $\bigcirc$
              \item How often do you think you would use this feature? (1) Never $\bigcirc$, (2) $\bigcirc$, (3)Occasionally $\bigcirc$, (4) $\bigcirc$, All the time $\bigcirc$
              \item Why or why not? Is there any way you would like to modify this feature to make it more useful?
          \end{enumerate}

          \vfill \pagebreak
    \item \textbf{In-context Docs} Imagine an editor feature which gives you access to the documentation while you are writing code. This might involve a tooltip that appears on hover (as in the image below) which describes the usage of a particular function. It could also involve showing the description in a dedicated pane on the side. While such features appear in some other editors it did not appear in our editor.

          \begin{figure}[h!]
              \centering
              \includegraphics[width=0.4\linewidth]{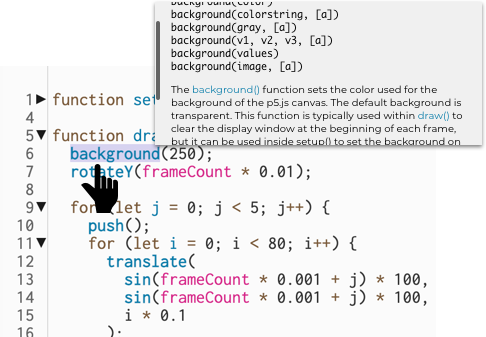}
          \end{figure}

          \begin{enumerate}[topsep=0pt,itemsep=0pt]
              \item Do you think this would be useful? (1) Not Very Useful $\bigcirc$, (2) $\bigcirc$, (3) Neither useful nor unuseful $\bigcirc$, (4) $\bigcirc$, Very Useful $\bigcirc$
              \item How often do you think you would use this feature? (1) Never $\bigcirc$, (2) $\bigcirc$, (3)Occasionally $\bigcirc$, (4) $\bigcirc$, All the time $\bigcirc$
              \item Why or why not? Is there any way you would like to modify this feature to make it more useful?
          \end{enumerate}

    \item \textbf{Code Snippet Templates} Imagine a feature which allows you to paste in common code snippets from a list. After clicking one of the desired options (such as in the image below) a piece of code achieving that functionality will be added to your code. These snippets could include small structures, such as for-loops, or larger structures, such as particular API uses or classes. This feature sometimes appears in other coding systems, but was not implemented in our system.

          \begin{figure}[h!]
              \centering
              \vspace{-0.1in}
              \includegraphics[width=0.4\linewidth]{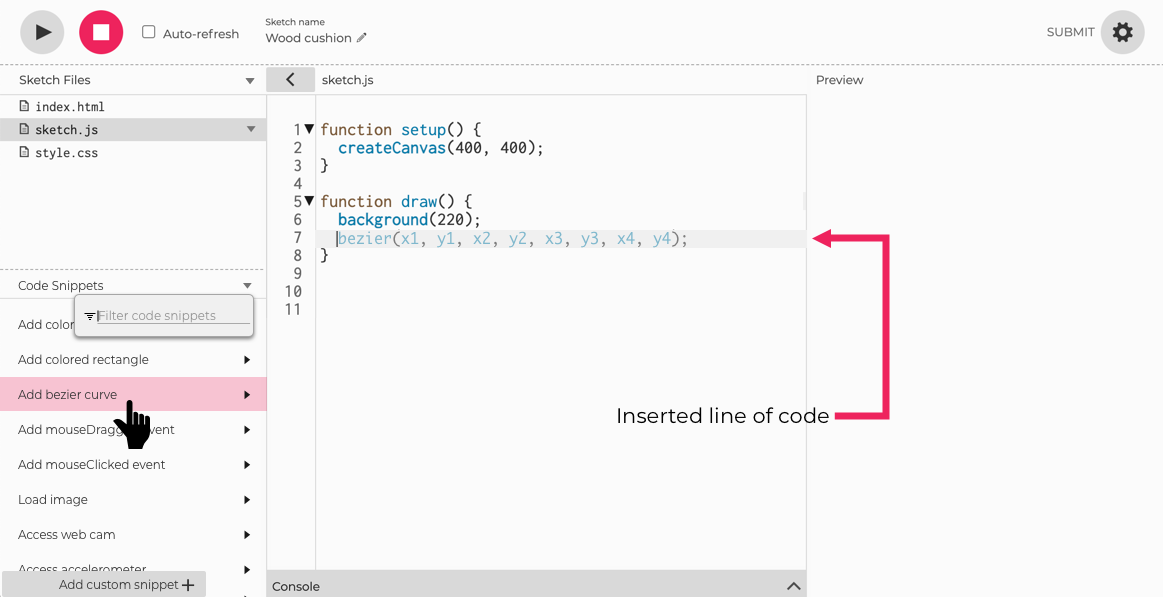}
          \end{figure}

          \begin{enumerate}[topsep=0pt,itemsep=0pt]
              \item Do you think this would be useful? (1) Not Very Useful $\bigcirc$, (2) $\bigcirc$, (3) Neither useful nor unuseful $\bigcirc$, (4) $\bigcirc$, Very Useful $\bigcirc$
              \item How often do you think you would use this feature? (1) Never $\bigcirc$, (2) $\bigcirc$, (3)Occasionally $\bigcirc$, (4) $\bigcirc$, All the time $\bigcirc$
              \item Why or why not? Is there any way you would like to modify this feature to make it more useful?
          \end{enumerate}

    \item \textbf{Coding by Drawing Tools} Imagine an editor feature which allows you to fill out the arguments to particular functions graphically. For instance, you might indicate to the editor that you are interested in drawing a bezier curve, and then draw each of the vertices in the curve directly on the editor, just as you would in a GUI-based tool like Illustrator, which in turn inserts a corresponding line of bezier command in your code. Unlike in the previous feature, which just inserted code templates, this feature allows you to specify the values of the inserted code with your mouse on the output canvas.

          \begin{figure}[h!]
              \centering
              \vspace{-0.1in}
              \includegraphics[width=0.4\linewidth]{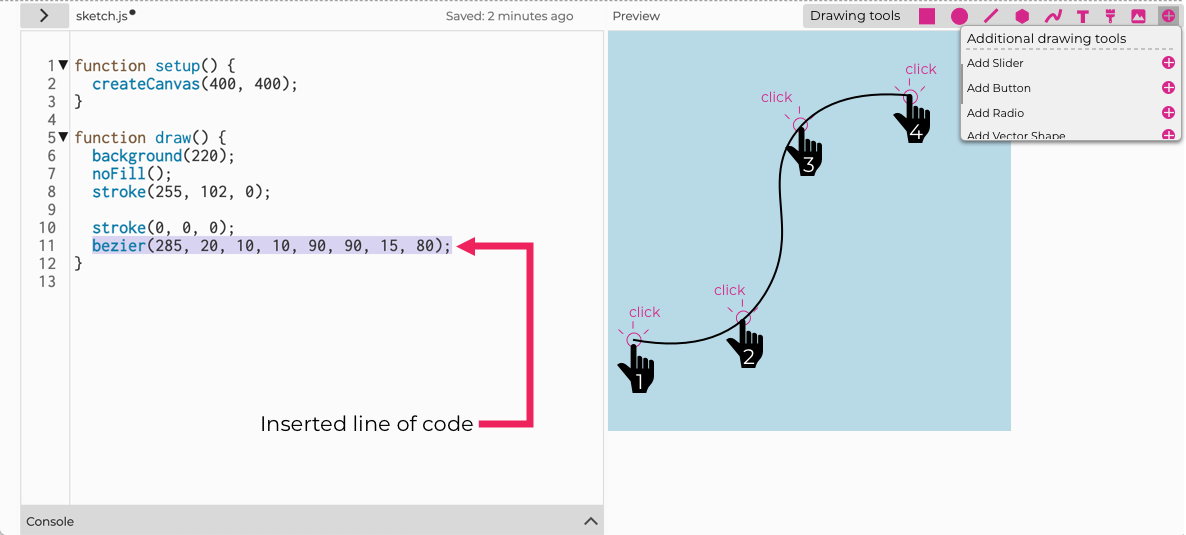}              \vspace{-0.1in}
          \end{figure}

          \begin{enumerate}[topsep=0pt,itemsep=0pt]
              \item Do you think this would be useful? (1) Not Very Useful $\bigcirc$, (2) $\bigcirc$, (3) Neither useful nor unuseful $\bigcirc$, (4) $\bigcirc$, Very Useful $\bigcirc$
              \item How often do you think you would use this feature? (1) Never $\bigcirc$, (2) $\bigcirc$, (3)Occasionally $\bigcirc$, (4) $\bigcirc$, All the time $\bigcirc$
              \item Why or why not? Is there any way you would like to modify this feature to make it more useful?
          \end{enumerate}

          \vfill \pagebreak

    \item \textbf{Linked Copy-and-Paste} A common abstraction mechanism that we used in class is to create variables or functions rather than copy-pasting chunks of code. While variables and functions are a useful form of computational thinking, there are other ways to approach this task. <br/><br/>Imagine a feature which keeps track of your copy-and-pastes: whenever you edit a value you’ve copied and pasted, all pieces of code which were copied are also changed. This special linked copy-paste can be selectively turned on and off so that you can make edits without changing all copies.

          \begin{figure}[h!]
              \centering
              \vspace{-0.1in}
              \includegraphics[width=0.45\linewidth]{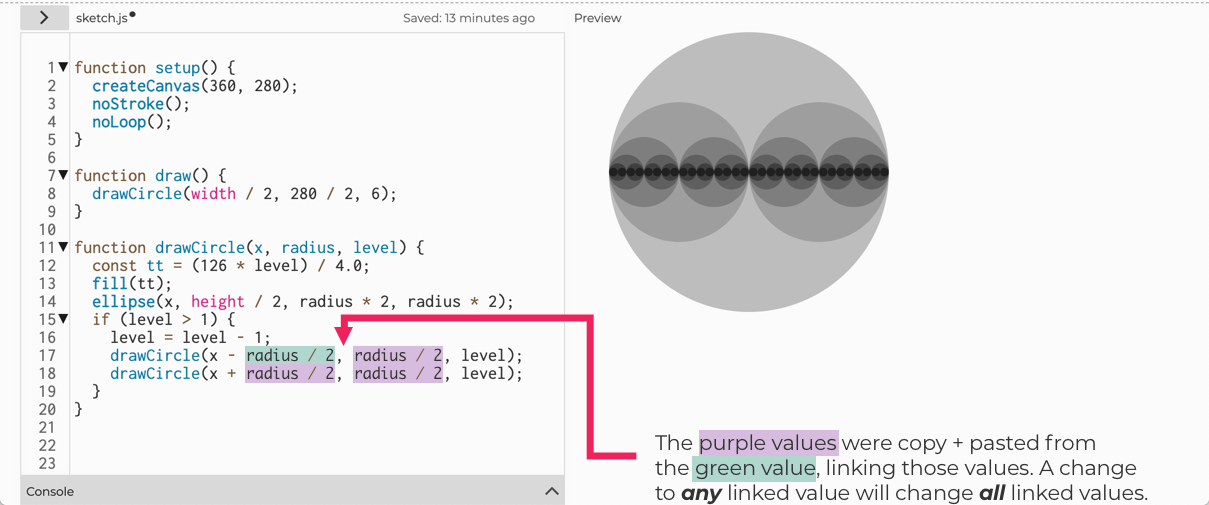}
              \vspace{-0.1in}
          \end{figure}

          \begin{enumerate}[topsep=0pt,itemsep=0pt]
              \item Do you think this would be useful? (1) Not Very Useful $\bigcirc$, (2) $\bigcirc$, (3) Neither useful nor unuseful $\bigcirc$, (4) $\bigcirc$, Very Useful $\bigcirc$
              \item How often do you think you would use this feature? (1) Never $\bigcirc$, (2) $\bigcirc$, (3)Occasionally $\bigcirc$, (4) $\bigcirc$, All the time $\bigcirc$
              \item Why or why not? Is there any way you would like to modify this feature to make it more useful?
          \end{enumerate}

    \item \textbf{Drag-and-Drop Refactoring} Refactoring is the process of changing the way a piece of code is organized such that the functionality remains the same, but the code is easier to work with. You probably did this during the course by making a variable to capture repeated code or by creating a function to represent some repeated functionality. <br/><br/>Imagine a feature which allows you to click and drag values to create arguments, variables, and functions. This might allow you to reorder lines of code by clicking and dragging them, or to highlight a series of repeated values and drag them to automatically create a new variable.

          \begin{figure}[h!]
              \centering
              \vspace{-0.1in}
              \includegraphics[width=0.45\linewidth]{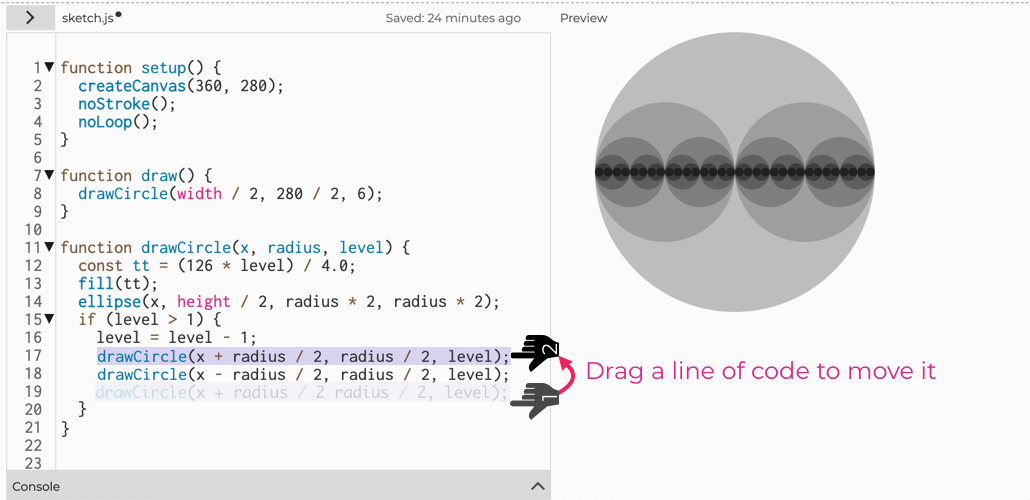}
              \vspace{-0.1in}
          \end{figure}

          \begin{enumerate}[topsep=0pt,itemsep=0pt]
              \item Do you think this would be useful? (1) Not Very Useful $\bigcirc$, (2) $\bigcirc$, (3) Neither useful nor unuseful $\bigcirc$, (4) $\bigcirc$, Very Useful $\bigcirc$
              \item How often do you think you would use this feature? (1) Never $\bigcirc$, (2) $\bigcirc$, (3)Occasionally $\bigcirc$, (4) $\bigcirc$, All the time $\bigcirc$
              \item Why or why not? Is there any way you would like to modify this feature to make it more useful?
          \end{enumerate}

    \item \textbf{Time Travel Slider} Imagine an editor feature which allows you to go back to earlier points in time of your code’s execution. With such a feature you’d press Play, as normal, and watch your code execute. If there was an intermediate state you were curious about you can pause the execution and go back (by dragging a slider) to an earlier state of the canvas. Once you are finished inspecting you can resume execution without rerunning the code. This would allow you to inspect how your code was adding shapes to the canvas over time.

          \begin{figure}[h!]
              \centering
              \vspace{-0.1in}
              \includegraphics[width=0.45\linewidth]{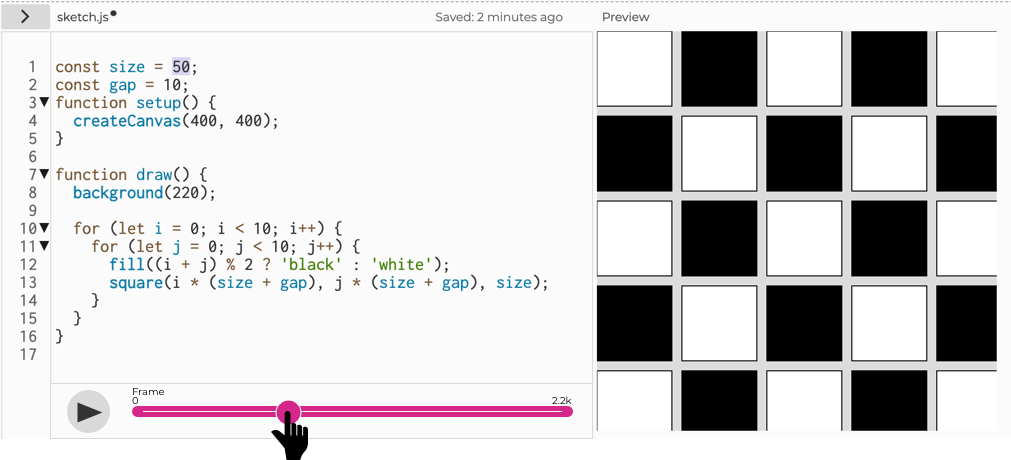}
              \vspace{-0.1in}
          \end{figure}

          \begin{enumerate}[topsep=0pt,itemsep=0pt]
              \item Do you think this would be useful? (1) Not Very Useful $\bigcirc$, (2) $\bigcirc$, (3) Neither useful nor unuseful $\bigcirc$, (4) $\bigcirc$, Very Useful $\bigcirc$
              \item How often do you think you would use this feature? (1) Never $\bigcirc$, (2) $\bigcirc$, (3)Occasionally $\bigcirc$, (4) $\bigcirc$, All the time $\bigcirc$
              \item Why or why not? Is there any way you would like to modify this feature to make it more useful?
          \end{enumerate}

          \vfill \pagebreak

    \item \textbf{Directly Manipulate Shape Attributes on Canvas} Imagine being able to edit the output canvas and have that change the corresponding JavaScript code. This would involve making changes to specific values graphically, such as changing the size of circle or end points of lines by dragging them to a desired position. This differs from the functionality of the previously described ""Code by Drawing Tools"" feature; that one allowed clicking and dragging to add new shapes to the code, whereas this one allows clicking and dragging to dynamically update and modify existing shapes in the code.

          \begin{figure}[h!]
              \centering
              \vspace{-0.1in}
              \includegraphics[width=0.7\linewidth]{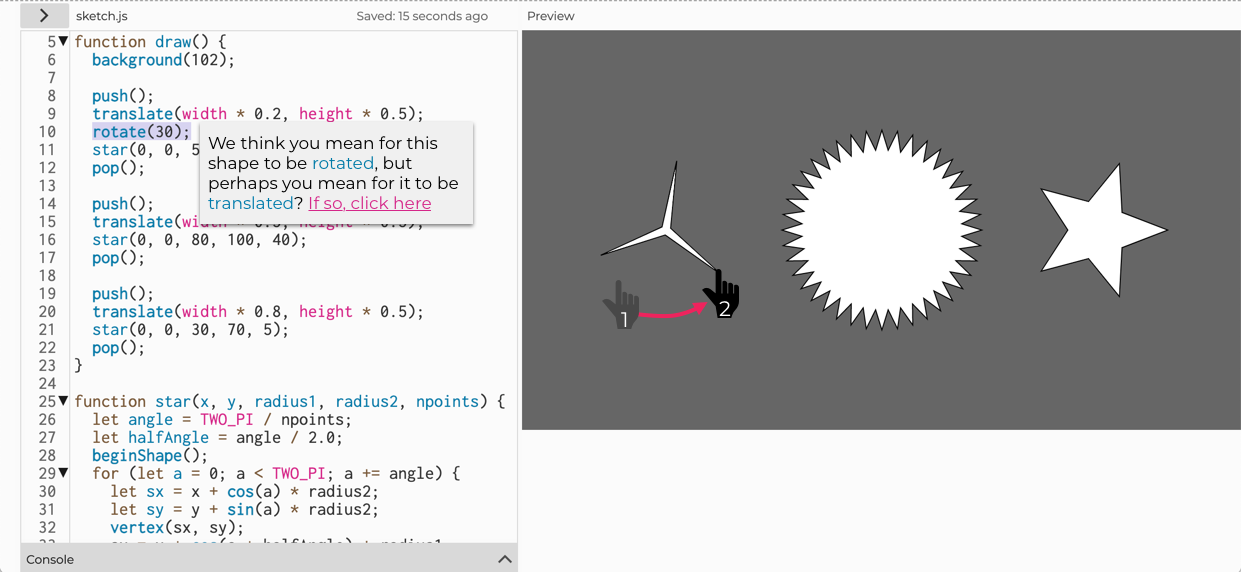}
              \vspace{-0.1in}
          \end{figure}

          \begin{enumerate}[topsep=0pt,itemsep=0pt]
              \item Do you think this would be useful? (1) Not Very Useful $\bigcirc$, (2) $\bigcirc$, (3) Neither useful nor unuseful $\bigcirc$, (4) $\bigcirc$, Very Useful $\bigcirc$
              \item How often do you think you would use this feature? (1) Never $\bigcirc$, (2) $\bigcirc$, (3)Occasionally $\bigcirc$, (4) $\bigcirc$, All the time $\bigcirc$
              \item Why or why not? Is there any way you would like to modify this feature to make it more useful?
          \end{enumerate}

\end{enumerate}

\newpage

\subsection{Page 19: Wrap up}

We just worked through a variety of potential editor features individually. To wrap up, we will recap those features and consider them collectively.

\begin{enumerate}[topsep=7pt,itemsep=7pt]

    \item \textbf{Feature Review}. Reflect on the features we've just been considering. Which of these hypothetical features we considered are you most excited about? (hover over (?) for details)

          \begin{itemize}[topsep=5pt,itemsep=0pt]
              \item \textbf{Autocomplete} Very Disinterested $\bigcirc$, Disinterested $\bigcirc$, Neutral $\bigcirc$, Interested $\bigcirc$, Very Interested $\bigcirc$
              \item \textbf{Linters} Very Disinterested $\bigcirc$, Disinterested $\bigcirc$, Neutral $\bigcirc$, Interested $\bigcirc$, Very Interested $\bigcirc$
              \item \textbf{Tidy Code} Very Disinterested $\bigcirc$, Disinterested $\bigcirc$, Neutral $\bigcirc$, Interested $\bigcirc$, Very Interested $\bigcirc$
              \item \textbf{Auto-refresh} Very Disinterested $\bigcirc$, Disinterested $\bigcirc$, Neutral $\bigcirc$, Interested $\bigcirc$, Very Interested $\bigcirc$
              \item \textbf{Code Folding} Very Disinterested $\bigcirc$, Disinterested $\bigcirc$, Neutral $\bigcirc$, Interested $\bigcirc$, Very Interested $\bigcirc$
              \item \textbf{Canvas Ruler} Very Disinterested $\bigcirc$, Disinterested $\bigcirc$, Neutral $\bigcirc$, Interested $\bigcirc$, Very Interested $\bigcirc$
              \item \textbf{Number Sliders} Very Disinterested $\bigcirc$, Disinterested $\bigcirc$, Neutral $\bigcirc$, Interested $\bigcirc$, Very Interested $\bigcirc$
              \item \textbf{Color Picker} Very Disinterested $\bigcirc$, Disinterested $\bigcirc$, Neutral $\bigcirc$, Interested $\bigcirc$, Very Interested $\bigcirc$
              \item \textbf{p5 State Displays} Very Disinterested $\bigcirc$, Disinterested $\bigcirc$, Neutral $\bigcirc$, Interested $\bigcirc$, Very Interested $\bigcirc$
              \item \textbf{Interactive Value Inspector} Very Disinterested $\bigcirc$, Disinterested $\bigcirc$, Neutral $\bigcirc$, Interested $\bigcirc$, Very Interested $\bigcirc$
              \item \textbf{In-context Docs} Very Disinterested $\bigcirc$, Disinterested $\bigcirc$, Neutral $\bigcirc$, Interested $\bigcirc$, Very Interested $\bigcirc$
              \item \textbf{Code Snippet Templates} Very Disinterested $\bigcirc$, Disinterested $\bigcirc$, Neutral $\bigcirc$, Interested $\bigcirc$, Very Interested $\bigcirc$
              \item \textbf{Coding by Drawing Tools} Very Disinterested $\bigcirc$, Disinterested $\bigcirc$, Neutral $\bigcirc$, Interested $\bigcirc$, Very Interested $\bigcirc$
              \item \textbf{Linked Copy-and-Paste} Very Disinterested $\bigcirc$, Disinterested $\bigcirc$, Neutral $\bigcirc$, Interested $\bigcirc$, Very Interested $\bigcirc$
              \item \textbf{Drag-and-Drop Refactoring} Very Disinterested $\bigcirc$, Disinterested $\bigcirc$, Neutral $\bigcirc$, Interested $\bigcirc$, Very Interested $\bigcirc$
              \item \textbf{Time Travel Slider} Very Disinterested $\bigcirc$, Disinterested $\bigcirc$, Neutral $\bigcirc$, Interested $\bigcirc$, Very Interested $\bigcirc$
              \item \textbf{Directly Manipulate Shape Attributes on Canvas} Very Disinterested $\bigcirc$, Disinterested $\bigcirc$, Neutral $\bigcirc$, Interested $\bigcirc$, Very Interested $\bigcirc$

          \end{itemize}

    \item \textbf{Course Experience}. Would any of these features have made the course easier for you? Which would you have ignored? Would any of them have helped you learn to program more easily?

    \item \textbf{Challenges}. In the first part of the survey we asked you to think about the assignment that gave you the most difficulty. Thinking about that assignment again, please describe in detail how some of these features may have changed your experience with that particular assignment.

    \item \textbf{Course Project}. Thinking about your course project, do you think any of these editor enhancements would have helped you reach your goals either more quickly or more expressively? Why or why not? Are there any particular editor features that would have made the process easier?

    \item \textbf{Creativity Tools}. If you have experience with creativity tools, such as Illustrator or Photoshop, are there any features you'd like to have as part of a coding editor? If so, what are they and how would they help you accomplish your goals?

    \item \textbf{Suggestions}. Do you have any ideas for editor features (beyond those you might have suggested in other places in this survey)?

    \item \textbf{Miscellaneous}. Is there anything else you would like us to know? Any additional feedback you’d like to share about the editor, or any other technical aspect of the course?

\end{enumerate}

\subsection{Page 20: Parting Questions}
We've now reached the end of the survey! Thanks for participating! Just two final questions.

\begin{enumerate}

    \item What is your CNetID? (It's the thing at the beginning of your @uchicago.edu email address.)

    \item What email address would you like your Amazon gift card delivered to?

\end{enumerate}
 \clearpage
\section{Survey Instrument for Follow-up Survey --- Year 2 (\protect\wiq{}, \protect\sumtwo{})}\label{appendix:survey-2}

\subsection{Page 1: Consent to Participate in a Research Study}

Study Number: IRB22-0158 \\
Study Title: Post-Course Survey of Students CS11111 (Winter 2022)\\
Researcher: Ravi Chugh\\
Graduate Student: Andrew McNutt

\textbf{Description}:
We are researchers at the University of Chicago doing a research study about the usability of editors for creative coding. In CS11111 WI22 we used an in-browser editor that featured a number of enhancements and augmentations to the publicly available p5 editor. We are interested in understanding what editor features are useful for someone learning to code (particularly in the context of a creative coding course) or otherwise making digital art works. To facilitate this inquiry we are asking you to complete a survey. This survey does not include questions about personal or sensitive information.  Participation should take about 10-20 minutes. Your participation is voluntary. You are eligible to participate in this survey because you are enrolled in CS11111 WI22, which is the sole criteria for eligibility.

\textbf{Incentives}:
In return for your participation, you will receive a small amount of extra credit, roughly equivalent to 1 exercise (~1%

\textbf{Risks and Benefits}:
The risks to your participation in the survey are those associated with basic computer tasks, including boredom, fatigue, or mild stress. The only benefit to you is the learning experience from participating in a research study. The benefit to society is the contribution to scientific knowledge.

\textbf{Confidentiality}:
Ultimately, this research may be published and presented at scientific conferences to improve the community's knowledge about editors for creative coding, and may be used to improve the editor used in future iterations of our course. Any reports and presentations about the findings from this study will not include your name or any other information that could identify you. If you decide to withdraw from this study, any data already collected will be destroyed.

\textbf{Use of Your Research Data}:
We will never share the data beyond the University of Chicago research team. However, an analysis of the data may be analyzed and published in scientific conference proceedings or journal articles. The free-text responses you provide to any portion of this survey may be quoted in part or in whole in this publication. We will remove any information from the analysis that could identify you before providing the analysis for publication.

\textbf{Voluntary Participation and Right to Refuse or Withdraw}:
Participation in this study is voluntary. The decision to participate in this study is entirely up to you. You may refuse to take part in the study at any time without prejudice or penalties to you and will not result in any loss of benefits to which you are otherwise entitled.

\textbf{Contact Information for Research Questions and Participation}:
If you have questions or concerns about the study, you can contact the researchers at

\noindent{}Principal Investigator\\
Ravi Chugh,\\
Associate Professor\\
John Crerar Library\\
University of Chicago\\
5730 S Ellis Ave\\
Chicago, IL 60637\\
Email: rchugh@uchicago.edu

\noindent{}Graduate Student\\
Andrew McNutt,\\
PhD student\\
John Crerar Library\\
University of Chicago\\
5730 S Ellis Ave\\
Chicago, IL 60637\\
Email: mcnutt@uchicago.edu

If you have any questions about your rights as a participant in this research, feel you have been harmed, or wish to discuss other study-related concerns with someone who is not part of the research team, you can contact the University of Chicago Institutional Review Board (IRB) Office by phone at (773) 702-2915, or by email at sbs-irb@uchicago.edu.

\textbf{Consent}:
Participation is voluntary. Refusal to participate or withdrawing from the research will involve no penalty or loss of benefits to which you might otherwise be entitled.  By clicking “Agree” below, you confirm that you have read the consent form, are at least 18 years old, and agree to participate in the research. Please print or save a copy of this page for your records.

\begin{enumerate}
    \item Full name
    \item GitHub username (same as used for homework submission in this class)
    \item Student Id (the username before your @uchicago.edu email address)
    \item I am a student, 18 years of age or older, I have read and understood this consent form, and give consent to participate in this study. Yes $\bigcirc$ no $\bigcirc$
\end{enumerate}

\newpage

\subsection{Page 2: Feature Questions}

The editor we used in the course included several notable features. We are interested in understanding your use and perception of these features.

\begin{enumerate}[topsep=7pt,itemsep=7pt]
    \item \textbf{Feature: Color Pickers}

          \begin{figure}[h!]
              \centering
              \includegraphics[width=0.6\linewidth]{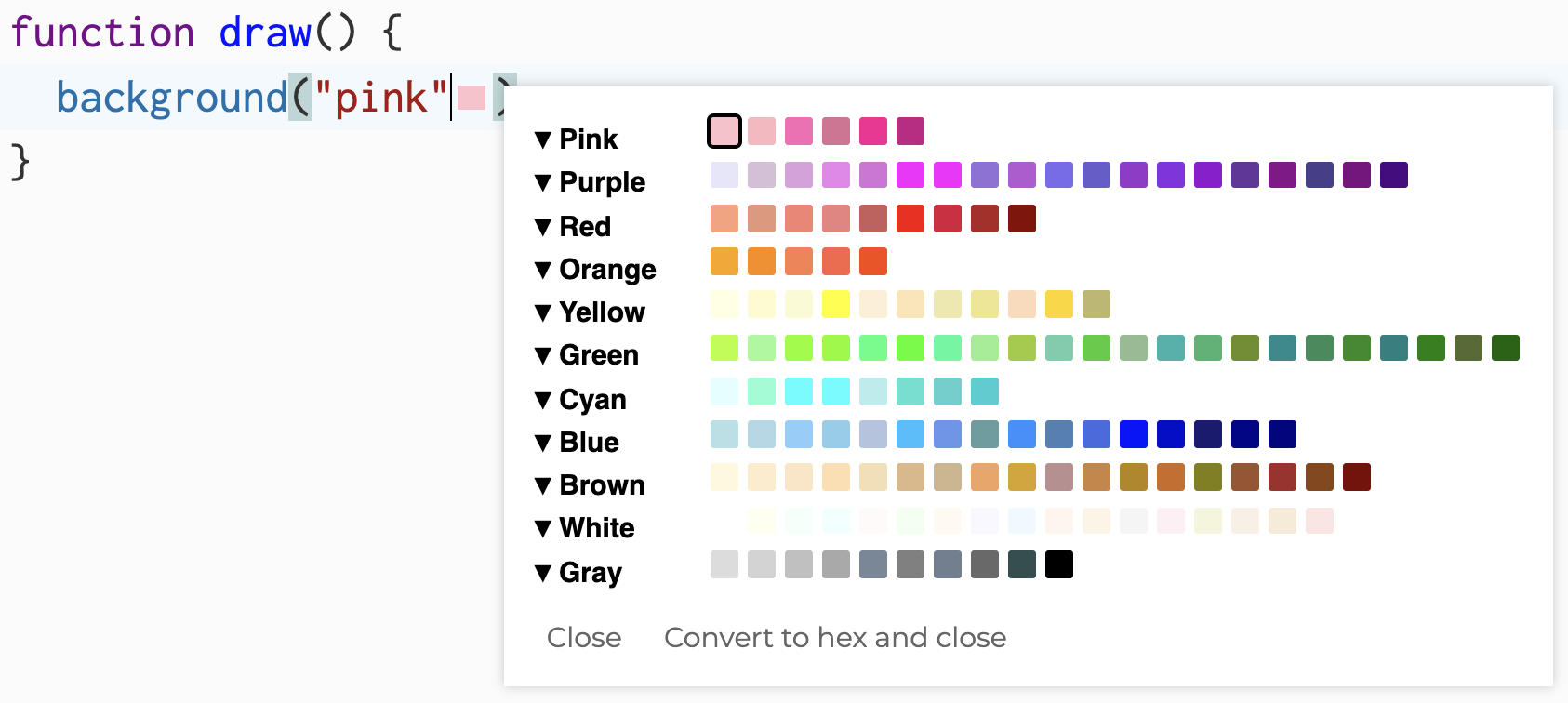}
          \end{figure}

          \begin{enumerate}[topsep=0pt,itemsep=0pt]
              \item How often did you use Color Pickers? $\bigcirc$ Never, $\bigcirc$ Once in a while, $\bigcirc$ Occasionally, $\bigcirc$ Frequently, $\bigcirc$ All the time

              \item Do you think Color Pickers are useful? $\bigcirc$ Not very useful, $\bigcirc$ Not useful, $\bigcirc$ Neither useful nor unuseful, $\bigcirc$ Useful, $\bigcirc$ Very Useful

              \item Do you have any comments about Color Pickers? For instance: How did you feel they affected your learning? Is there any way you would modify them to make them more useful?

          \end{enumerate}

    \item \textbf{Feature: Number Pickers}

          \begin{figure}[h!]
              \centering
              \includegraphics[width=0.4\linewidth]{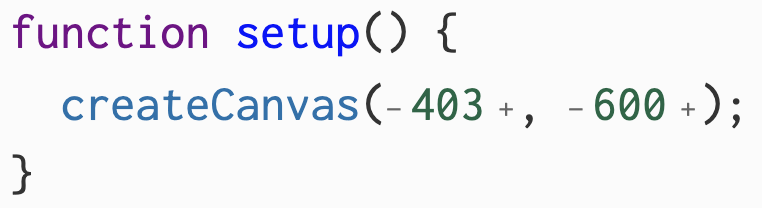}
          \end{figure}

          \begin{enumerate}[topsep=0pt,itemsep=0pt]
              \item How often did you use Number Pickers? $\bigcirc$ Never, $\bigcirc$ Once in a while, $\bigcirc$ Occasionally, $\bigcirc$ Frequently, $\bigcirc$ All the time

              \item Do you think Number Pickers are useful? $\bigcirc$ Not very useful, $\bigcirc$ Not useful, $\bigcirc$ Neither useful nor unuseful, $\bigcirc$ Useful, $\bigcirc$ Very Useful

              \item Do you have any comments about Number Pickers? For instance: How did you feel they affected your learning? Is there any way you would modify them to make them more useful?

          \end{enumerate}

    \item \textbf{Feature: Number Sliders}

          \begin{figure}[h!]
              \centering
              \includegraphics[width=0.6\linewidth]{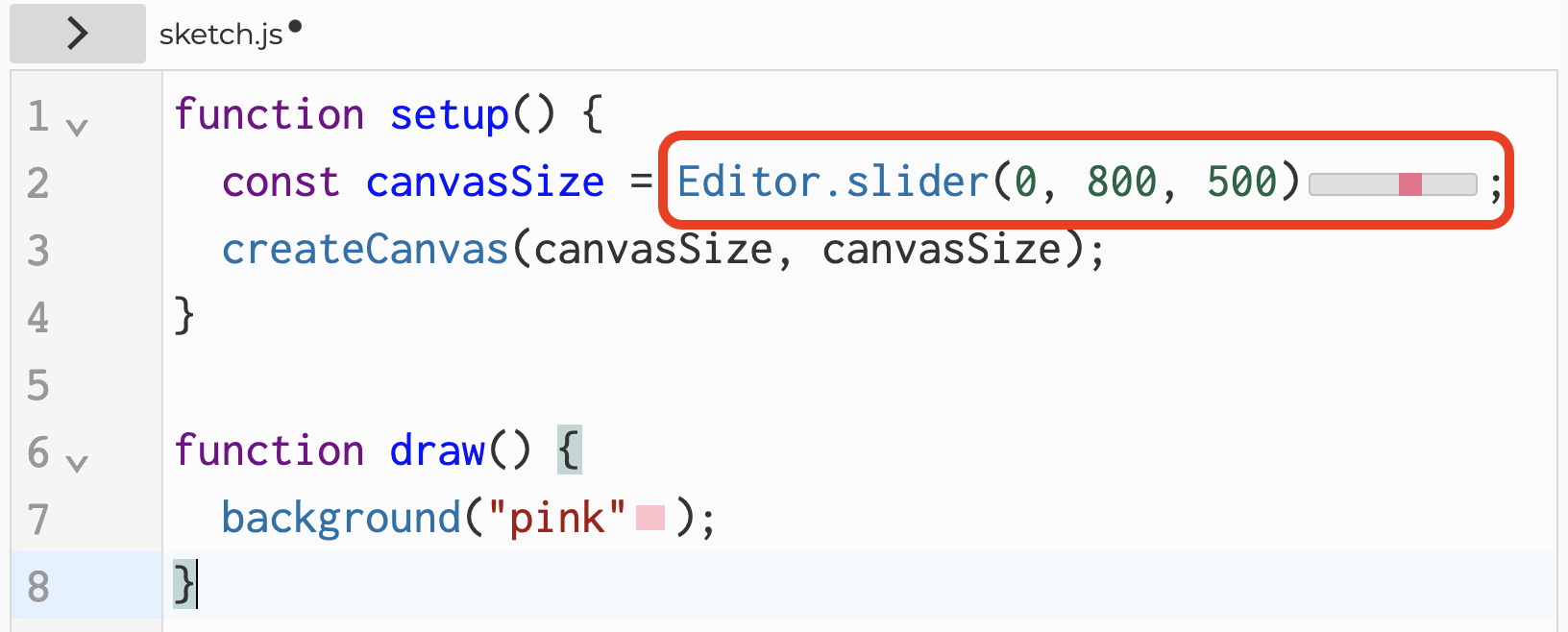}
          \end{figure}

          \begin{enumerate}[topsep=0pt,itemsep=0pt]
              \item How often did you use Number Sliders? $\bigcirc$ Never, $\bigcirc$ Once in a while, $\bigcirc$ Occasionally, $\bigcirc$ Frequently, $\bigcirc$ All the time

              \item Do you think Number Sliders are useful? $\bigcirc$ Not very useful, $\bigcirc$ Not useful, $\bigcirc$ Neither useful nor unuseful, $\bigcirc$ Useful, $\bigcirc$ Very Useful

              \item Do you have any comments about Number Sliders? For instance: How did you feel they affected your learning? Is there any way you would modify them to make them more useful?

          \end{enumerate}

          \vfill\pagebreak
    \item \textbf{Feature: Linting}

          \begin{figure}[h!]
              \centering
              \includegraphics[width=0.4\linewidth]{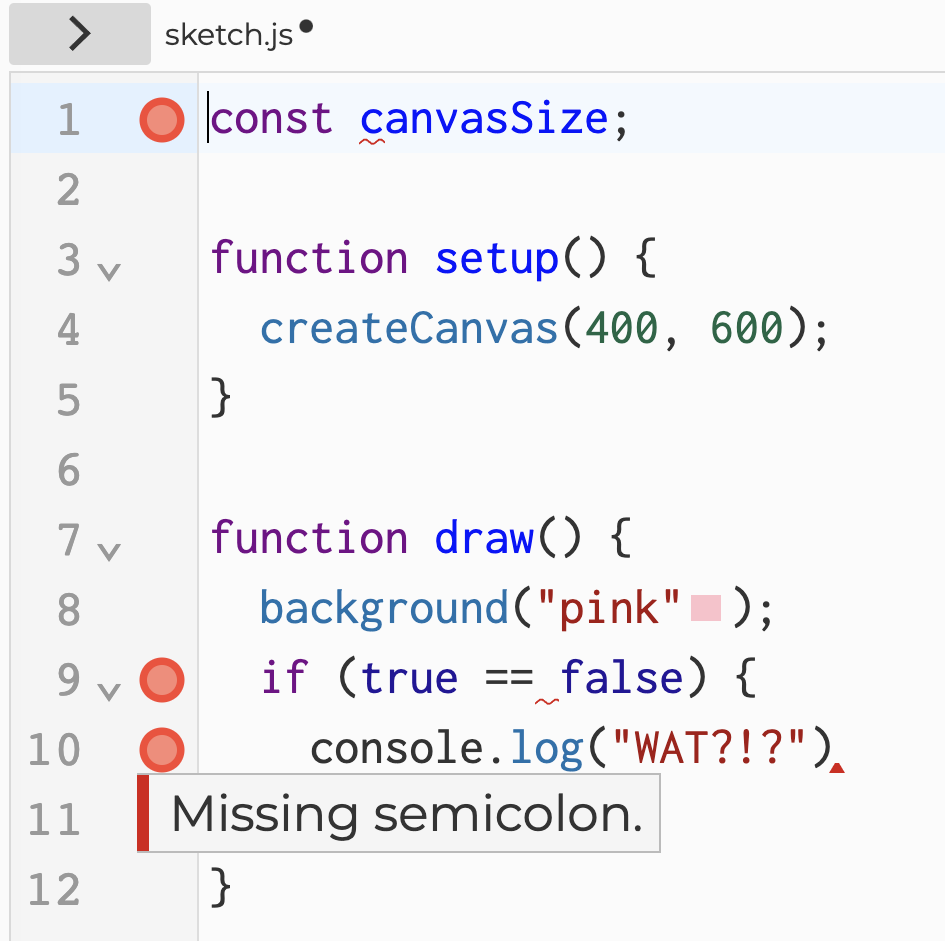}
          \end{figure}

          \begin{enumerate}[topsep=0pt,itemsep=0pt]
              \item How often did you use Linting? $\bigcirc$ Never, $\bigcirc$ Once in a while, $\bigcirc$ Occasionally, $\bigcirc$ Frequently, $\bigcirc$ All the time

              \item Do you think Linting is useful? $\bigcirc$ Not very useful, $\bigcirc$ Not useful, $\bigcirc$ Neither useful nor unuseful, $\bigcirc$ Useful, $\bigcirc$ Very Useful

              \item Do you have any comments about Linting? For instance: How did you feel it affected your learning? Is there any way you would modify it to make it more useful?

          \end{enumerate}

    \item \textbf{Feature: Tidy Code}

          \begin{figure}[h!]
              \centering
              \includegraphics[width=0.7\linewidth]{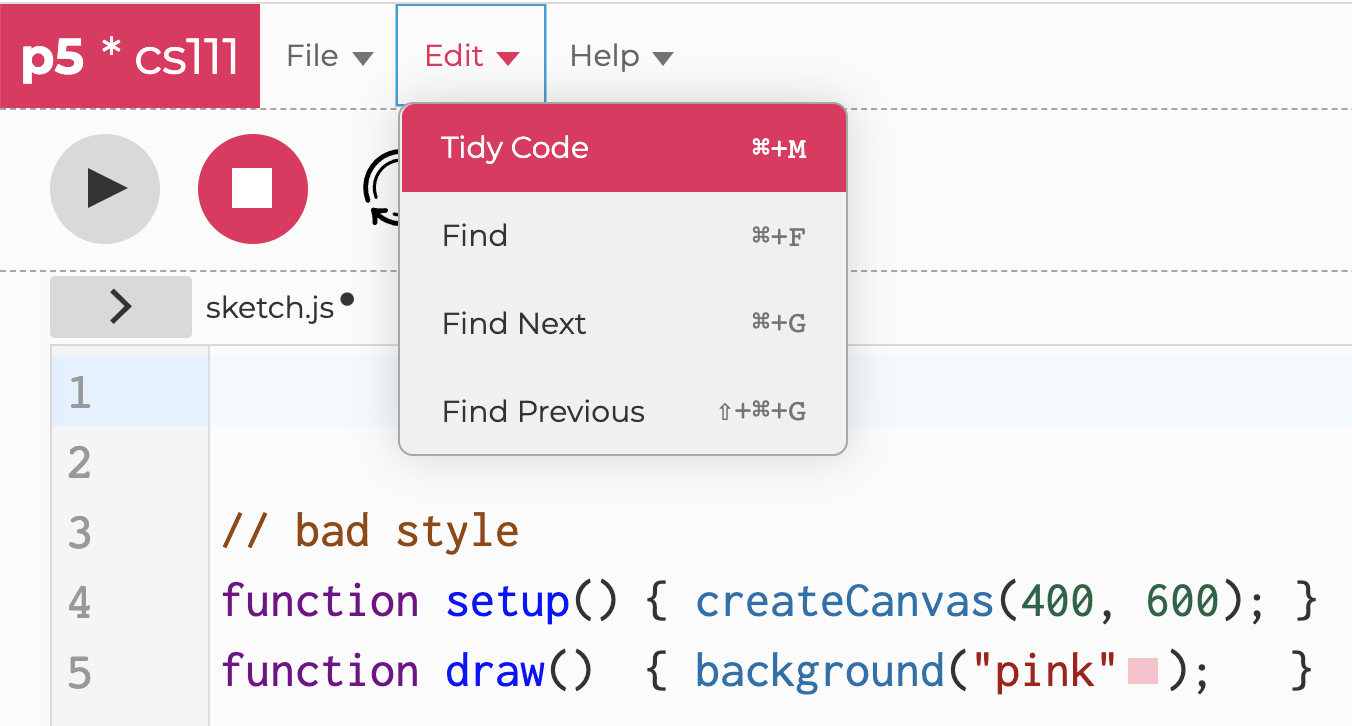}
          \end{figure}

          \begin{enumerate}[topsep=0pt,itemsep=0pt]
              \item How often did you use Tidy Code? $\bigcirc$ Never, $\bigcirc$ Once in a while, $\bigcirc$ Occasionally, $\bigcirc$ Frequently, $\bigcirc$ All the time

              \item Do you think Tidy Code is useful? $\bigcirc$ Not very useful, $\bigcirc$ Not useful, $\bigcirc$ Neither useful nor unuseful, $\bigcirc$ Useful, $\bigcirc$ Very Useful

              \item Do you have any comments about Tidy Code? For instance: How did you feel it affected your learning? Is there any way you would modify it to make it more useful?

          \end{enumerate}

          \vfill\pagebreak
    \item \textbf{Feature: Auto-Refresh}

          \begin{figure}[h!]
              \centering
              \includegraphics[width=0.3\linewidth]{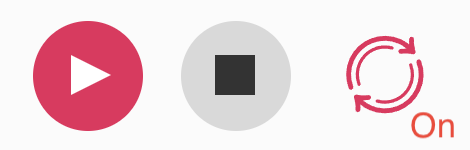}
          \end{figure}

          \begin{enumerate}[topsep=0pt,itemsep=0pt]
              \item How often did you use Auto-Refresh? $\bigcirc$ Never, $\bigcirc$ Once in a while, $\bigcirc$ Occasionally, $\bigcirc$ Frequently, $\bigcirc$ All the time

              \item Do you think Auto-Refresh is useful? $\bigcirc$ Not very useful, $\bigcirc$ Not useful, $\bigcirc$ Neither useful nor unuseful, $\bigcirc$ Useful, $\bigcirc$ Very Useful

              \item Do you have any comments about Auto-Refresh? For instance: How did you feel it affected your learning? Is there any way you would modify it to make it more useful?

          \end{enumerate}

    \item \textbf{Feature: Shape Toolbox}

          \begin{figure}[h!]
              \centering
              \includegraphics[width=0.7\linewidth]{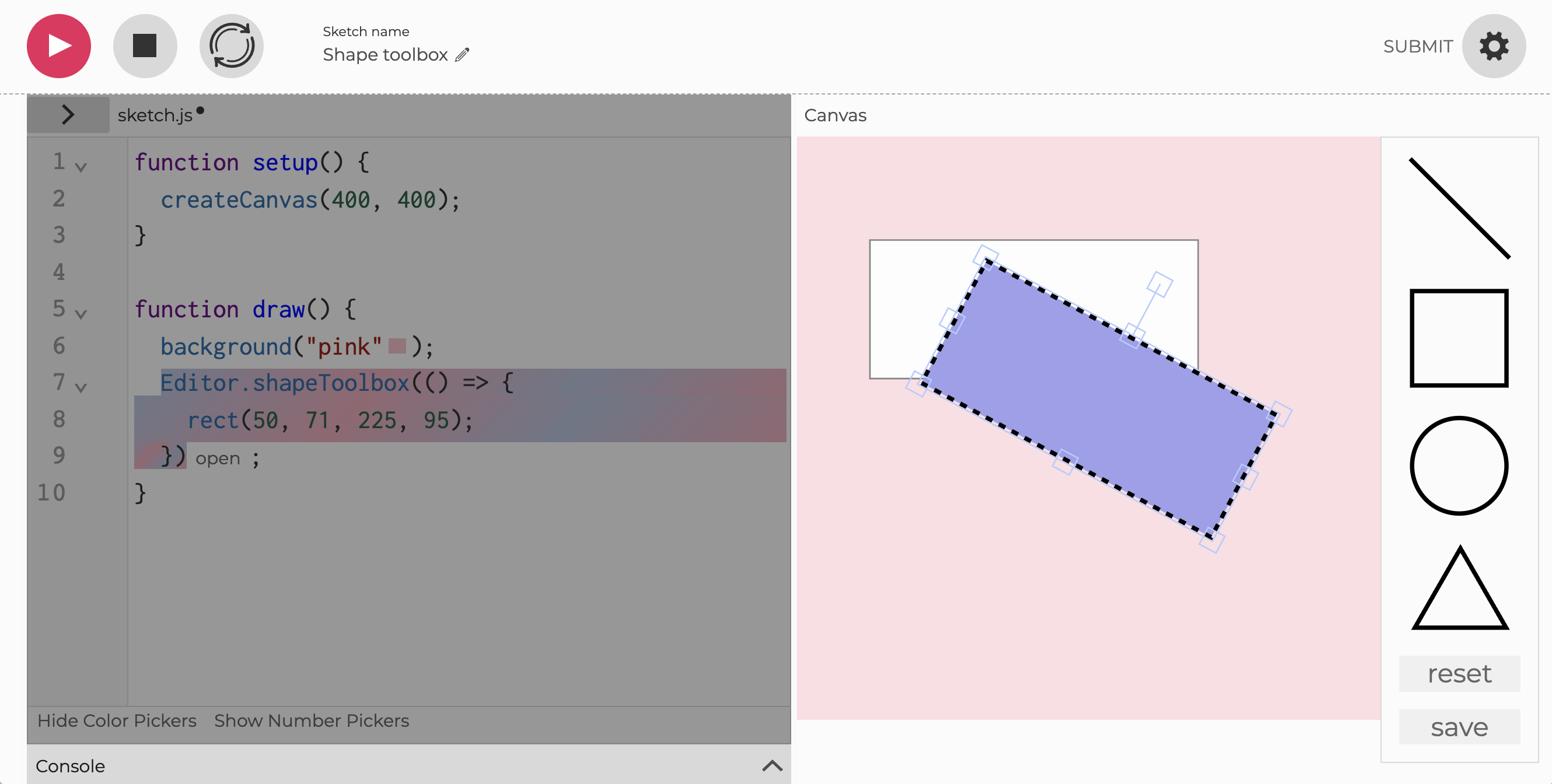}
          \end{figure}

          \begin{enumerate}[topsep=0pt,itemsep=0pt]
              \item How often did you use Shape Toolbox? $\bigcirc$ Never, $\bigcirc$ Once in a while, $\bigcirc$ Occasionally, $\bigcirc$ Frequently, $\bigcirc$ All the time

              \item Do you think Shape Toolbox is useful? $\bigcirc$ Not very useful, $\bigcirc$ Not useful, $\bigcirc$ Neither useful nor unuseful, $\bigcirc$ Useful, $\bigcirc$ Very Useful

              \item Do you have any comments about Shape Toolbox? For instance: How did you feel it affected your learning? Is there any way you would modify it to make it more useful?

          \end{enumerate}

    \item \textbf{Feature: Autocomplete}

          \begin{figure}[h!]
              \centering
              \includegraphics[width=0.5\linewidth]{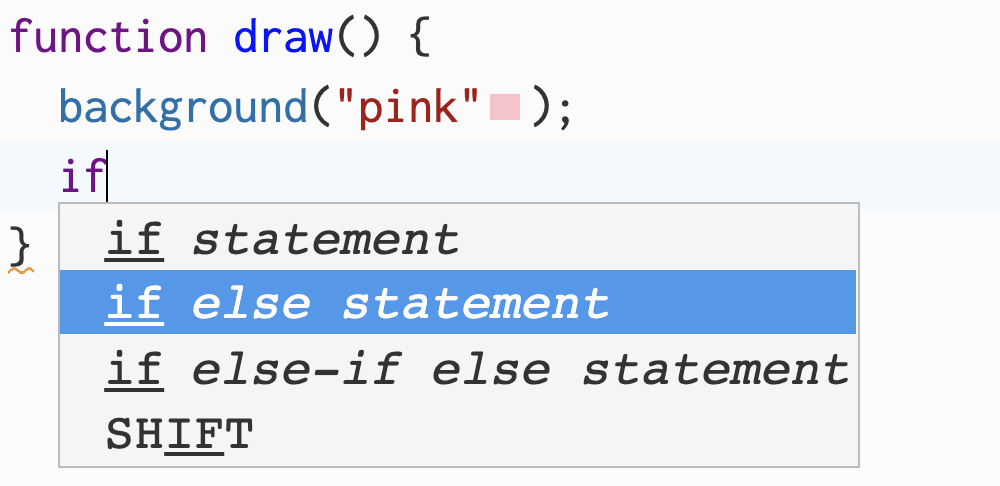}
          \end{figure}

          \begin{enumerate}[topsep=0pt,itemsep=0pt]
              \item How often did you use Autocomplete? $\bigcirc$ Never, $\bigcirc$ Once in a while, $\bigcirc$ Occasionally, $\bigcirc$ Frequently, $\bigcirc$ All the time

              \item Do you think Autocomplete is useful? $\bigcirc$ Not very useful, $\bigcirc$ Not useful, $\bigcirc$ Neither useful nor unuseful, $\bigcirc$ Useful, $\bigcirc$ Very Useful

              \item Do you have any comments about Autocomplete? For instance: How did you feel it affected your learning? Is there any way you would modify it to make it more useful?

          \end{enumerate}

\end{enumerate}

\subsection{Page: Reflection Questions}

Next, we'd like you to reflect on a few additional aspects of the course, how they might be improved, and ways in which the editor might be modified to meet those challenges.

\begin{enumerate}[topsep=7pt,itemsep=7pt]
    \item \textbf{Feature Review}  Reflect on the features we've just been considering. Which of the features we considered are you most excited about?

          \begin{itemize}[topsep=7pt,itemsep=0pt]
              \item \textbf{Color Pickers} Very Disinterested $\bigcirc$, Disinterested $\bigcirc$, Neutral $\bigcirc$, Interested $\bigcirc$, Very Interested $\bigcirc$
              \item \textbf{Number Pickers} Very Disinterested $\bigcirc$, Disinterested $\bigcirc$, Neutral $\bigcirc$, Interested $\bigcirc$, Very Interested $\bigcirc$
              \item \textbf{Number Sliders} Very Disinterested $\bigcirc$, Disinterested $\bigcirc$, Neutral $\bigcirc$, Interested $\bigcirc$, Very Interested $\bigcirc$
              \item \textbf{Linting} Very Disinterested $\bigcirc$, Disinterested $\bigcirc$, Neutral $\bigcirc$, Interested $\bigcirc$, Very Interested $\bigcirc$
              \item \textbf{Tidy Code} Very Disinterested $\bigcirc$, Disinterested $\bigcirc$, Neutral $\bigcirc$, Interested $\bigcirc$, Very Interested $\bigcirc$
              \item \textbf{Auto-Refresh} Very Disinterested $\bigcirc$, Disinterested $\bigcirc$, Neutral $\bigcirc$, Interested $\bigcirc$, Very Interested $\bigcirc$
              \item \textbf{Shape Toolbox} Very Disinterested $\bigcirc$, Disinterested $\bigcirc$, Neutral $\bigcirc$, Interested $\bigcirc$, Very Interested $\bigcirc$
              \item \textbf{Autocomplete} Very Disinterested $\bigcirc$, Disinterested $\bigcirc$, Neutral $\bigcirc$, Interested $\bigcirc$, Very Interested $\bigcirc$
          \end{itemize}

    \item \textbf{Effect on Learning} While it is hard to compare with something you didn’t do, how do you think your experience in the course would have been different had these features not been part of the editor? Do you feel you would have learned more or less than you did?

    \item \textbf{Art Tools} Now that you've learned some programming for creative coding, how does that affect your perspective of art making? How might a code editor help or hinder the art making process?
    \item \textbf{Challenges} What aspect of coding or learning to program gave you the most trouble? As a way to help organize your thinking, consider the assignment that you had the most difficulty with. Could the editor have done anything to help you with that?
    \item \textbf{External Tools} It's natural to use other tools as part of the programming process, such as color eye droppers or p5's online documentation. Do you think it would be useful to integrate these tools as part of the editor?  What other tools can you imagine wanting to be part of your in-editor coding workflow?
    \item \textbf{Desired Features} What sorts of editor features might have allowed you to be more effective in your coding? What sorts of editor features might have allowed you to be more creative?
    \item \textbf{Miscellaneous} Is there anything else you would like us to know? Any additional feedback you’d like to share about the editor, or any other technical aspect of the course?
\end{enumerate}
  
\end{document}